\documentclass[sigplan,screen, nonacm]{acmart}

\usepackage{microtype}
\usepackage{subcaption}
\usepackage{mathtools}
\usepackage{amsthm}
\usepackage{soul}
\usepackage{xspace}
\usepackage{enumitem}
\usepackage{float}
\usepackage{placeins}           
\usepackage{cuted}             
\usepackage{caption}            
\usepackage[textsize=tiny]{todonotes} 
\usepackage{algorithm}
\usepackage{algorithmic}
\usepackage{afterpage}
\usepackage{booktabs}
\usepackage{siunitx}

\sethlcolor{yellow}

\newcommand{\delete}[1]{}                     
          
\newcommand\zxz[1]{{\color{black}#1}}        
\newcommand\NAME{Kareto\xspace}             

\theoremstyle{plain}

\theoremstyle{definition}

\theoremstyle{remark}

\AtBeginDocument{%
  }

\setcopyright{none}
\begin{document}

\title{Adaptive Multi-Objective Tiered Storage Configuration for KV Cache in LLM Service}

\author{%
Xianzhe Zheng\textsuperscript{1},
Zhengheng Wang\textsuperscript{2},
Ruiyan Ma\textsuperscript{2},
Rui Wang\textsuperscript{1},
Xiyu Wang\textsuperscript{2},
Rui Chen\textsuperscript{2},
Peng Zhang\textsuperscript{2},
Sicheng Pan\textsuperscript{2},
Zhangheng Huang\textsuperscript{2},
Chenxin Wu\textsuperscript{2},
Yi Zhang\textsuperscript{2},
Bo Cai\textsuperscript{2},
Kan Liu\textsuperscript{2},
Teng Ma\textsuperscript{2},
Yin Du\textsuperscript{2},
Dong Deng\textsuperscript{2},
Sai Wu\textsuperscript{1},
Guoyun Zhu\textsuperscript{2},
Wei Zhang\textsuperscript{2},
Feifei Li\textsuperscript{2}\\[6pt]
\normalsize
\begin{tabular}{@{}c@{\hspace{1.2em}}c@{}}
\textsuperscript{1}Zhejiang University, Hangzhou, China &
\textsuperscript{2}Alibaba Group, Hangzhou, China
\end{tabular}
}

\renewcommand{\shortauthors}{Zheng, Wang, Ma, Wang, et al.}


\begin{abstract}
The memory-for-computation paradigm of KV caching is essential for accelerating large language model (LLM) inference service, but limited GPU high-bandwidth memory (HBM) capacity motivates offloading the KV cache to cheaper external storage tiers. 
While this expands capacity, it introduces the challenge of dynamically managing heterogeneous storage resources to balance cost, throughput, and latency under varying workloads. 
We formulate this as a multi-objective optimization problem: identifying the Pareto frontier across these metrics within the storage configuration space. 
Using a high-fidelity end-to-end simulator, we observe that the objective functions are non-analytic and exhibit complex variable coupling, making the Pareto frontier difficult to approximate analytically. 
To obtain the frontier, we introduce Kareto, a \underline{K}V-cache \underline{A}daptive \underline{RE}source managemen\underline{T} \underline{O}ptimizer.
Kareto leverages a diminishing-return-guided pruning method to efficiently navigate the large configuration space and approximate the Pareto frontier. 
Additionally, it incorporates a fine-grained adaptive tuner that uses eviction policies in tier storage and KV block access patterns for group-specific cache management, improving cache efficiency. 
Experiments on real-world traces show that Kareto adapts to workload and can identify configurations of better cost efficiency, covering static strategies. Compared to the fixed setup with 1024 GB DRAM, Kareto can improve throughput by up to 9.3\%, or reduce latency by up to 58.3\%, or lower cost by up to 20.2\% under respective optimization objectives.
\end{abstract}

\maketitle

\section{Introduction}

The widespread deployment of large language models (LLMs) \cite{attention, gpt3, instruct-gpt, codex, agent, text} has made efficient inference serving of high throughput and low latency a critical challenge~\cite{efficient}. 
A key acceleration technique is KV caching, which stores previous tokens' key and value vectors to avoid redundant computations during autoregressive generation, representing a {\em memory-for-computation paradigm} trade-off. 
However, the memory footprint of the KV cache grows linearly with concurrent requests and sequence length, often exceeding GPU high-bandwidth memory (HBM) capacity and creating a significant bottleneck.

A common solution is to use a tiered storage architecture, offloading the cache to host DRAM or disk to expand capacity~\cite{hicache, lmcache, mooncake, nvidia-dynamo, memserve}.
While systems like vLLM~\cite{vllm} and SGLang~\cite{sglang}, along with specialized frameworks, e.g., Mooncake~\cite{mooncake}, LMCache~\cite{lmcache}, MemServe~\cite{memserve}, have implemented this paradigm, they primarily focus on cache replacement policies for a fixed capacity. 
A fundamental yet often overlooked challenge is the effective capacity configuration across these tiers. 
Simply expanding storage does not always yield performance gains; beyond a certain point, storage cost can outweigh computation savings, leading to inefficient resource utilization.

Existing approaches are typically suboptimal. 
Some mainly focus on a fixed allocation (e.g., provisioning 1 TB of DRAM per 8 GPUs)~\cite{vllm, orca, cached-attention, sglang, memserve, mooncake, lmcache}, which often results in over-provisioning or under-provisioning and cannot adapt to dynamic workload characteristics. 
Mainstream cloud platforms offer user-configured storages~\cite{aws, aliyun, google-cloud}, but this places a high burden on users, requiring deep expertise to analyze workload patterns and understand complex performance-cost trade-offs across heterogeneous storage tiers ~\cite{storage-configure-challenge-for-user}. 
Even with this expertise, achieving an optimal setup is exceedingly difficult without precise tools to analyze the multi-dimensional trade-offs between cost, latency, and throughput.
Crucially, both models depend on a statically predefined capacity that cannot dynamically respond to changing workload patterns. 
Consequently, the lack of an intelligent, adaptive configuration mechanism fundamentally limits the ability of existing tiered storage solutions to balance these key trade-offs optimally in production environments.

To address these limitations, we revisit KV cache configuration and management through the lens of elastic, adaptive storage provisioning. 
We aim to dynamically determine cost-effective storage configurations across GPU HBM, host DRAM, and disk tiers that adapt to the performance requirements and KV cache reuse patterns of real-world workloads. 
Through trace-driven analysis of production LLM serving workloads, we derived key insights into the relationships between cache policy, storage tiering, and performance.

Informed by these findings, we propose \NAME, an adaptive framework for KV cache storage management. 
\NAME shifts the paradigm from static provisioning to an elastic, workload-aware orchestration system. 
By leveraging simulation-driven Pareto optimization and fine-grained adaptive cache management, \NAME autonomously navigates the complex trade-off space between cost, throughput, and latency to achieve superior operational efficiency in LLM inference serving.
Our key contributions are as follows:

\begin{itemize}
    \item We formalize the throughput-latency-cost tradeoff as a Pareto optimization problem, demonstrating its lack of closed-form solutions and non-convexity, necessitating simulation-based optimization.
    
    \item We develop a high-fidelity, end-to-end simulator that replays historical traces to accurately model inference engine behavior under arbitrary configurations, enabling efficient workload analysis and insight into cache policy and storage tiering effects.

    \item We introduce an adaptive search strategy that detects diminishing marginal returns to reduce the configuration search space while improving sensitivity to critical regions, aiding in Pareto frontier identification. 

    \item We propose a fine-grained adaptive tuner that uses prefix tree analysis to assign group-specific TTL configurations, enhancing cache efficiency and storage utilization compared to one-size-fits-all strategies.

    \item We implement and evaluate \NAME using real-world traces, demonstrating its ability to automatically adapt to different workload patterns and user requirements,  achieving superior resource utilization.
\end{itemize}

\section{Background}

\subsection{KV Cache Management in LLM Inference Services}
Modern LLM serving systems predominantly employ decoder-only Transformers~\cite{attention}, where inference proceeds in two phases: (1) \textit{prefill}, processing the full input prompt in parallel to generate the first token; and (2) \textit{decode}, autoregressively generating subsequent tokens one at a time. 
During decode, each new token depends on all prior tokens. To avoid redundant computation, key (K) and value (V) vectors from each decoder layer are cached across steps, known as the KV Cache~\cite{vllm}. 
This caching mechanism significantly reduces latency but consumes substantial memory.
KV Cache reuse is also common across different requests due to shared prefixes, such as in multi-turn dialogues where the request of subsequent turns is concatenated after full conversation history, or in the case that a fixed system prompt is prepended to user inputs~\cite{prompt-openai, prompt-tuning, prompt-tuning2}. 
Serving systems can retain the KV Cache of completed requests to accelerate future inferences by reusing cached K/V vectors from identical prefixes.

However, the KV cache is a major memory bottleneck in production environments. 
Per-token KVCache may occupy several to hundreds of kilobytes, and the total KV cache size grows linearly with sequence length and batch size, and under high concurrency or long-context workloads, it often exceeds GPU memory capacity~\cite{vllm, orca}. 
To address this, modern systems employ a tiered storage hierarchy, utilizing GPU memory for active cache blocks, CPU DRAM for less frequently accessed data, and even local disk or remote storage (e.g., Redis) for cold blocks~\cite{cached-attention, hicache, memserve, mooncake, lmcache}. 
This architecture balances cost, capacity, and access latency. 
Cache blocks are dynamically managed across tiers using eviction policies like LRU, aiming to expand effective capacity and improve cache hit rates while managing access costs~\cite{pensieve, kvcache-wild, continuum}.

\subsection{Challenges in KVCache Configuration}

Existing KV cache optimization research has largely focused on refining eviction and scheduling policies under a fixed, pre-configured storage budget~\cite{orca, cached-attention, sglang, kvcache-wild, pensieve}, typically set either via a default cluster setting or through manual user specification.
While effective for managing a given capacity, this static paradigm fails to address the dynamic nature of real-world LLM serving workloads, which exhibit significant intra-day and weekly variations in request patterns~\cite{burst-gpt, kvcache-wild}.

The inherent diversity of workloads, characterized by fluctuating concurrency, varying prompt/response lengths, and different prefix-sharing ratios~\cite{pensieve}, further complicates capacity planning. 
Current systems place the burden on users to manually configure capacities, a process requiring deep expertise and often resulting in costly over-provisioning or performance-degrading under-provisioning.

Moreover, a critical limitation is the treatment of storage as a homogeneous resource. 
Modern infrastructure offers a hierarchy of storage tiers, e.g., GPU HBM, host DRAM, SSD, each with distinct cost and bandwidth profiles. 
However, existing KV cache management systems lack the intelligence to dynamically adjust both the size and distribution across these economically heterogeneous tiers for better cost-performance trade-offs.
Consequently, while current methods excel at KV Cache management within a fixed budget, they fall short in setting the budget in response to evolving workload demands and infrastructure economics, leaving substantial gains in resource efficiency untapped.

\newcommand{\E}{\mathbb{E}}          
\newcommand{\ind}{\mathbb{I}}        
\newcommand{\cost}{\mathcal{C}}      
\newcommand{\latency}{\mathcal{L}}   
\newcommand{\throughput}{\mathcal{T}} 
\newcommand{\hw}{\text{hw}}          
\newcommand{\io}{\text{I/O}}         
\newcommand{\eff}{\text{eff}}        
\newcommand{\smax}{s^*}              
\newcommand{\workload}{\mathcal{D}_r} 
\newcommand{\policymap}{\epsilon}    

\section{Cost-Aware KVCache Optimization: Problem Formulations}
\label{sec:problem_formulation}

\subsection{Beyond Performance: The Tripartite Optimization Landscape}
\label{subsec:tripartite}

\subsubsection{Decomposition of Performance Factors}

Prior work on KV cache optimization has predominantly focused on \textit{latency} and \textit{throughput} under fixed hardware constraints. 
We pioneer a holistic optimization framework that explicitly incorporates \textit{economic cost} as a critical objective, establishing a Pareto frontier across latency, throughput, and operational expenditure (OPEX). 
This integration is critical for real-world deployment where LLM inference services on cloud platforms constitutes a significant operational budget. 

Formally, we define the optimization problem as:  
\begin{equation}
\min_{\mathbf{x} \in \mathcal{X}} \left( 
    \underbrace{\E[\latency(\mathbf{x})]}_{\text{Latency}}, 
    \underbrace{-\E[\throughput(\mathbf{x})]}_{\text{Throughput}}, 
    \underbrace{\cost(\mathbf{x})}_{\text{Cost}} 
\right)
\label{eq:optimization_problem}
\end{equation}
where $\mathbf{x} = [x_1, x_2, x_3, x_4]$ denotes the decision vector over domains $\mathcal{X}_i$, and $\cost(\mathbf{x})$ decomposes as:  
\begin{equation}
\cost(\mathbf{x}) = c_{\hw} \cdot \text{GPU-Hours}(\mathbf{x}) + \sum_{k \in \mathcal{K}} \phi_k(s_k(\mathbf{x}))
\label{eq:cost_decomposition}
\end{equation}
The first term of $\cost(\mathbf{x})$ captures \textit{compute expenditure}, scaled by hardware cost rate $c_{\hw}$ and GPU-hours consumed during execution. The second term aggregates \textit{resource-specific operational costs} across dimensions $k \in \mathcal{K}$ (e.g., memory bandwidth, network), where $s_k(\mathbf{x})$ denotes resource utilization and $\phi_k(\cdot)$ models nonlinear pricing effects (e.g., tiered billing, saturation penalties).   

We dissect the system into four core decision variables:  
\begin{itemize}[leftmargin=*, nosep]
    \item \textbf{X\textsubscript{1}: User workload Patterns:} Request distribution, sequence length variance, temporal locality
    \item \textbf{X\textsubscript{2}: Compute Configuration:} GPU architecture, inference engine scheduling, model parameters
    \item \textbf{X\textsubscript{3}: Storage Medium:} Latency profile, bandwidth limits, pricing tiers
    \item \textbf{X\textsubscript{4}: Storage Management Policy:} Eviction algorithms, TTL management, capacity allocation, access prediction
\end{itemize}

\vspace{\baselineskip}

\zxz{We identify three derived metrics that directly translate decision variables into inference efficiency and cost outcomes.}

\begin{itemize}
    \item \textbf{$X_5 = O_1(X_1,X_4)$: Dynamic Storage Capacity} \\
    The actual storage consumption resulting from the interaction between workload patterns ($X_1$) and storage policies ($X_4$). \autoref{fig:storage_capacity} demonstrates the significant impact of TTL optimization on storage requirements, comparing an oracle TTL policy (where TTL is set to zero if a KVCache block will never be accessed again) against conventional non-TTL approaches. 
    \begin{figure}[htbp]
  \centering
  \includegraphics[width=\columnwidth]{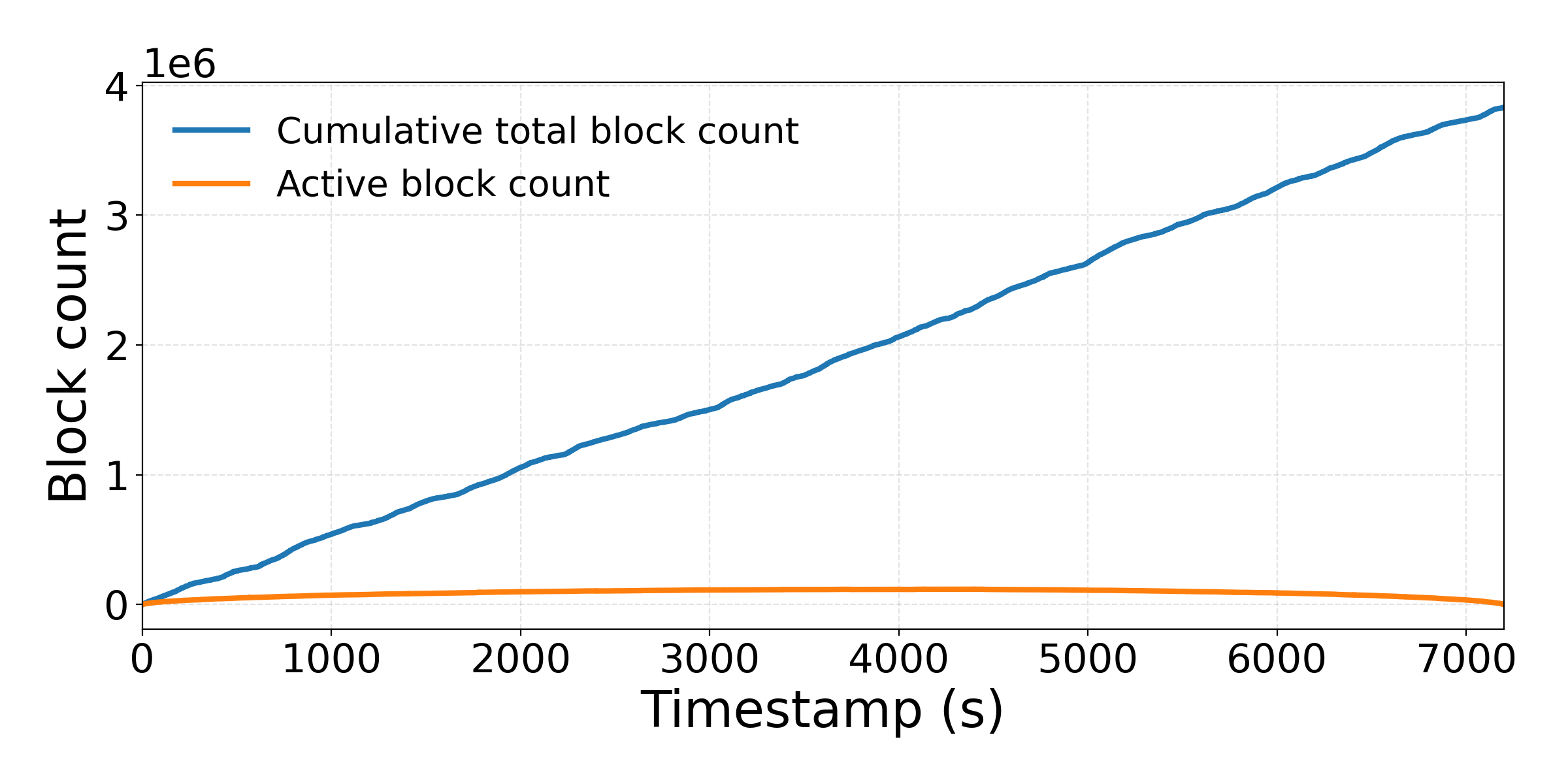}
  \caption{Cumulative and active block counts under the oracle TTL policy for traceB}
  \label{fig:storage_capacity}
\end{figure} 
    
    \item \textbf{$X_6 = O_2(X_1,X_3,X_4)$: Hit Rate} \\
    Analysis of KV reuse in production traces reveals highly skewed reuse patterns across tokens, where a small fraction of tokens contribute disproportionately to reuse counts. The degree of skewness varies significantly across workloads: 31.95\% of blocks account for 90\% of hits in Trace A, compared to only 0.67\% in Trace B (which contains more system prompts). This variation implies that the size of the effective working set dominating reuse is workload-dependent. Consequently, fixed-capacity provisioning without accounting for intrinsic reuse skew inevitably leads to either wasted resources or degraded performance.
\begin{figure}[!htbp]
\centering
\begin{subfigure}{0.48\linewidth}
    \centering
    \includegraphics[width=\linewidth]{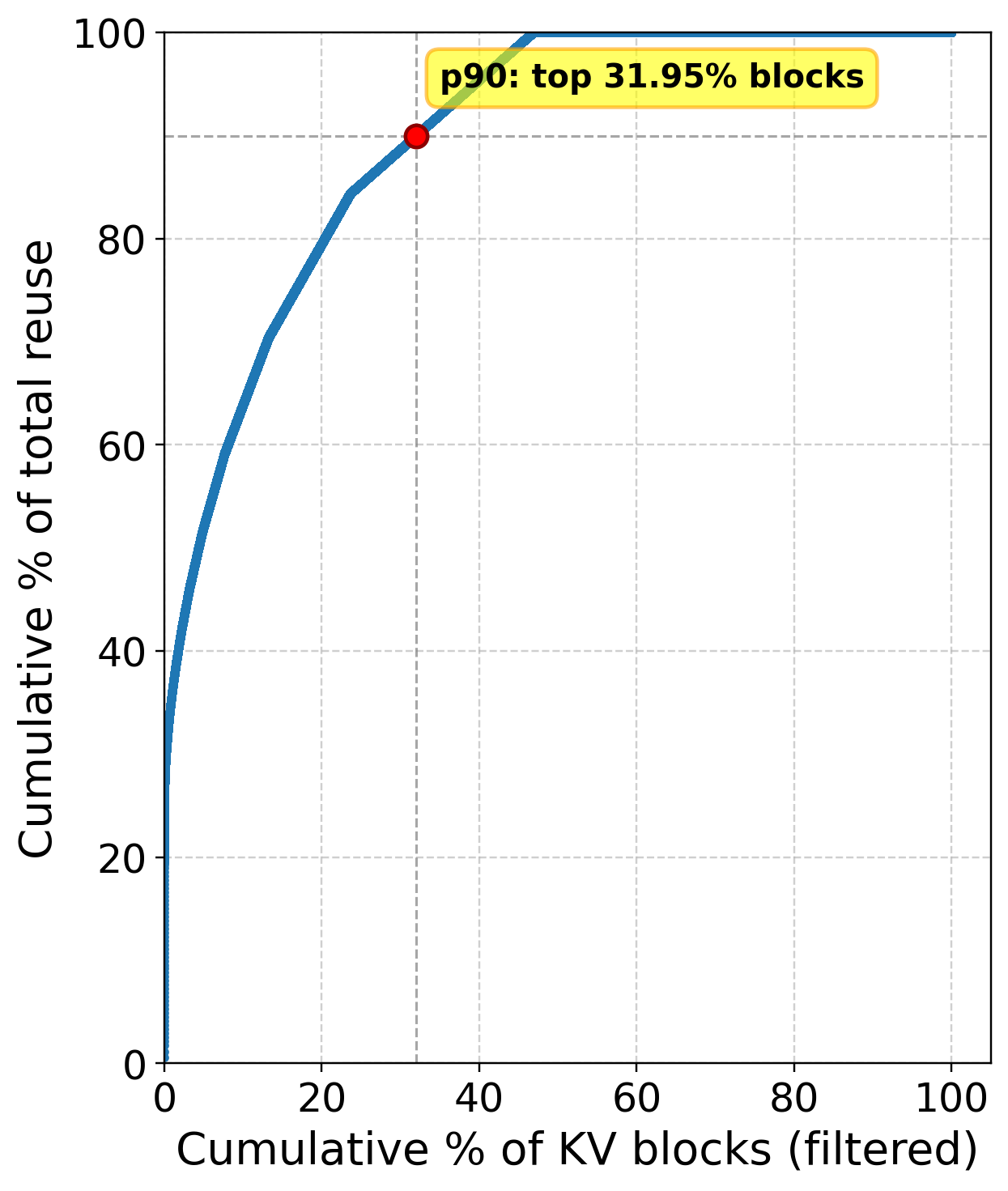}
    \caption{Trace A}
    \label{fig:traceA}
\end{subfigure}
\hfill
\begin{subfigure}{0.48\linewidth}
    \centering
    \includegraphics[width=\linewidth]{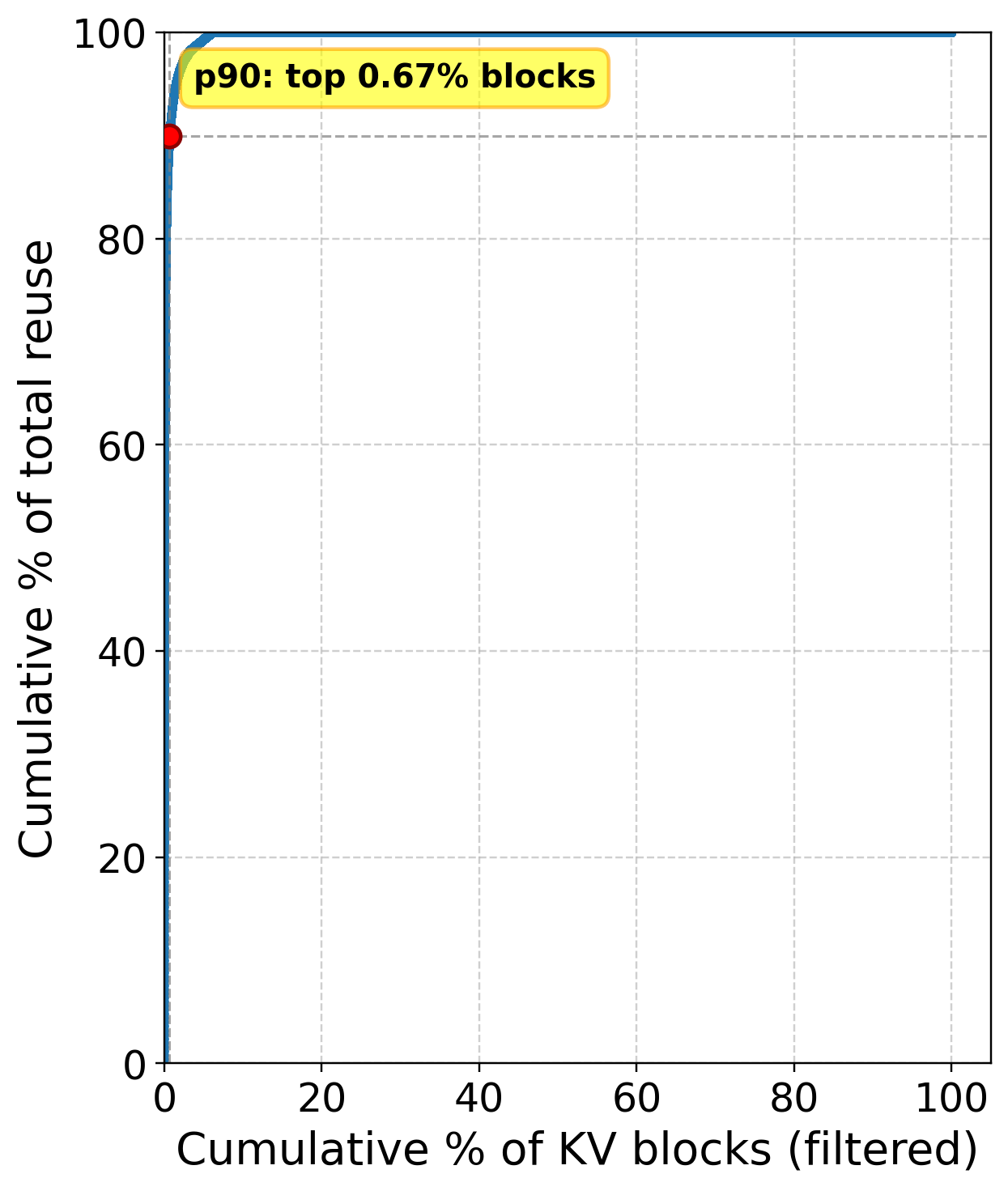}
    \caption{Trace B}
    \label{fig:traceB}
\end{subfigure}
\caption{Reuse distribution (lorenz curve) of traceA and traceB.}
\label{fig:analysis_skew}
\end{figure}
    
    \item \textbf{$X_7 = O_3(X_3,X_4)$: Effective I/O Bandwidth} \\
    While primarily constrained by storage media characteristics, effective bandwidth is significantly modulated by policy decisions ($X_3$, $X_4$). \autoref{tab:io_bw} shows simulation results where increasing bandwidth from 350\,MB/s to 40\,GB/s substantially reduces Time To First Token (TTFT), with diminishing returns beyond certain thresholds. In cloud environments, provisioned bandwidth is often coupled non-linearly with storage capacity tiers (e.g., AWS EBS gp3 volumes), creating complex tradeoffs between allocated space and achievable transfer rates that must be explicitly modeled.
     \begin{table}[htbp]
  \centering
  \caption{Impact of DRAM Bandwidth on Time to First Token (TTFT)}
  \label{tab:io_bw}
  \begin{tabular}{
    l
    S[table-format=7.2]   
    S[table-format=8.2]   
  }
    \toprule
    {DRAM Bandwidth} & {Mean TTFT (ms)} & {P90 TTFT (ms)} \\
    \midrule
    350\,MB/s  & 6873432.52 & 12765228.29 \\
    1\,GB/s    & 1083642.90 & 2024294.99 \\
    5\,GB/s    & 2995.42 & 9252.34 \\
    20\,GB/s    & 1915.50 & 4381.57 \\
    40\,GB/s    & 1787.81 & 3795.69 \\
    60\,GB/s   &    1757.50 &  3619.81 \\
    100\,GB/s   &    1744.31 &   3573.42 \\
    \bottomrule
  \end{tabular}
  
  \smallskip
  \footnotesize
  Note: TTFT = Time to First Token.
\end{table}
\end{itemize}

\zxz{
\subsubsection{The Inherent Inadequacy of Closed-form Models}

A natural approach might seek closed-form equations mapping decision variables $(X_1,\dots,X_7)$ to performance outcomes (latency, throughput, cost). We rigorously demonstrate why this is infeasible for production KVCache systems due to two fundamental properties:

\textbf{Nonlinear State Coupling.} Key metrics exhibit interdependent nonlinearities:
\begin{itemize}
    \item $X_5$ (Storage Capacity) depends nonlinearly on $X_1$ (workload) and $X_4$ (TTL policy)
    \item $X_6$ (Hit Ratio) depends on $X_1$, $X_3$ (capacity), and $X_4$
    \item $X_7$ (I/O Bandwidth) depends on $X_3$ and $X_4$
\end{itemize}
This creates high-order feedback loops where marginal changes cascade unpredictably. \textit{Example:} Increasing DRAM capacity ($X_3$) improves hit rate ($X_6$), but beyond saturation points, further increases yield diminishing returns while the cost still increases linearly.

\textbf{Critical Discontinuities.} System behavior exhibits sharp phase transitions that defy smooth approximation:
\begin{itemize}
    \item \textbf{Capacity Saturation:} Hit rate gains follow power-law decay with capacity. Initial capacity increase from 250 GB to 1000 GB yield about 21\% increase in reuse ratio, while subsequent increase to 2000 GB yield <7\% improvement (\autoref{fig:reuse_ratio_vs_storage}).
     \begin{figure}[htbp]
  \centering
  \includegraphics[width=\columnwidth]{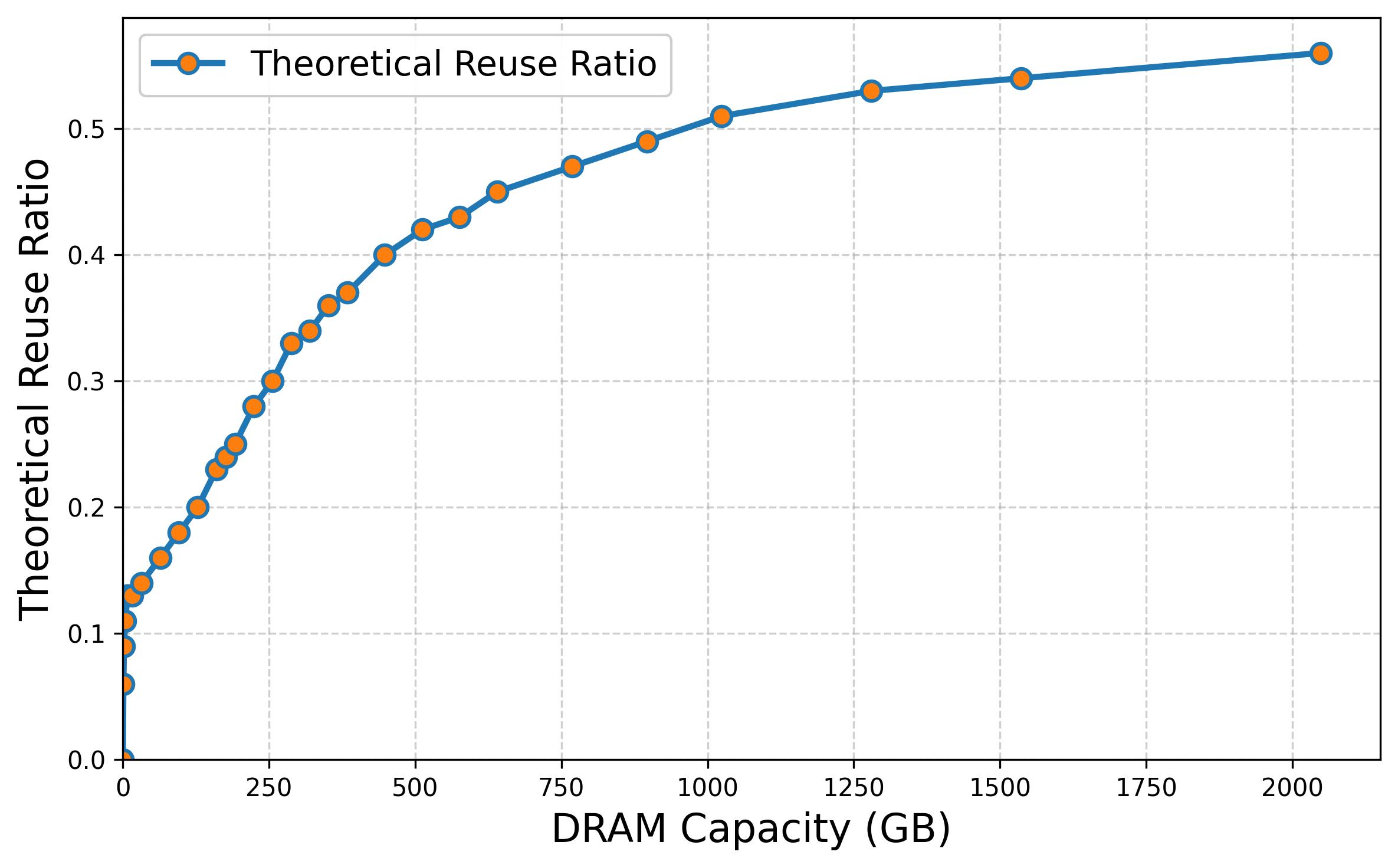}
  \caption{Storage capacity vs. reuse ratio.}
  \label{fig:reuse_ratio_vs_storage}
\end{figure}
    
    
    \item \textbf{Cloud Pricing Cliff Edges:} Analysis of AWS EBS pricing reveals non-linear cost jumps at performance thresholds: (1) at 3,000 IOPS where \texttt{gp3} volumes transition from free baseline to \$0.005/IOPS charges, and (2) at 32,000 IOPS where migration to \texttt{io1}/\texttt{io2} volumes causes IOPS costs to surge 13$\times$ (from \$0.005 to \$0.065 per IOPS)~\cite{aws}.
\end{itemize}

As discussed above, closed-form models cannot capture such multidimensional discontinuities without workload-specific heuristics. This fundamentally limits their generality for production KVCache optimization.
To overcome this challenge, we introduce a high-fidelity simulator-based optimization framework that accurately models the interplay among our four decision variables while respecting real-world cost structures.

\begin{figure}
\centering
\begin{subfigure}[b]{\linewidth}
    \centering
    \includegraphics[width=\linewidth]{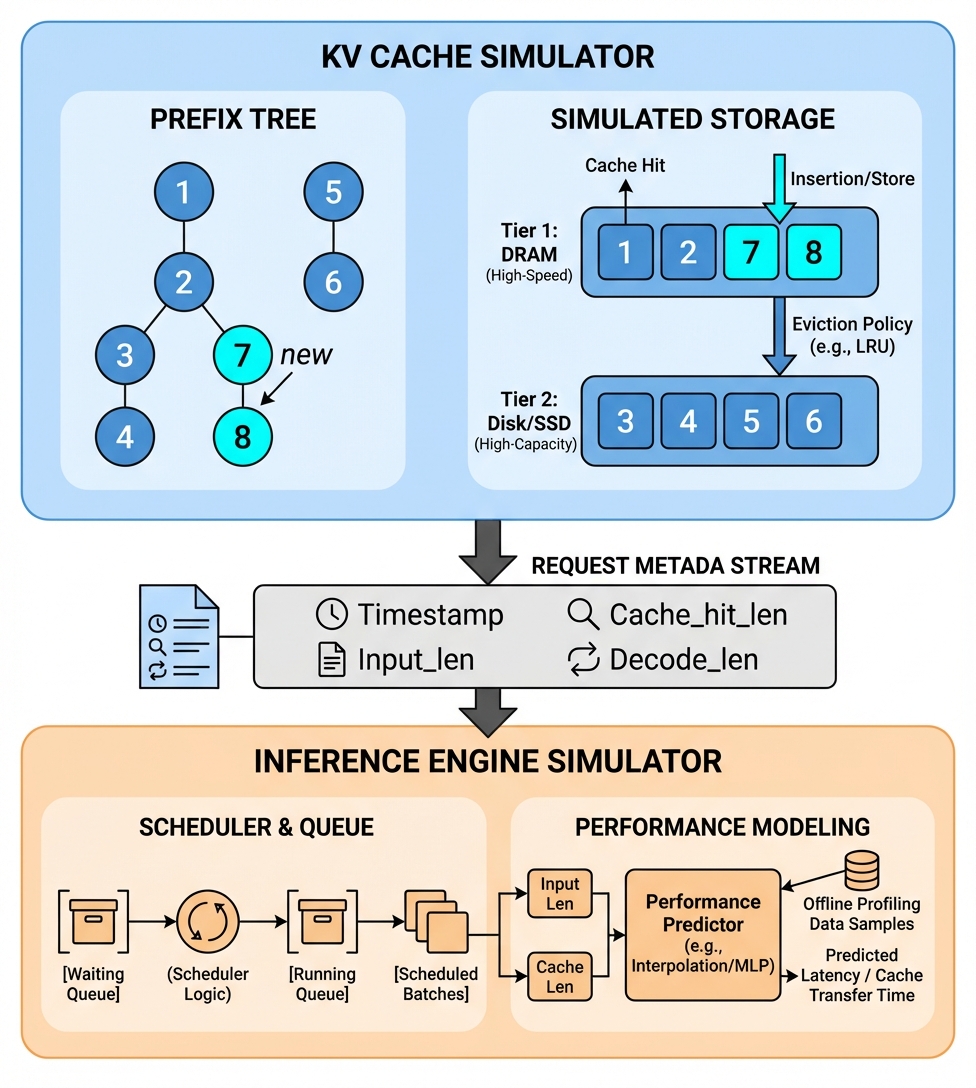}
\end{subfigure}
\caption{Design of simulator.}
\label{fig:simulator_design}
\vskip -15pt
\end{figure}

\subsection{Simulation-Driven Optimization}
\label{subsec:simulation_optimization}

The KVCache simulator integrates explicit discontinuity modeling with fine-grained coupling capture to overcome the limitations of analytical approaches. 
As shown in \autoref{fig:simulator_design}, 
it models a multi-tier storage hierarchy (HBM, DRAM, remote storage), accurately simulating KVCache data placement and access hit rates across tiers while precisely reflecting cloud pricing structures. 
A discrete-event inference engine simulator captures nonlinear pipeline interactions: variations in cache hit rates dynamically propagate to GPU utilization and I/O stall durations, whereas adjustments to storage policies concurrently influence storage capacity and access latency. 
Kernel execution times for attention and feed-forward network (FFN) operations are estimated through empirical profiling on target GPUs, with interpolation across input lengths and context sizes. 
The simulator faithfully reproduces production-grade mechanisms, such as shared prefix based on radix-tree for cross-request reuse and hierarchical layer-wise KV prefetching to overlap transfer latency. 

\subsection{Workload Density-Dependent KV Cache Performance Characteristics}
\label{sec:density_characteristics}

High-fidelity simulation reveals that workload density (the ratio of request arrival rate to baseline system capacity) fundamentally shapes hierarchical KV cache performance. We identify density-dependent behaviors across DRAM and disk storage configurations, validated on multiple representative traces.
We validate these behaviors using the simulator on three representative production traces:

\begin{itemize}[leftmargin=*,nosep]
    \item \textbf{Trace A}: interactive chatbot workloads featuring multi-turn dialogues;
    \item \textbf{Trace B}: programmatic API workloads (e.g., batch document processing);
    \item \textbf{Trace C}: agent-based workloads.
\end{itemize}

Traces A and B ~\cite{kvcache-wild} are derived from open source datasets. Trace C is synthetically generated to reflect the request distribution characteristics observed in real internal clusters. Each trace spans 2 hours and contains between 40,000 and 170,000 requests, containing salted hash blocks (16 tokens per block) of each requests. Despite their diversity, these traces exhibit several common characteristics in our analysis.

\textbf{Observation 1:} \textit{Low-density workloads exhibit immediate throughput saturation, making latency the primary optimization target.} Throughput plateaus at the arrival rate regardless of storage configuration; additional capacity yields diminishing latency improvements beyond moderate cache sizes, as computational resources remain underutilized.

\textbf{Observation 2:} \textit{Under low-density workloads, Disk-based KV caching only provides minimal benefit due to timing constraint.} 
\zxz{
In contrary to DRAM, disk's limited bandwidth often makes KV reloading slower than recomputation. The common strategy~\cite{hicache} is to perform disk-based KV reloading exclusively during queuing time of requests.
In low-density workloads, requests spend minimal time in queue, consequently leaving an extremely narrow window for disk prefetching,
}
causing effective hit rates to fall below theoretical capacity-based predictions.

\textbf{Observation 3:} \textit{High-density workloads realize multiplicative benefits from increased KV cache capacity, improving both throughput and latency.} Larger caches reduce computational redundancy, expanding effective system capacity. Throughput scales nearly linearly with cache size.

\textbf{Observation 4:} \textit{Extended queuing times enable effective disk utilization despite bandwidth limitations.} 
\zxz{
By contrast, high-density workloads exhibit prolonged queuing periods that provide sufficient time for disk I/O to complete before execution begins, thereby enabling effective KV block reuse. }
Queuing delays provide sufficient prefetching windows; a substantial majority of theoretically cacheable KV blocks become available before computation, aligning effective hit rates with capacity-based predictions.

\begin{figure}[!htbp]
\centering
\begin{subfigure}[!htbp]{\linewidth}
    \centering
    \hspace*{0.3cm}\includegraphics[width=0.95\linewidth]{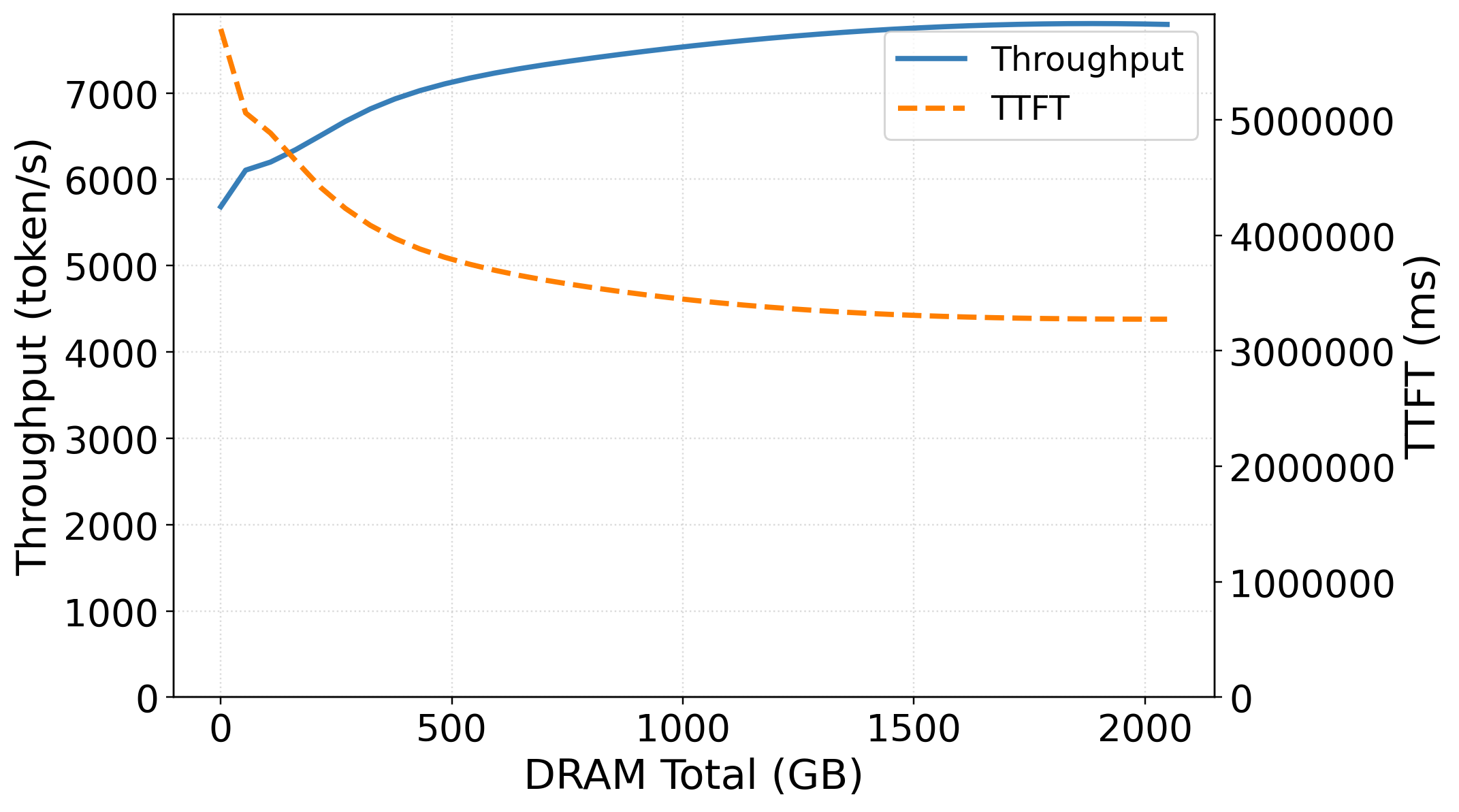}
    \caption{compute-bound condition}
\end{subfigure}

\begin{subfigure}[!htbp]{\linewidth}
    \centering
    \includegraphics[width=0.94\linewidth]{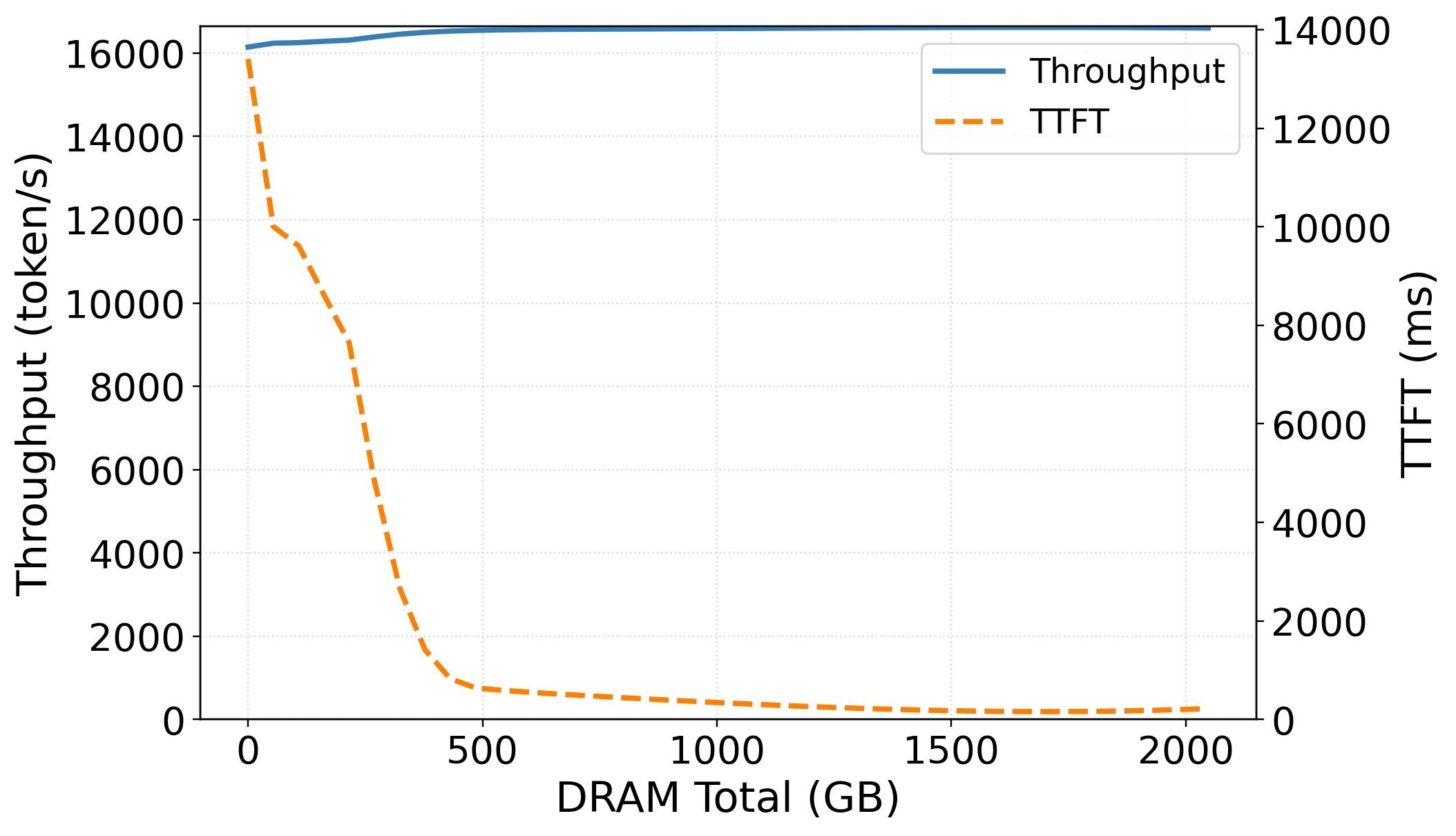}
    \caption{compute-abundant condition}
\end{subfigure}

\caption{Impact of DRAM capacity on reuse ratio, throughput, and mean TTFT.}
\label{fig:analysis_capacity}
\end{figure}
\begin{figure*}[!t]
\centering
\begin{subfigure}[b]{0.24\linewidth}
    \centering
    \includegraphics[width=\linewidth]{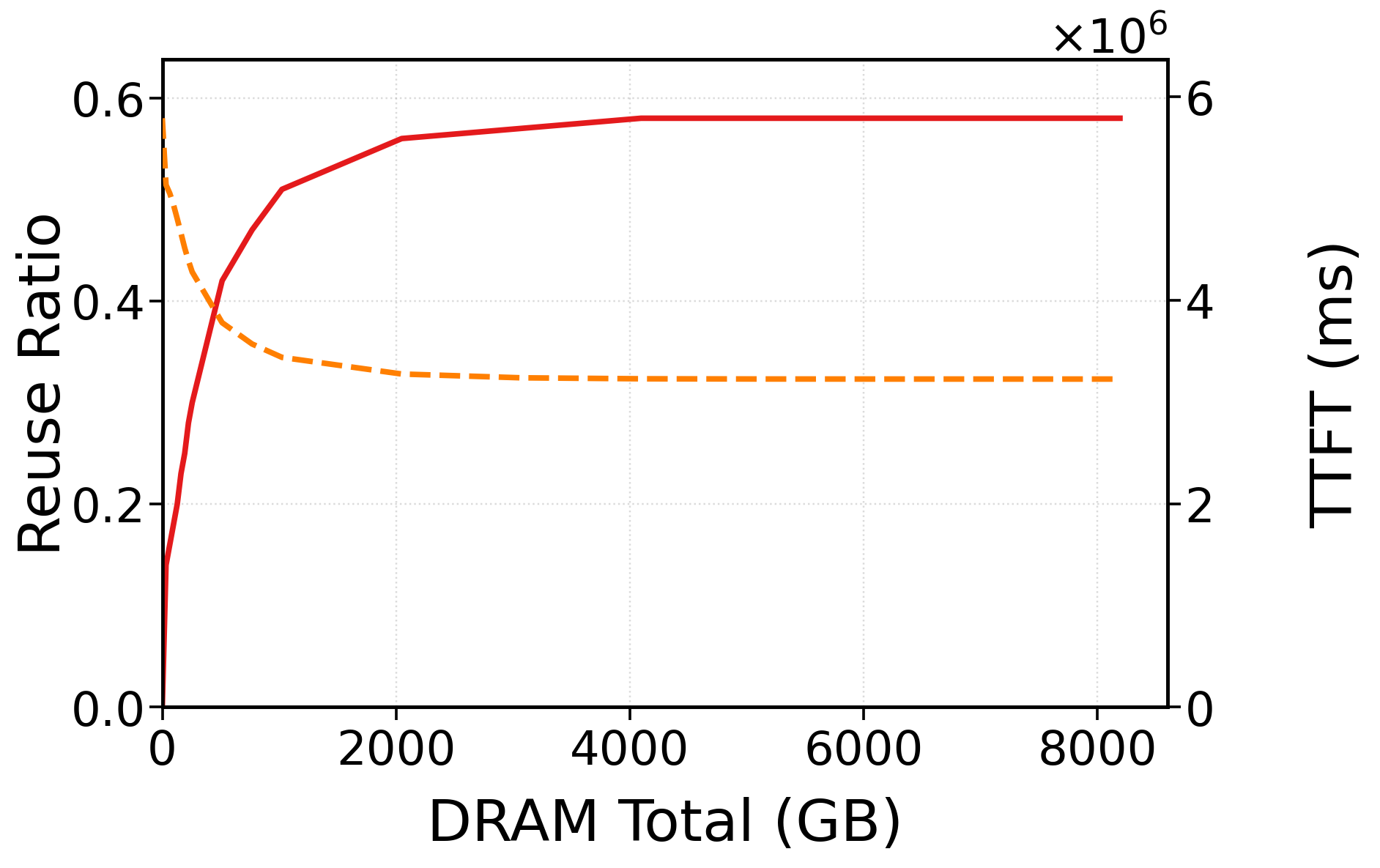}
    \caption{}
\end{subfigure}%
\hspace{-0.2em}
\begin{subfigure}[b]{0.24\linewidth}
    \centering
    \includegraphics[width=\linewidth]{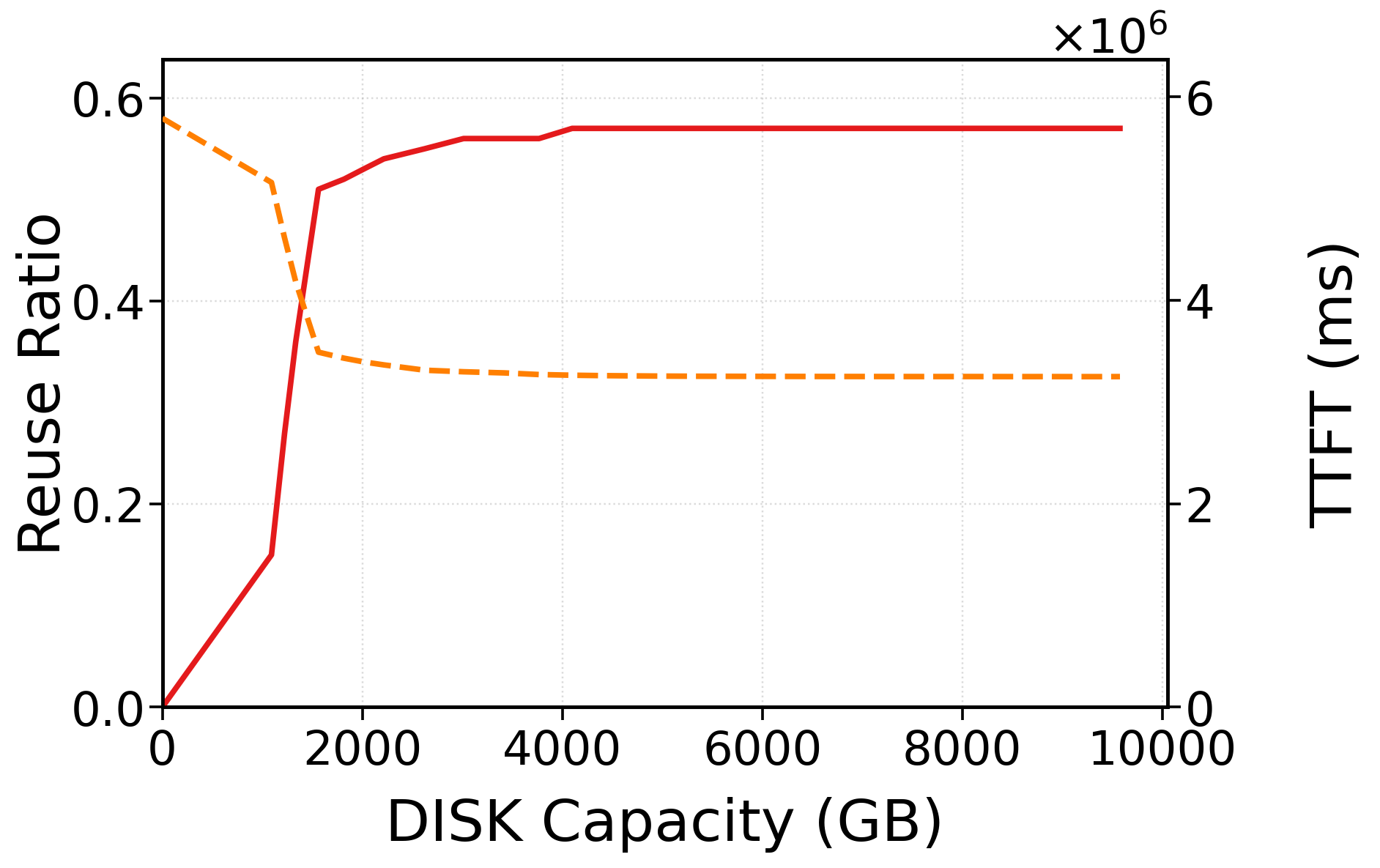}
    \caption{}
\end{subfigure}%
\hspace{-0.2em}
\begin{subfigure}[b]{0.24\linewidth}
    \centering
    \includegraphics[width=\linewidth]{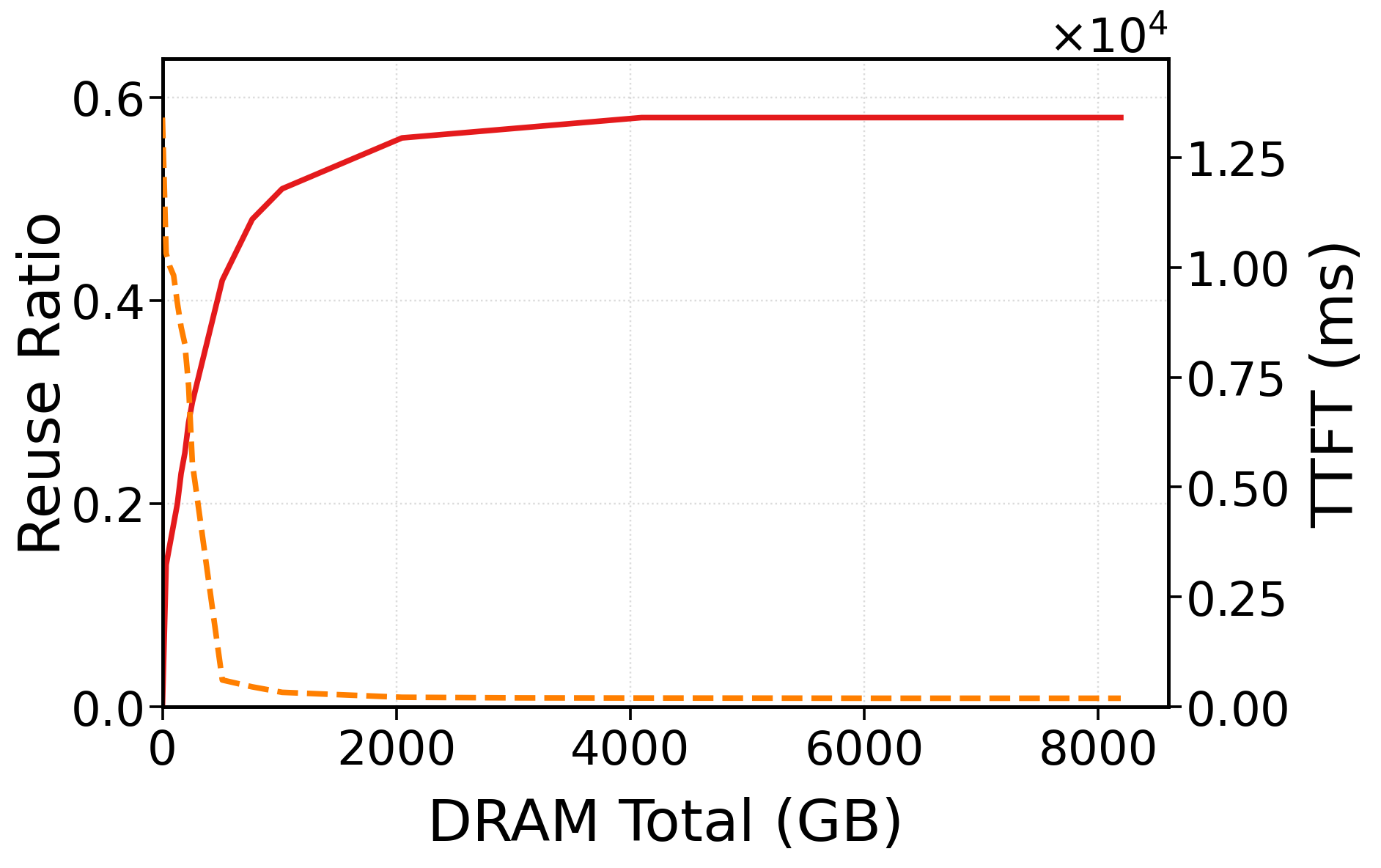}
    \caption{}
\end{subfigure}%
\hspace{-0.2em}
\begin{subfigure}[b]{0.24\linewidth}
    \centering
    \includegraphics[width=\linewidth]{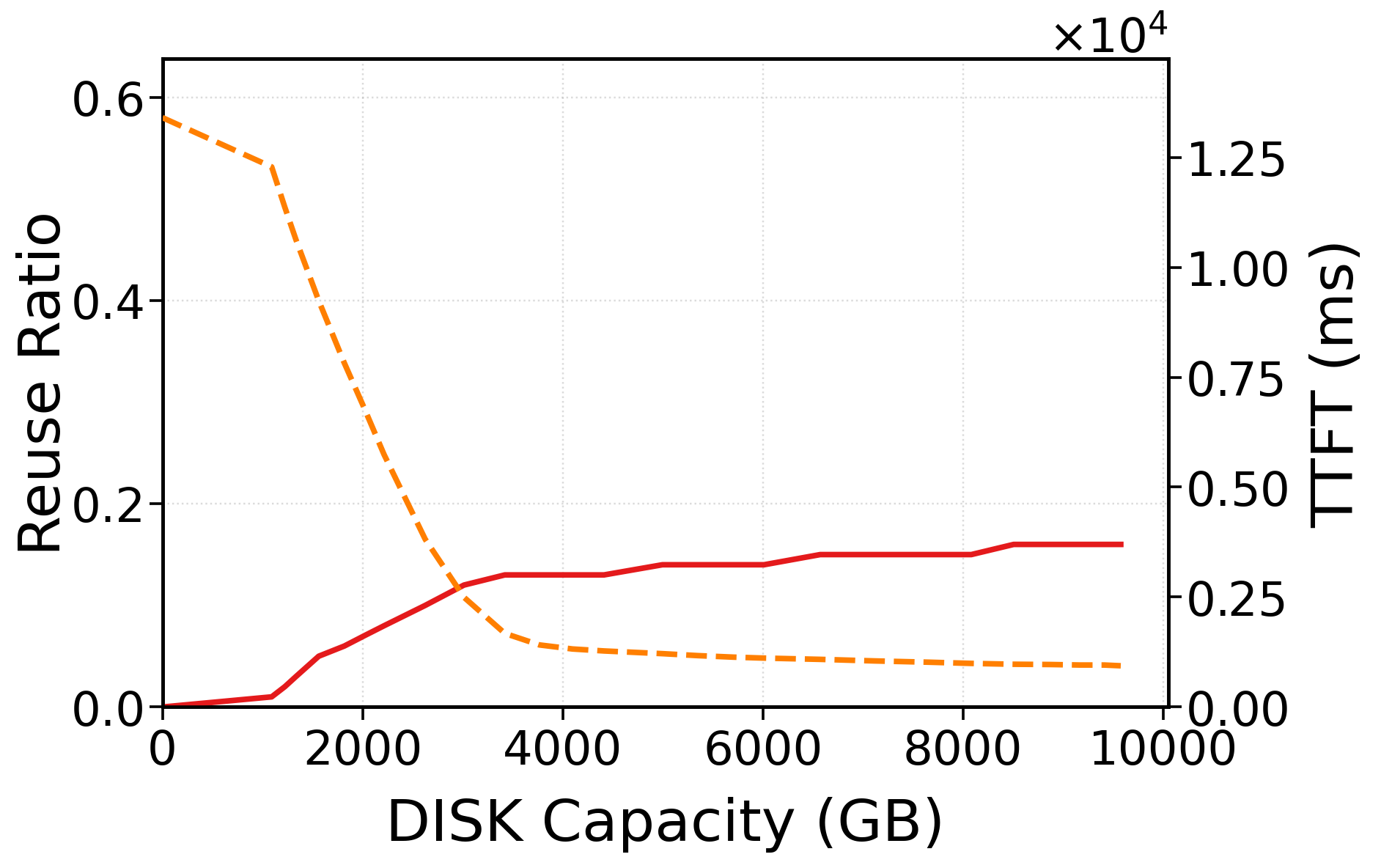}
    \caption{}
\end{subfigure}%

\caption{Reuse ratio of DRAM and disk in different workloads varying with capacity. (a) DRAM, traceA, ins1(high-density workload). (b) Disk, traceA, ins1(high-density workload). (c) DRAM, traceA, ins4(low-density workload). (d) Disk, traceA, ins4(low-density workload).}
\label{fig:dram_disk_comparison}
\end{figure*}

\begin{figure}[!htbp]
\centering
\begin{subfigure}[b]{0.95\linewidth}
    \centering
    \includegraphics[width=\linewidth, height=0.2\textheight, keepaspectratio]{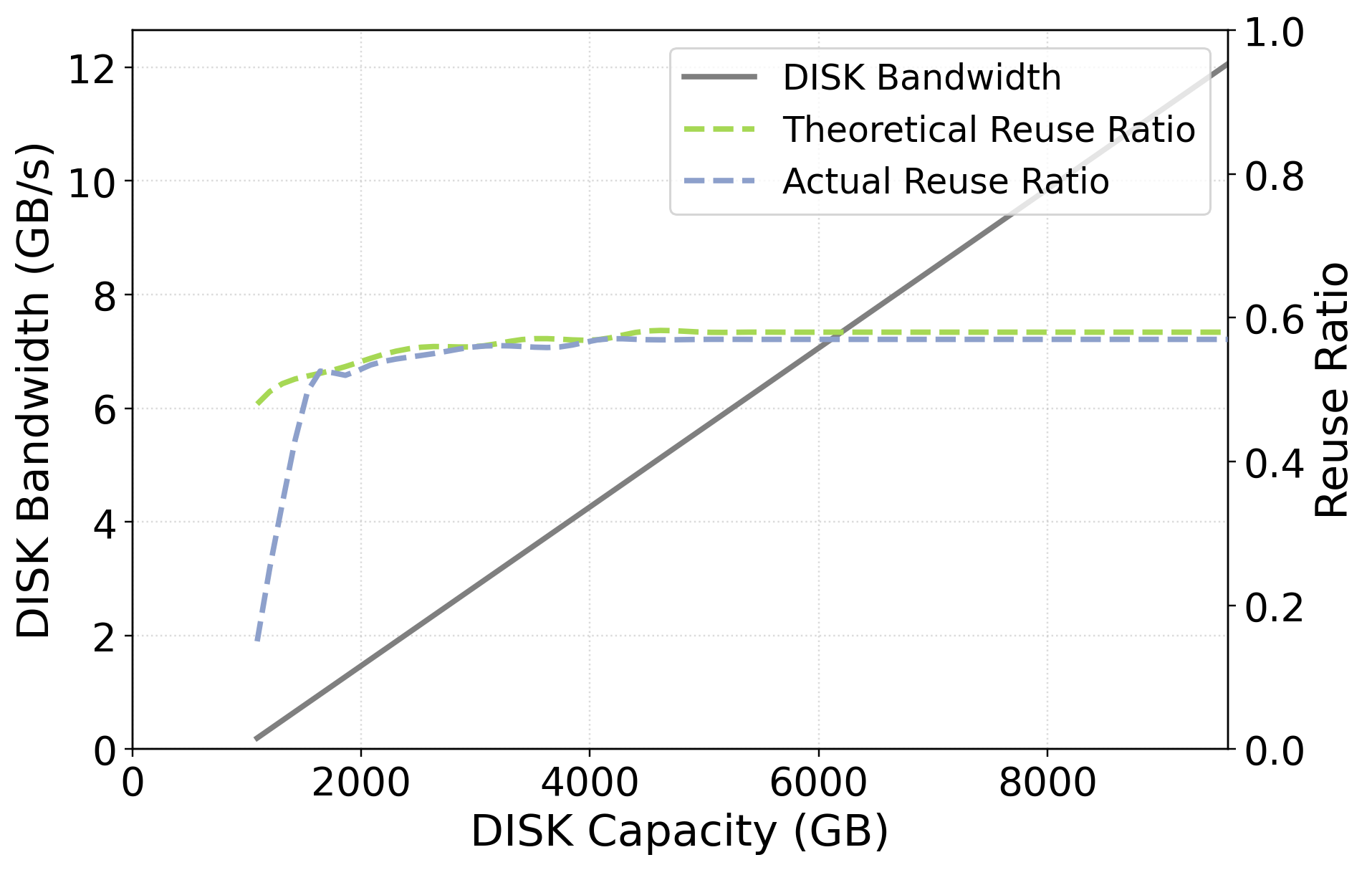}
    \caption{TraceA (Compute Bound)}
\end{subfigure}


\begin{subfigure}[b]{0.95\linewidth}
    \centering
    \includegraphics[width=\linewidth, height=0.2\textheight, keepaspectratio]{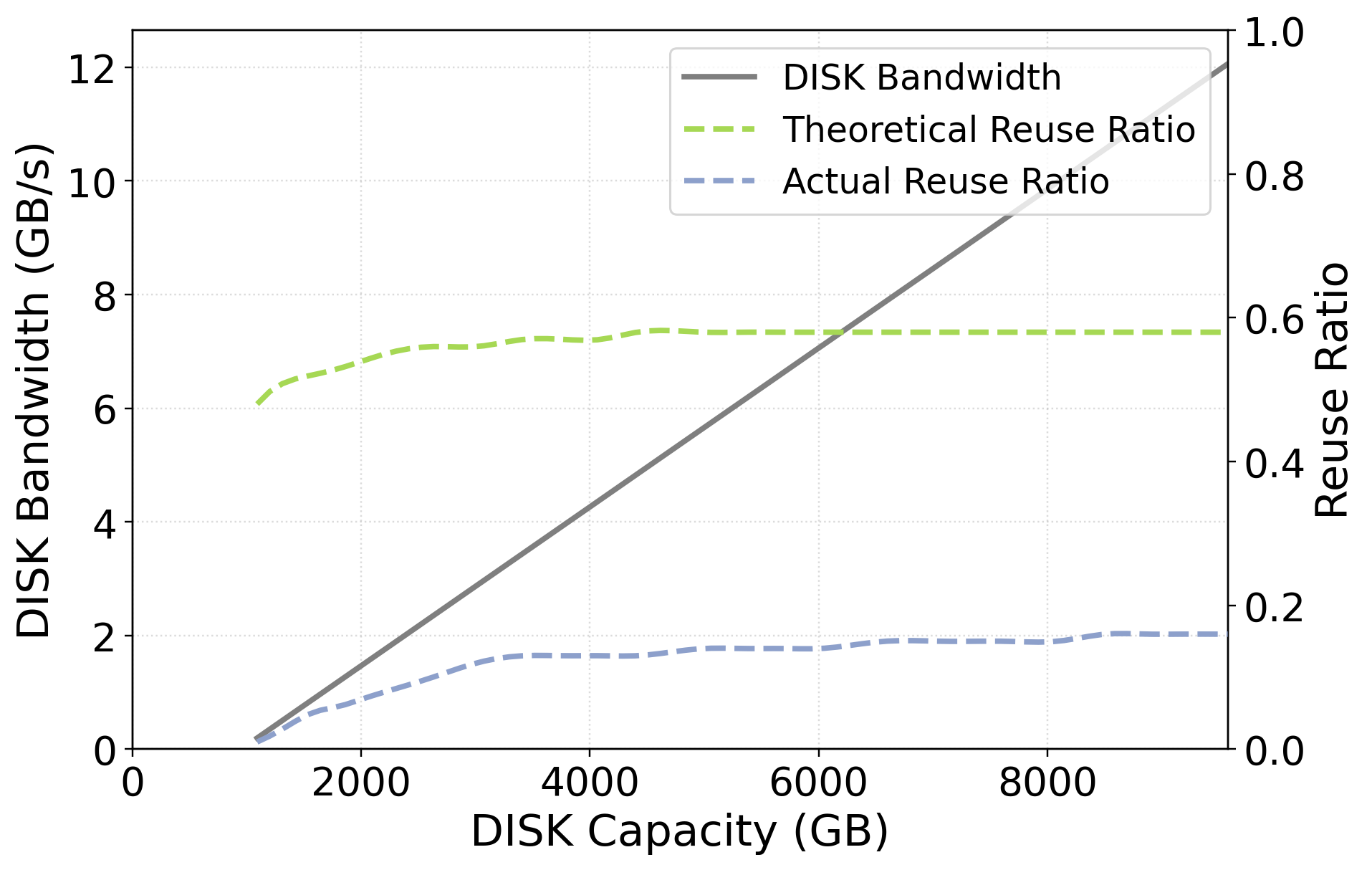}
    \caption{TraceA (Compute Abundant)}
\end{subfigure}
\caption{Disk bandwidth and reuse ratio in different DISK capacity}
\label{fig:analysis_3_bw_cp}
\end{figure}

\vspace{-10pt}

\vspace{10pt}

\textbf{Observation 5:} \textit{Performance on disk exhibits non-linear scaling due to capacity-bandwidth coupling.} 
\zxz{
Cloud-provisioned remote disks impose two key constraints: (1) providers typically scale disk bandwidth linearly with allocated storage capacity, and (2) writes (from DRAM eviction) and reads (for KV reloading) compete for the same I/O channel.
Consequently, even after capacity exceeds the total KV footprint, further increases continue to improve performance, and a minimum disk capacity is demanded for DRAM eviction(\autoref{fig:analysis_3_bw_cp}). This entanglement makes the performance–cost trade-off for disk-resident KV caches more complicated as the marginal benefit of additional disk capacity becomes non-monotonic and highly workload-dependent.
}

\zxz{
\textbf{Observation 6:} \textit{Tiered storage achieves Pareto-optimal cost-performance trade-offs.} 
DRAM is expensive but offers high bandwidth, whereas disk is cost-effective but suffers from low bandwidth.
Hybrid configurations combining them can occupy a compelling point on the cost–performance Pareto frontier.
As an example illustrates(~\autoref{fig:analysis_combine_better}), augmenting 256 GB of DRAM with disk achieves substantially lower latency than a disk-only setup while costing far less than scaling DRAM alone, where DRAM is allocated to the most frequently accessed blocks, while disk economically retains blocks with moderate reuse potential.
}

\begin{figure*}[!t]
\centering
\includegraphics[width=\linewidth]{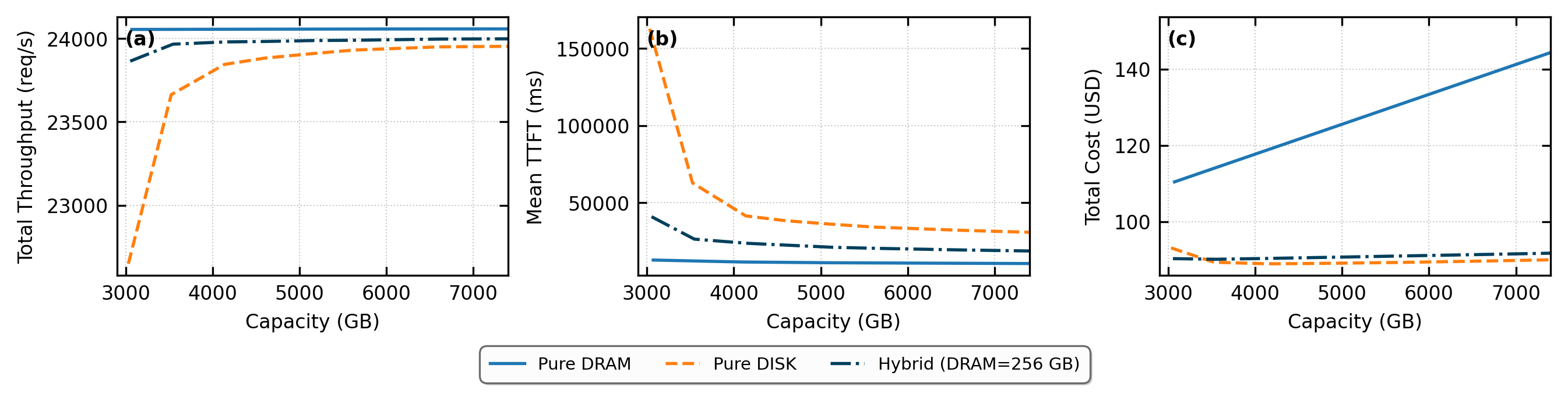}
\caption{Performance comparison of three resource allocation strategies under 2 instances: pure DRAM (blue solid), pure DISK (orange dashed), and hybrid with 256 GB DRAM + varying DISK (dark blue dash-dot). The hybrid strategy achieves a balance between cost and latency while maintaining high throughput.}
\label{fig:analysis_combine_better}
\end{figure*}

Above factors make manual or heuristic-based provisioning highly unreliable. Instead, we advocate for a simulation-driven approach: by replaying historical workloads through a high-fidelity simulator, we can systematically explore the configuration space to identify storage allocations that optimally balance between cost, latency, and throughput under the real-world constraints.
}
\begin{figure*}[ht!]
\centering
    \includegraphics[width=\linewidth]{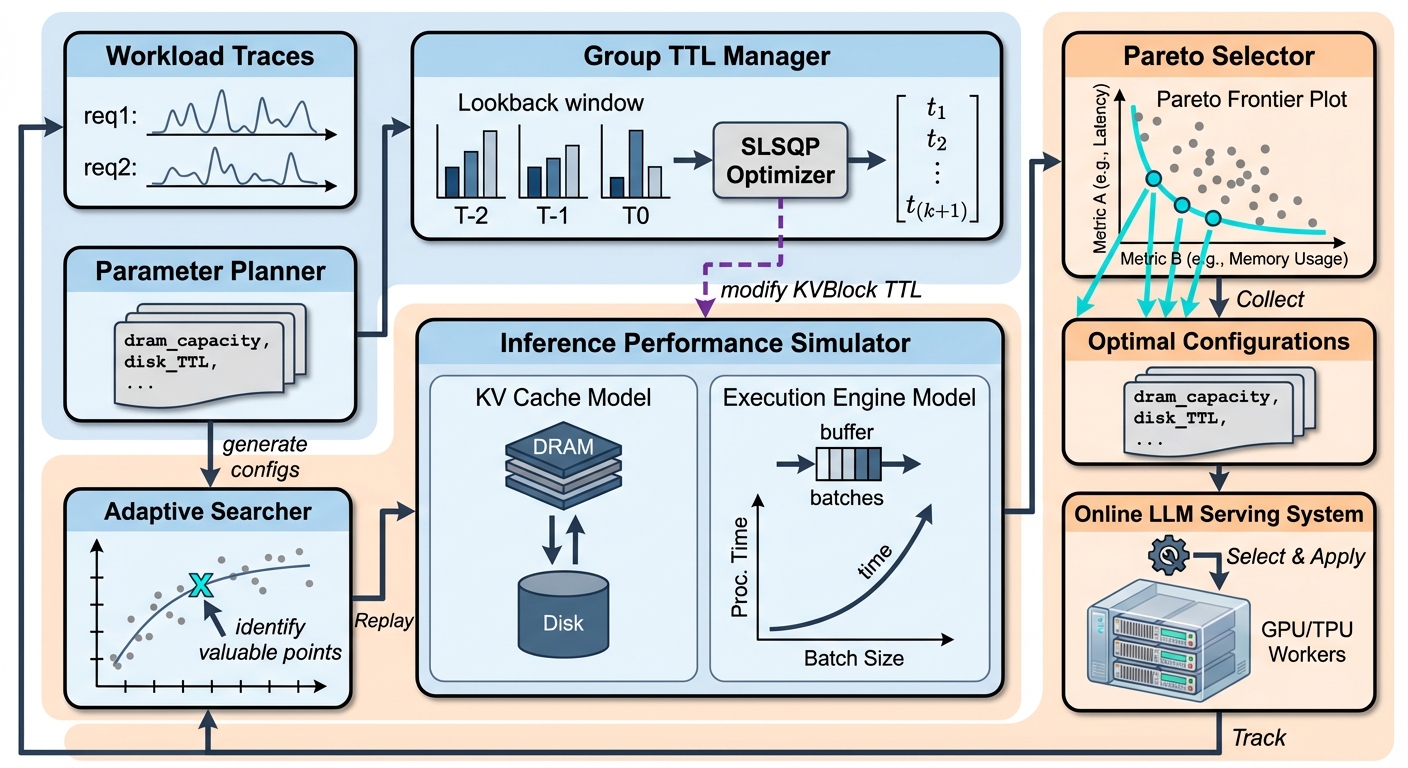}
\caption{Overview of Kareto}
\label{fig:kareto_overview}
\end{figure*}

\section{Adaptive Storage Configuration and Management}
\label{sec:Design}

\subsection{Overview}

Prior studies have revealed notable inter-day and inter-week periodicity~\cite{kvcache-wild, burst-gpt}: within a single day or between weekdays and holidays, request patterns show significant dynamics; however, strong periodicity emerges across corresponding days (e.g., Mondays to Mondays) and across weeks. Building upon this observation, we exploit such periodicity by replaying workload traces from a recent historical period (e.g., the same time slot of the preceding week) across diverse storage configurations through our high-fidelity simulator. The Pareto-optimal configurations identified from this simulation-driven analysis are then applied to guide resource provisioning for the subsequent period, enabling proactive adaptation to various and dynamic workload characteristics.

To address the complex interplay between diverse workloads and system configurations in LLM serving, we propose an integrated system comprising three key components: a \emph{planner}, a \emph{simulator}, and a \emph{Pareto-based configuration selector}.

The \textbf{planner} generates a set of candidate configurations without assuming prior knowledge of user requirements. Each configuration consists of: (1) DRAM capacity allocated for KV-cache, (2) TTL (Time-to-Live) for KV blocks stored on disk, and (3) the type of disk storage medium (e.g., cloud SSD tiers such as Alibaba Cloud ESSD PL1, PL2, or PL3, which differ in price, capacity, and I/O bandwidth).

The \textbf{simulator} evaluates each candidate configuration by replaying historical request traces comprising request arrival times, input/output token lengths, and token-level KV-block hashes through a high-fidelity model of the full LLM serving pipeline. It produces key performance metrics, including KV-cache hit rates, throughput (tokens/s), latency (e.g., TTFT), and estimated cloud cost.

The \textbf{Pareto-based configuration selector} then takes user-specified performance or cost constraints (e.g., P99 TTFT $\leq$ 2s) and filters the simulated results accordingly. Since the underlying optimization is inherently multi-objective, balancing cost, throughput, and latency, we adopt a Pareto-optimal selection criterion: a configuration is dominated if another configuration achieves better or equal performance in all objectives and strictly better in at least one. The selector returns all non-dominated configurations that satisfy the user’s constraints, enabling informed trade-offs based on specific service-level needs.

Despite this framework’s promise, two challenges remain:

(1) Simulating a single candidate configuration on a two-hour trace typically consumes several minutes of computation time. Given that a single search can involve evaluating hundreds of configurations, the overall search process becomes prohibitively expensive, limiting the density and coverage of the explored Pareto frontier. On the other hand, a coarse-grained grid risks overlooking critical regions of the Pareto frontier, particularly those exhibiting high sensitivity or sharp transitions in the trade-off surface.

(2) While a fixed TTL can adapt to changes in request concurrency, it treats all KV blocks uniformly. This uniformity prevents the system from discerning the distinct reuse characteristics of individual blocks Consequently, the resulting Pareto frontier remains suboptimal and leaves room for the potential further improvement.

To address these challenges, we propose two complementary techniques: adaptive Pareto search and ROI-aware group TTL configuration. 
Their design details are presented next.

\subsection{Adaptive Pareto Search}
\label{subsec:pareto_pruning}

To mitigate the high time consumption of exhaustive grid search, we adopt the key insights from our empirical analysis (\S\ref{sec:density_characteristics}) to guide a more efficient exploration.

The performance-cost trade-off exhibits long-tail diminishing returns beyond a critical capacity threshold. An exhaustive search over the full configuration space is extremely time-consuming, as each simulation can take several minutes and the combinatorial space can spans hundreds or even thousands of points. Our analysis reveals that while the objective functions (e.g., throughput, latency, cost) are non-convex, they consistently enter a long-tail regime of diminishing returns once resource capacities (such as DRAM or disk) exceed a certain workload-dependent threshold. In this regime, further capacity increases yield only marginal performance gains at a significant cost. We leverage this insight to design a pruning strategy: during a coarse-to-fine grid expansion, we monitor the marginal performance gain. If it falls below the preset threshold, indicating entry into the long-tail diminishing-return region, we terminate further exploration along that dimension, thereby avoiding costly simulations of configurations with negligible practical value.
Meanwhile, We first perform an initial exploration using a coarse grid to efficiently cover the configuration space. We then identify subregions where performance or cost metrics exhibit significant variation and refine these regions with additional sampling points. This iterative refinement enables us to accurately capture critical trade-off regions on the Pareto frontier while avoiding the prohibitive cost of uniform fine-grained evaluation.
The detailed algorithm is described in algorithm 1.




\begin{algorithm}[ht]
  \caption{Adaptive Pareto Exploration}
  \label{alg:adaptive_pareto_search}
  \begin{algorithmic}[1]
    \REQUIRE DRAM initial capacity range $[d_\text{min}, d_\text{max}]$ with step $\Delta_d$; TTL initial range $[t_\text{min}, t_\text{max}]$ with step $\Delta_t$; expand threshold $\tau_\text{e}$, refinement thresholds $\tau_{\text{perf}}$, $\tau_{\text{cost}}$
    \STATE $P \gets \emptyset$
    \STATE $C \gets$ uniform grid over $[d_{\mathrm{min}},d_{\mathrm{max}}] \times [t_{\mathrm{min}},t_{\mathrm{max}}]$
    \STATE $S \gets \emptyset$  \hfill\text{//simulated configurations}
    \REPEAT
    \FOR{each unvisited $(d,t) \in C$}
    \STATE $R \gets \text{Simulate}(d,t)$
    \STATE append $(d,t)$ to $S$, append $R$ to $P$
    \ENDFOR

    \STATE $C \gets \emptyset$
  
    \text{//DRAM expansion (focus on TTL=0)}
    \STATE $d_\text{max} \gets \max\{ d \mid (d, 0) \in S \}$
    \IF{$\Delta\text{latency}((d_\text{max}-\Delta_d, 0), (d_\text{max}, 0)) > \tau_{\text{e}}$}
    \STATE append $(d\!+\!\Delta_d, t)$ to $C$ for {$t \gets t_{\min}$ to $t_{\max}$ step $\Delta_t$}
    \ENDIF
  
    \text{//Refinement in high-curvature regions}
    \FOR{each adjacent pair $(d_1,t_1), (d_2,t_2)$ in $S$}
    \IF{($\Delta\text{latency} > \tau_{\text{perf}}$ or $\Delta\text{throughput} > \tau_{\text{perf}}$) and $\Delta\text{cost} > \tau_{\text{cost}}$}
    \STATE append $midpoint (\frac{d_1+d_2}{2}, \frac{t_1+t_2}{2})$ to $C$
    \ENDIF
    \ENDFOR
    \UNTIL $C = \emptyset$
    \STATE {\bfseries Return} ParetoFilter(P)
\end{algorithmic}
\end{algorithm}

\begin{algorithm}[ht]
\caption{ROI-Aware TTL Allocation}
\label{alg:ttl-opt}
\begin{algorithmic}[1]
\REQUIRE Block access trace $T$, budget $B$, top-$K$ groups
\ENSURE Optimal TTLs $\mathbf{t}^* = [t_1^*, \dots, t_{K+1}^*]$
\STATE Partition requests into groups $\{\mathcal{G}_g\}_{g=1}^{K+1}$: top-$K$ frequent prefix subtrees and residual group $\mathcal{G}_{K+1}$
\FOR{each group $\mathcal{G}_g$}
    \STATE Extract inter-arrival times $\Delta_g \gets \{\delta_i^{(g)}\}$ from $T$
    \STATE Compute ROI curve $R_g(t) \gets H_g(t)/C_g(t)$ for $t \in [0, T_{\max}]$
    \STATE $t_g^{\text{roi}} \gets \arg\max R_g(t)$ \hfill\text{//Per-group ROI-optimal TTL}
\ENDFOR
\STATE $\mathbf{t}^{\text{roi}} \gets [t_1^{\text{roi}}, \dots, t_{K+1}^{\text{roi}}]$
\STATE Compute unscaled cost: $C_{\text{unscaled}} \gets \sum_{g=1}^{K+1} C_g(t_g^{\text{roi}})$
\STATE $\alpha \gets \frac{B}{C_{\text{unscaled}}}$ \hfill\text{//Budget-aware scaling factor}
\STATE $\mathbf{t}^{\text{init}} \gets \alpha \cdot \mathbf{t}^{\text{roi}}$ \hfill\text{//Scale to respect budget}
\STATE Initialize candidate set $P \gets \{ \mathbf{t}^{\text{init}} \}$
\STATE Randomly $\lfloor \sqrt{K} \rfloor + 1$ perturbed points around $\mathbf{t}^{\text{init}}$
\STATE Add perturbed points to $P$, ensuring non-negativity ($t_g \geq 0$)
\STATE $\mathbf{t}^* \gets \mathbf{0}$, $\text{best\_hits} \gets 0$
\FOR{each $\mathbf{t}^{\text{start}} \in P$}
    \STATE $\mathbf{t}^{\text{sol}} \gets \textsc{Slsqp}\big( \max_{\mathbf{t}} \sum_g H_g(t_g) \;\; \text{s.t.} \;\; \sum_g C_g(t_g) \leq B,\, \mathbf{t} \geq 0 \big)$ initialized at $\mathbf{t}^{\text{start}}$
    \STATE $\text{hits} \gets \sum_{g=1}^{K+1} H_g(t_g^{\text{sol}})$
    \IF{$\text{hits} > \text{best\_hits}$}
        \STATE $\mathbf{t}^* \gets \mathbf{t}^{\text{sol}}$
        \STATE $\text{best\_hits} \gets \text{hits}$
    \ENDIF
\ENDFOR
\STATE {\bfseries Return} $\mathbf{t}^*$
\end{algorithmic}
\end{algorithm}

\subsection{ROI-Aware Group TTL Based on Access Pattern}
\label{subsec:dynamic_ttl}

Although TTL-based management adapts to fluctuations in request rate, its effectiveness is limited when a uniform fixed TTL is applied uniformly across all KV blocks.
If TTL is set too short, blocks with long reuse intervals are evicted prematurely, degrading hit rates and increasing recomputation overhead.
If TTL is set too long, low-reuse or non-reused cold blocks unnecessarily occupy storage, wasting precious DRAM or disk capacity and inflating system cost.


\begin{figure}[t]
  \centering
  \captionsetup{aboveskip=0pt}
  \includegraphics[width=\columnwidth]{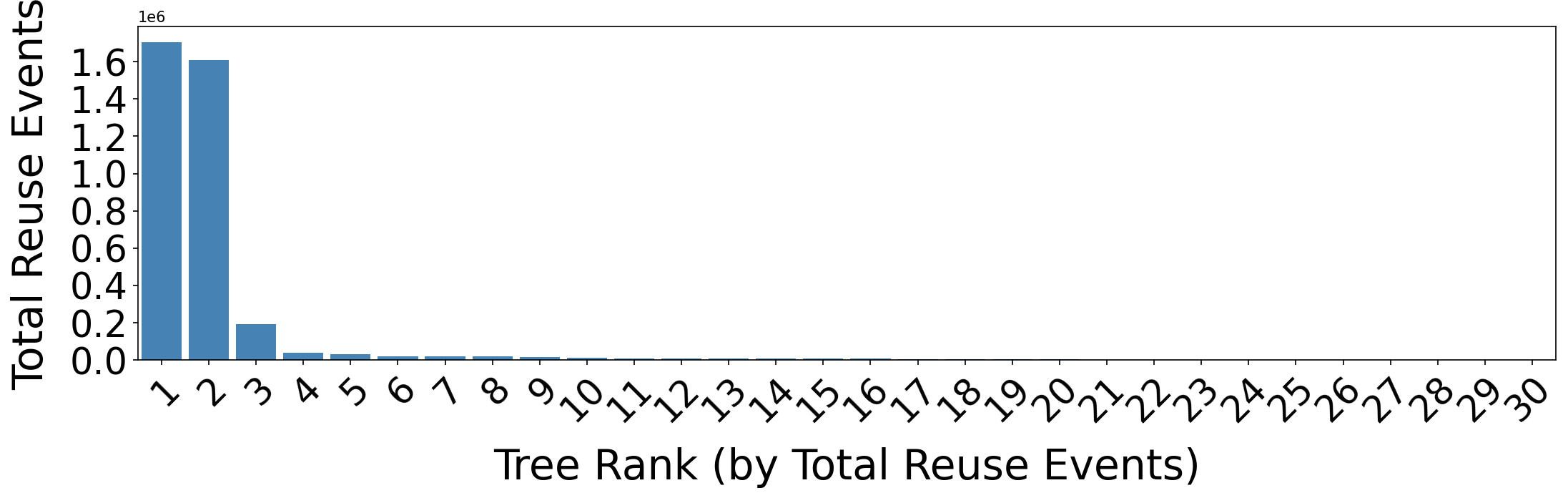}
  \caption{Ranked block reuse counts of each subtree}
  \label{fig:design_group_ttl_subtree_rank}
  
  \vspace{-3pt} 
  
  \begin{subfigure}[b]{0.48\columnwidth}
    \centering
    \captionsetup{aboveskip=0pt}
    \includegraphics[width=\linewidth]{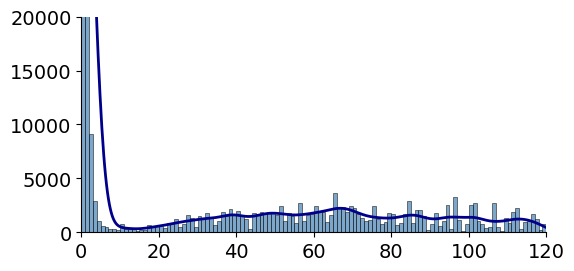}
    \caption{Subtree 1, 0--3600s}
    \label{fig:a1}
  \end{subfigure}
  \hfill
  \begin{subfigure}[b]{0.48\columnwidth}
    \centering
    \captionsetup{aboveskip=0pt}
    \includegraphics[width=\linewidth]{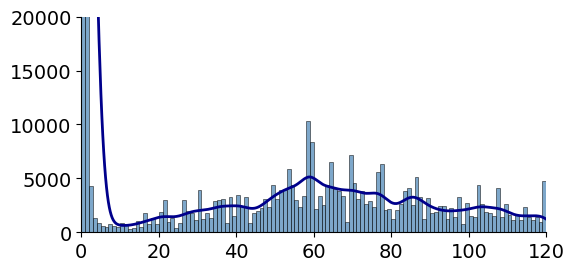}
    \caption{Subtree 1, 3600--7200s}
    \label{fig:a4}
  \end{subfigure}
  
  \vspace{-2pt} 
  
  \begin{subfigure}[b]{0.48\columnwidth}
    \centering
    \captionsetup{aboveskip=0pt}
    \includegraphics[width=\linewidth]{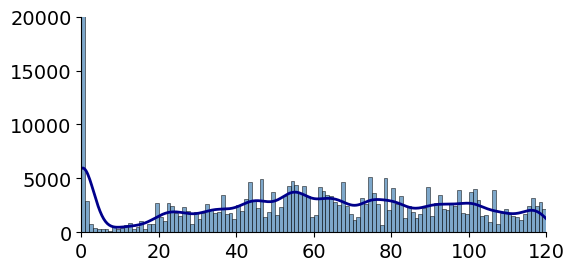}
    \caption{Subtree 2, 0--3600s}
    \label{fig:b1}
  \end{subfigure}
  \hfill
  \begin{subfigure}[b]{0.48\columnwidth}
    \centering
    \captionsetup{aboveskip=0pt}
    \includegraphics[width=\linewidth]{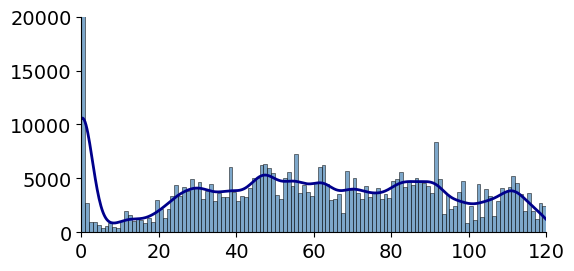}
    \caption{Subtree 2, 3600--7200s}
    \label{fig:b4}
  \end{subfigure}
  
  \vspace{-2pt}
  
  \begin{subfigure}[b]{0.48\columnwidth}
    \centering
    \captionsetup{aboveskip=0pt}
    \includegraphics[width=\linewidth]{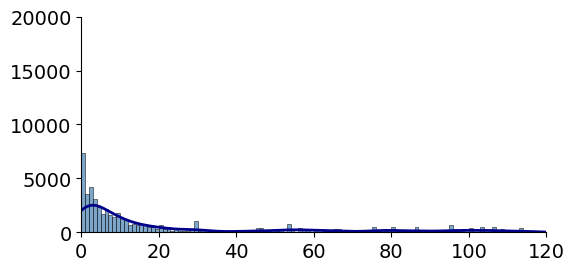}
    \caption{Subtree 3, 0--3600s}
    \label{fig:c1}
  \end{subfigure}
  \hfill
  \begin{subfigure}[b]{0.48\columnwidth}
    \centering
    \captionsetup{aboveskip=0pt}
    \includegraphics[width=\linewidth]{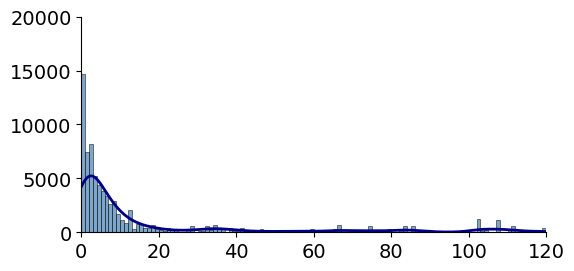}
    \caption{Subtree 3, 3600--7200s}
    \label{fig:c4}
  \end{subfigure}
  
  \caption{Reuse interval distribution of the top-3 most frequently reused subtrees in traceA during the early phase (first 3600s, left column) and the late phase (last 3600s, right column). Rows correspond to subtrees ranked from \#1--\#3 by reuse count.}
  \label{fig:design_group_ttl_subtree_distribution}
\end{figure}

\begin{figure*}[!htbp]
\centering
\begin{subfigure}[!htbp]{\textwidth}
\includegraphics[width=0.32\linewidth]{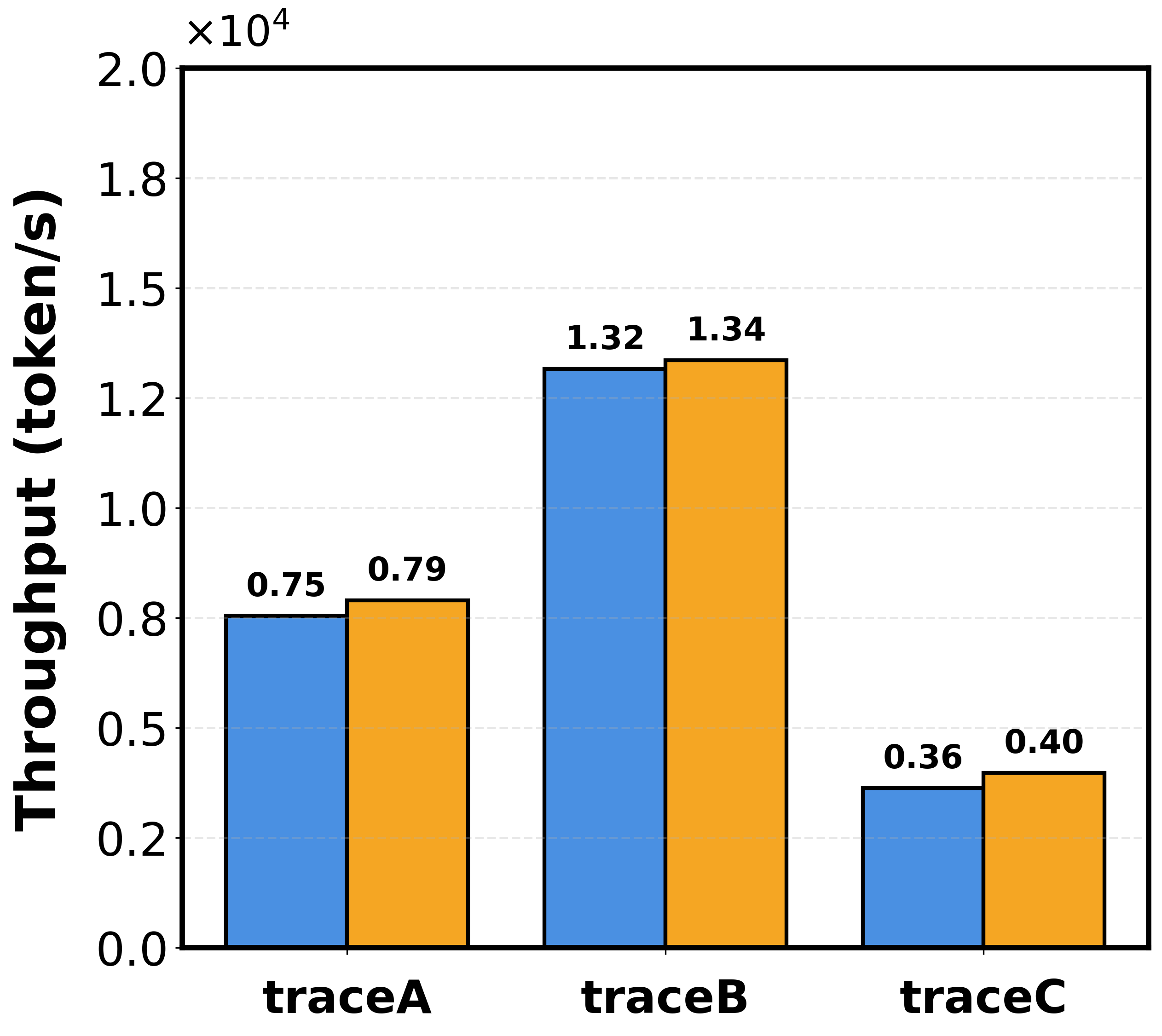} \includegraphics[width=0.32\linewidth]{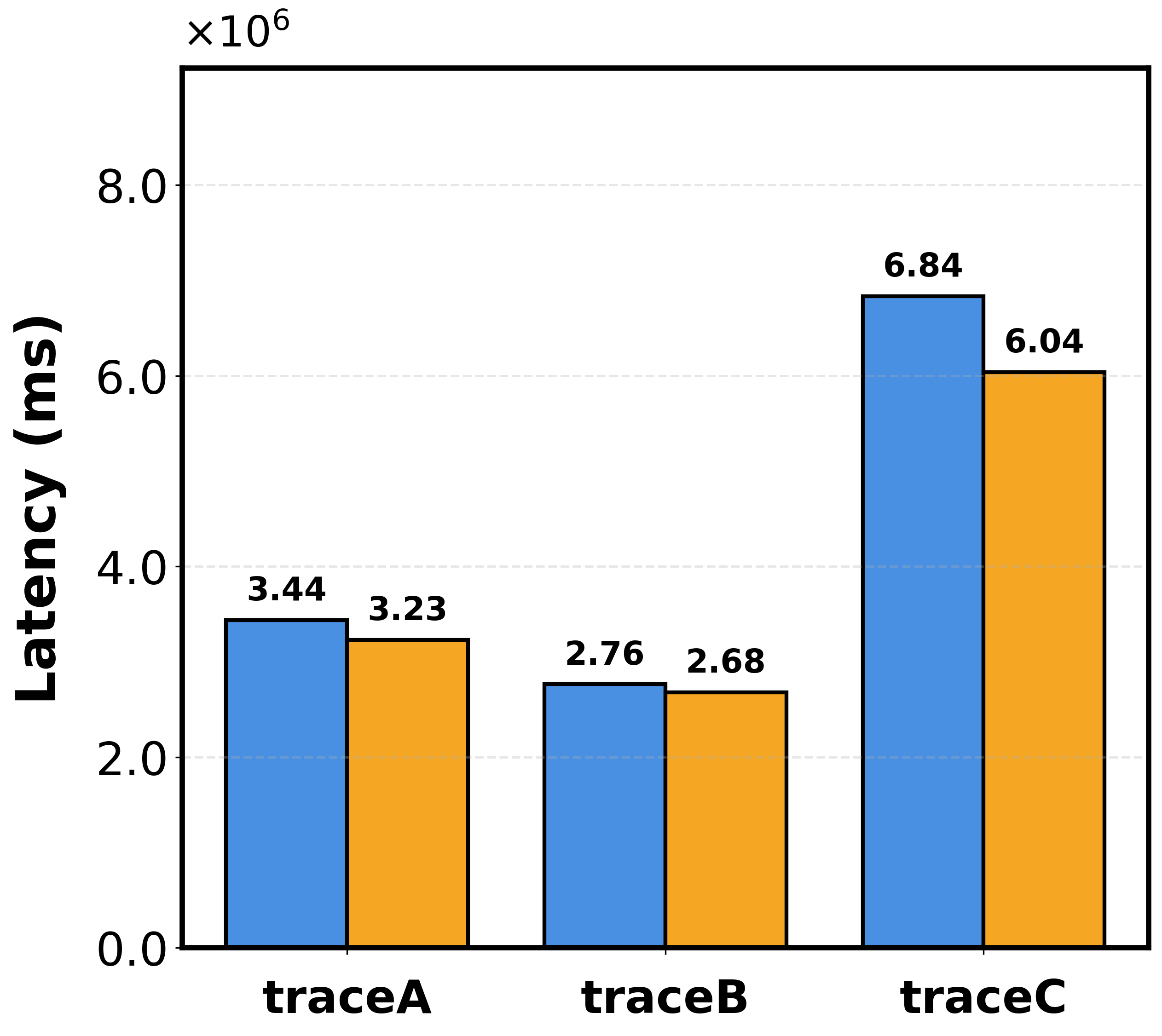} \includegraphics[width=0.32\linewidth]{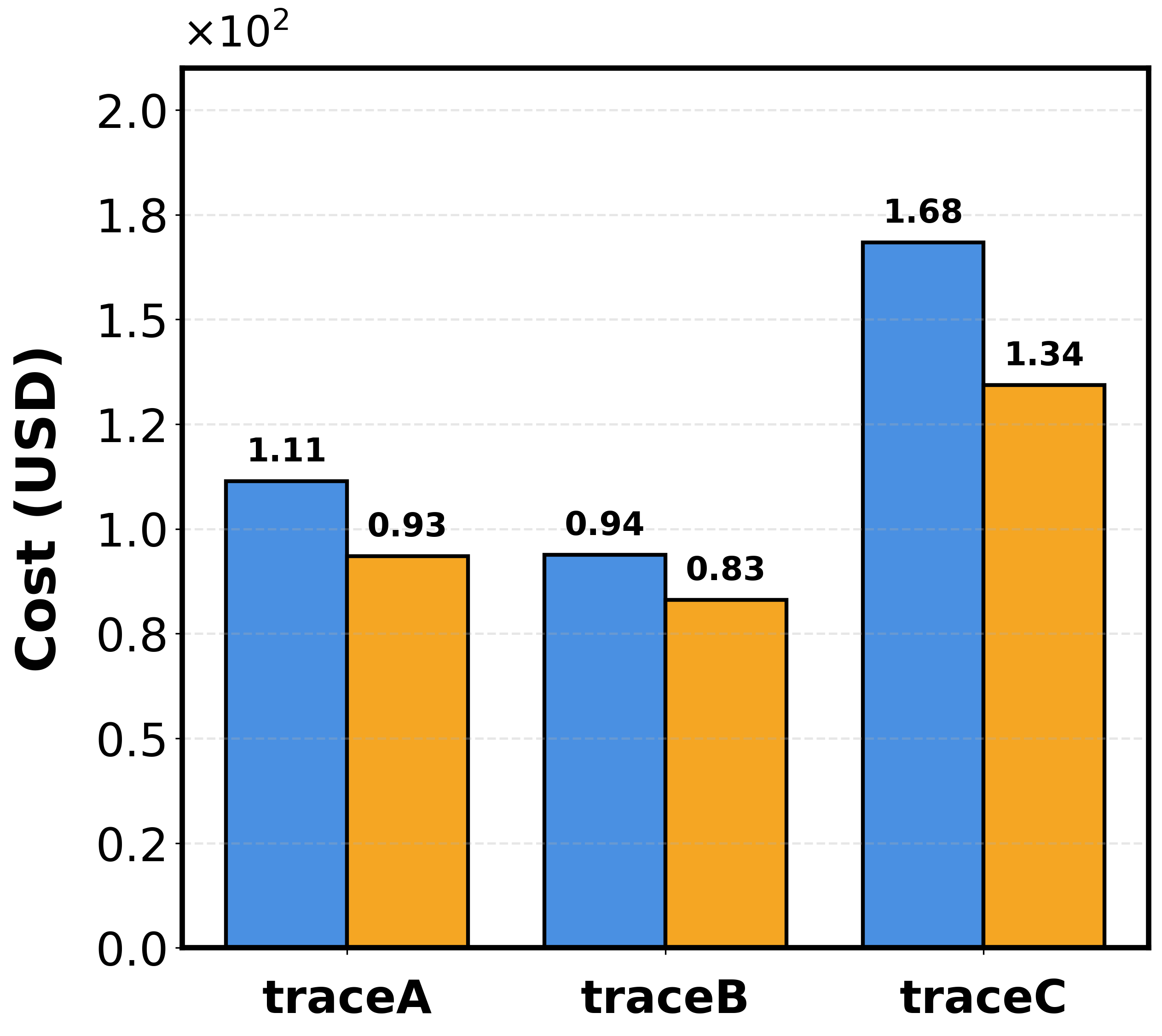}
\caption*{(ins1)}
\includegraphics[width=0.32\linewidth]{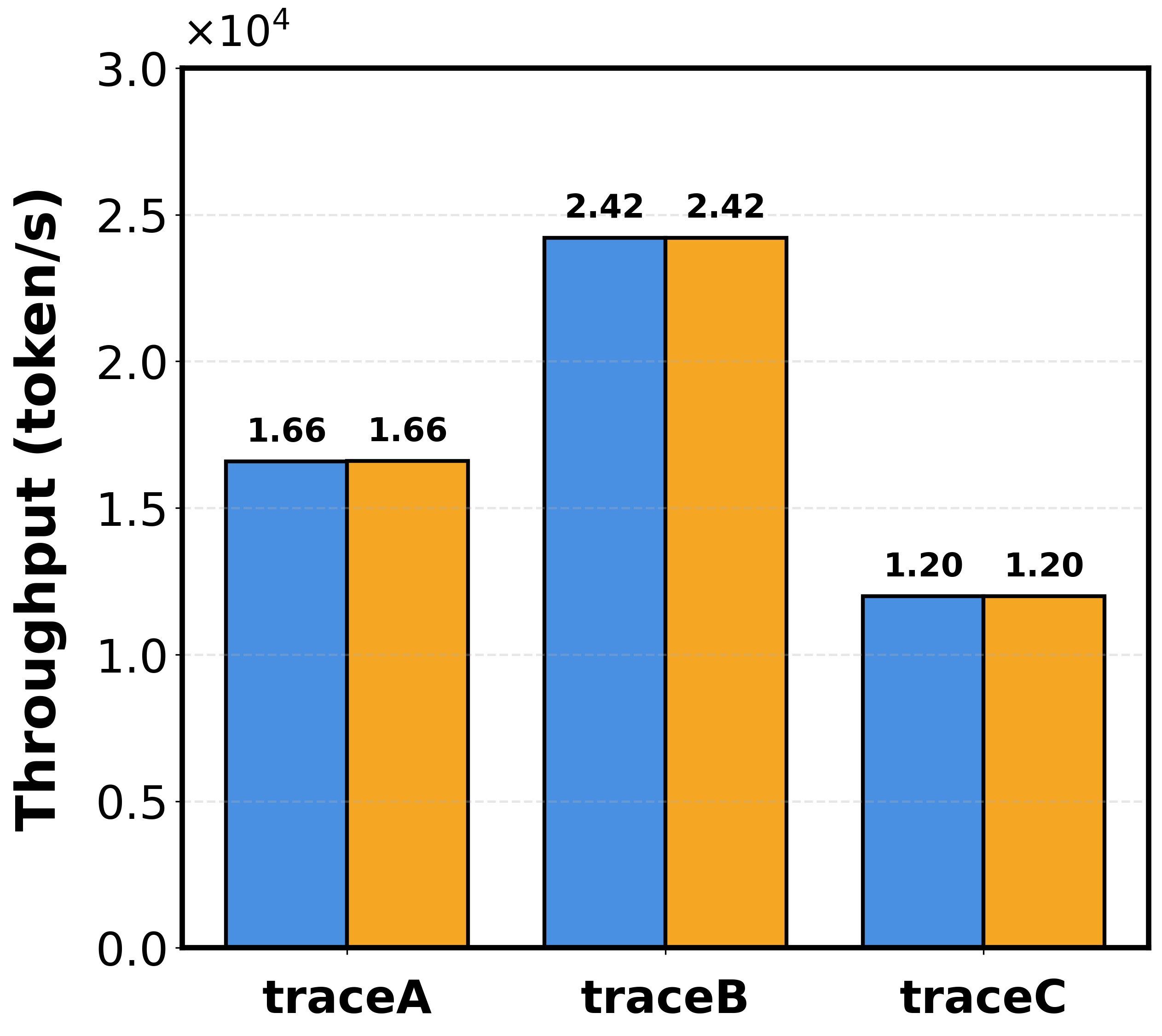} \includegraphics[width=0.32\linewidth]{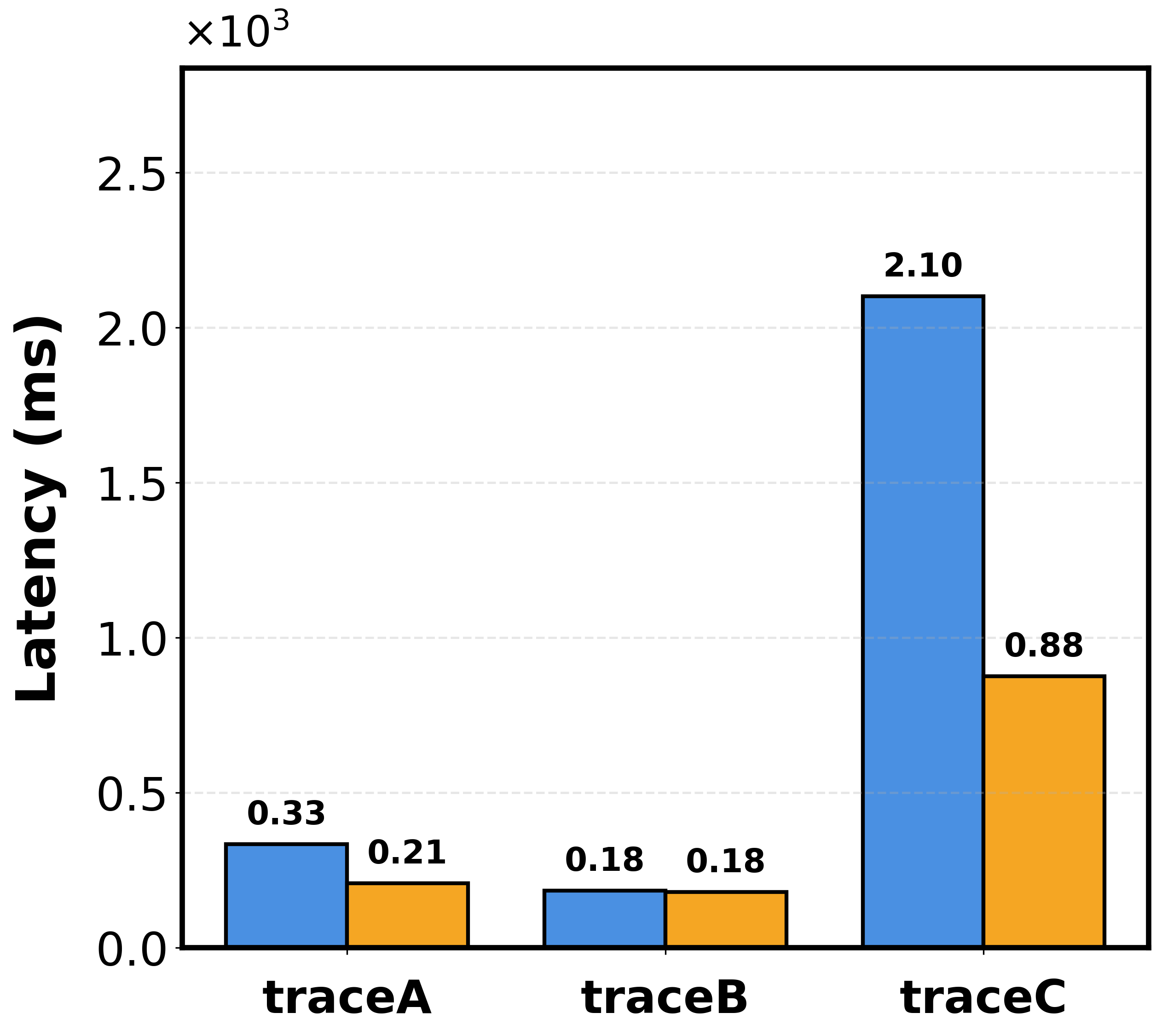} \includegraphics[width=0.32\linewidth]{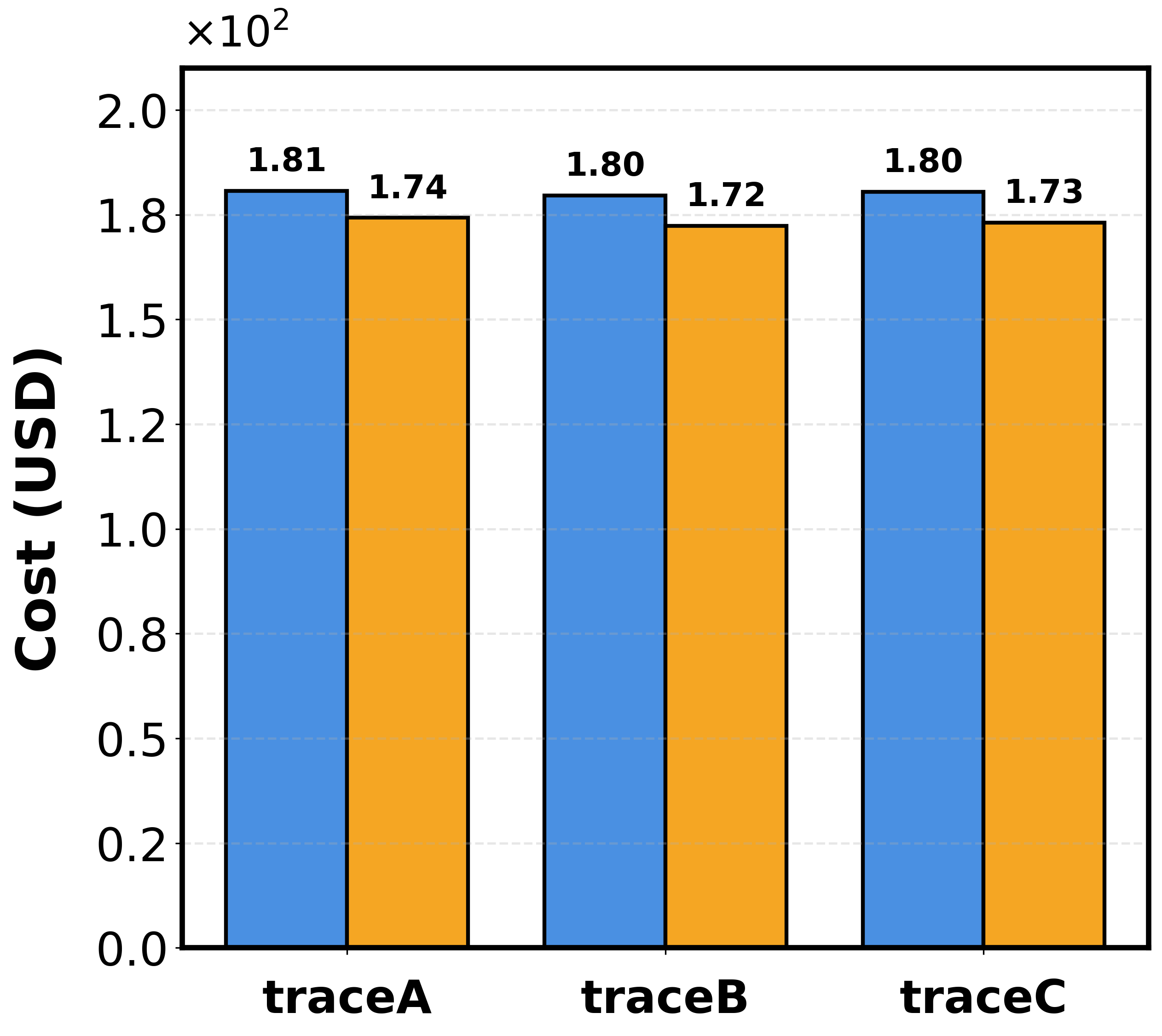}
\caption*{(ins4)}
\end{subfigure}

\caption{Performance comparison. The three subplots show the extreme points on the pareto frontier maximizing throughput, minimizing TTFT latency, and minimizing cost, each compared against the fixed configuration (1024GB DRAM only).}
\label{fig:eval_overview_traceBC}
\end{figure*}

To address this, we propose leveraging the information embedded within the prefix tree's structure to assign subtree-specific TTL values. \autoref{fig:design_group_ttl_subtree_rank} shows the reuse distribution of subtrees grouped by their initial prefixes in the trace, indicating that a small number of subtrees account for the majority of reuse and thus deserve special attention. Moreover, \autoref{fig:design_group_ttl_subtree_distribution} demonstrates substantial divergence in reuse behavior across subtrees, underscoring the need for distinct TTL assignments per subtree.
However, The same subtree exhibits similar reuse distribution characteristics across the different time periods.

Our objective is to improve cache hit rate under any given capacity budget $X$ through fine-grained, adaptive TTL configuration. This design integrates naturally with our Pareto frontier search framework.

To configure a reasonable TTL during online serving, we search for the optimal configuration using historical traces from a recent time window and apply it to set the TTL for the subsequent period.

Under our TTL mechanism, the total storage cost can be defined as
\[Cost_{disk}\cdot \sum_{\mathrm{block}}{(Capacity_{block}\cdot TTL_{block})}\]
Consequently, our optimization goal can be formally expressed as \[maximizing{\sum_{\mathrm{block}} ReuseCount_{block}}\] in budget storage cost.

Given block access patterns in a trace, we partition requests into $K+1$ groups: the top-$K$ frequent prefix subtrees $\{\mathcal{G}_k\}_{k=1}^K$ and a residual group $\mathcal{G}_{K+1}$ containing all other subtrees. For each group $\mathcal{G}_g$, let $\Delta_g = \{\delta_i^{(g)}\}$ denote the multiset of inter-arrival times of its constituent blocks. We define:

\begin{itemize}[leftmargin=*,nosep]
    \item \textbf{Cumulative hits} $H_g(t) = \bigl| \{ \delta \in \Delta_g \mid \delta \leq t \} \bigr|$: number of KVCache block hits under TTL $t$
    \item \textbf{Storage cost} $C_g(t) = |\mathcal{B}_g| \cdot t + \sum_{\delta \in \Delta_g} \min(t, \delta)$, where $|\mathcal{B}_g|$ is the number of unique blocks in $\mathcal{G}_g$
\end{itemize}

Let $\Delta_g = \{\delta_i^{(g)}\}_{i=1}^{N_g}$ denote the sequence of inter-arrival times 
for group $g$, where $N_g$ is the total number of reuse events (allowing duplicates).

Given a global storage budget $B$, we solve the constrained optimization problem:
\begin{equation}
\begin{aligned}
\max_{\mathbf{t} \in \mathbb{R}_+^{K+1}} & \quad \sum_{g=1}^{K+1} H_g(t_g) \\
\text{s.t.} & \quad \sum_{g=1}^{K+1} C_g(t_g) \leq B
\end{aligned}
\label{eq:ttl-opt}
\end{equation}

The objective is non-convex. We solve it via Sequential Least Squares Programming (SLSQP), which can efficiently navigates the non-convex landscape while respecting the given budget constraints.

To address the non-convexity in capacity scaling, which necessitates exploring multiple local optima, we adopt a multi-start strategy with a modest number of initial points, specifically $\lfloor \sqrt{K} \rfloor + 1$. 
We obtain an initial point \[\mathbf{t} = \{t_1, t_2, \dots, t_{K+1}\}\] by performing online statistics over a recent workload history window, where each $t_g$ maximizes the ROI (ratio of reuse to storage cost $\frac{H_g(t_g)}{C_g(t_g)}$) for its corresponding subtree. We then scale this initial point according to the total budget and generate multiple starting points by sampling around it.
The detailed algorithm is described in algorithm 2.

\section{Evaluation}
\label{sec:Evaluation}


\label{sec:experiment}

We conduct a comprehensive simulation-based evaluation comprising two aspects: (1) the overview effectiveness of Kareto, (2) ablation studies of two important design components: adaptive Pareto search and group TTL. Meanwhile, we conduct a simulator fidelity validation on a real GPU server to compare with the simulated results.

Our evaluation mainly focuses on three key metrics: total throughput (tokens per second), mean Time-To-First-Token (TTFT) in milliseconds, and total cost (including the GPU cost and the storage cost).

\subsection{Experimental Setup}
Most experiments are conducted on a dual-socket server equipped with two Intel Xeon 8369B CPUs (32 cores each), running our simulator that emulates an inference system based on Alibaba Cloud ecs.gn8v-8x.48xlarge instances~\cite{ali-baremetal} serving the Qwen3-480B model~\cite{qwen} via SGLang~\cite{sglang} with Alibaba Cloud enterprise SSD storage~\cite{ali-storage}. 

We evaluate Kareto on all three traces described in \S\ref{sec:density_characteristics}: trace A (interactive chatbot), trace B (API calls) and trace C (agent application). They cover distinct KVCache reuse patterns including traditional multi-turn chat where cache hits occur more stochastically(trace A and B), API calls where a high reuse ratio is contributed by shared prompt(trace B), and agent application where reuse intervals are affected by tool invocation durations~\cite{continuum}(trace C).

We simulate two cluster scales for two distinct scenarios:

\textbf{1-instance deployment:} a compute-constrained setting (e.g., batch processing application with limited GPUs, but can tolerate higher latency);

\textbf{4-instance deployment:} a compute-abundant setting (e.g., chatbot service that imposes strict low-latency requirements and is equipped with abundant GPU resources).

For each trace and scale, our planner explores configurations over two dimensions: DRAM capacity and TTL. We fix the storage medium type to enterprise SSD (Alibaba Cloud ESSD PL1) and focus our analysis on how DRAM/disk capacity provisioning shapes the Pareto frontier. The simulator outputs estimates of throughput, mean TTFT, and hourly cost calculated based on mainstream cloud pricing.

\subsection{Overview Effectiveness}
\label{subsec:overview}

We compare the Pareto-optimal configurations identified by our system against a practical baseline with 1024 GB DRAM, which represents a common provisioning practice in LLM inference clusters.
For each workload (Traces A/B/C) under both compute-constrained (1-instance) and compute-abundant (4-instances) settings, we select three representative configurations from Kareto's Pareto frontier that respectively maximize throughput, minimize mean TTFT latency, and minimize total cost to compare against the fixed baseline.

In terms of throughput, under the compute-constrained setting (1-instance) Kareto can achieve up to 9.3\% higher throughput. Under the compute-abundant setting (4-instances), increasing storage capacity provides almost no throughput benefit because throughput is limited by the request rate. Kareto captures this behavior and provides more cost-effective configurations.

In terms of latency, Kareto achieves up to 58.3\% lower mean TTFT by searching the configuration space. Notably, trace B shows more modest improvements, as its reuse is highly concentrated in a small fraction of blocks, so additional storage capacity yields little benefit. Kareto can also correctly identifies this regime and avoids over-provisioning capacity and provides configurations with lower storage capacity in order to reduce the cost.

For cost optimization, Kareto can reduce cost by up to 20.2\%. This is achieved by searching the configuration space to identify jointly right-sized DRAM–disk configurations. Under the compute-constrained (1-instance) setting, the main savings come from right-sizing to reduce GPU-hour cost while avoiding over-provisioning of DRAM and disk. Under the compute-abundant (4-instances) setting, the savings are primarily driven by avoiding over-provisioned storage capacity, thereby reducing storage cost.


\subsection{Ablation Experiment}

In this section, we evaluate two design components of Kareto.

\subsubsection{Adaptive Pareto Search}

\begin{figure}[!htbp]
\centering
\includegraphics[width=\linewidth]{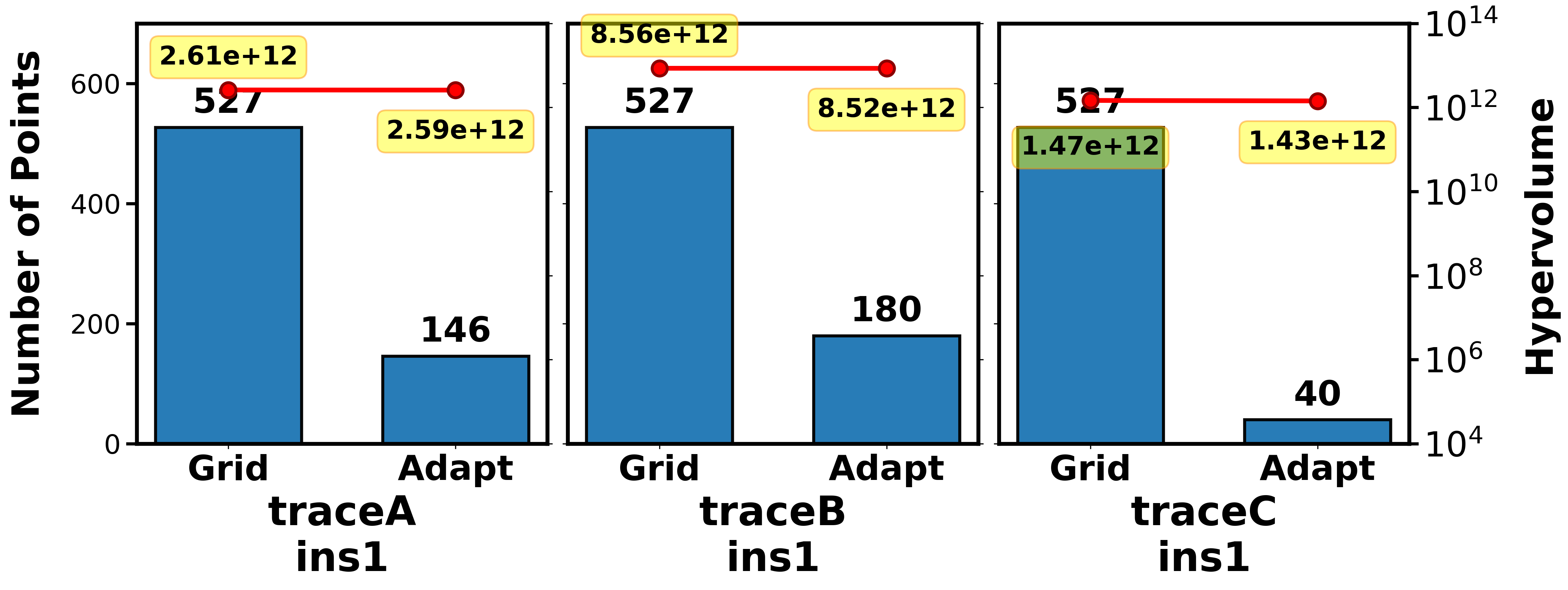}
\includegraphics[width=\linewidth]{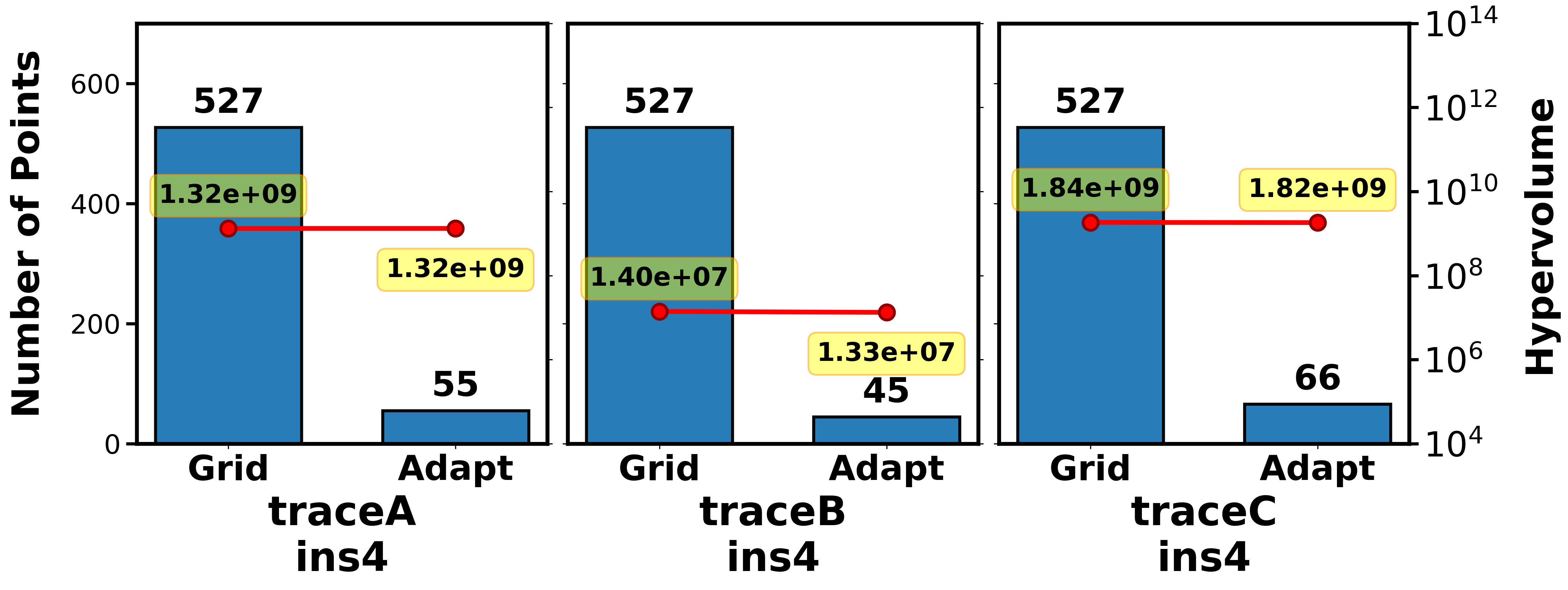}
\caption{Comparison of adaptive search and grid search in terms of search cost (number of evaluations) and solution quality (hypervolume).}
\label{fig:eval_ablation_adaptive}
\end{figure}

We evaluate the number of search points and the hypervolume of the Pareto front obtained by two search strategies: grid search and our adaptive search method. We compute hypervolume using an identical dominated reference point for both grid search and adaptive search under the same workload. The reference point is constructed to be strictly worse than all emulated configurations across the three objectives, allowing fair comparison. Grid search employs a uniform grid (DRAM: 0–4096 GB, step 256; disk: 0–3600 GB, step 120), while adaptive search initializes with a coarser grid (DRAM: 0–2048 GB, step 512; disk: 0–2400 GB, step 600) and adaptively refines subsequent exploration.

As presented in \autoref{fig:eval_ablation_adaptive}, our method requires lower number of search points
as it avoids excessive exploration in long-tail, where benefits diminish, and plain regions, where the impact of storage configuration is relatively minor.
Though sample on much fewer points, by refining the search more densely in high-gradient regions, our method still achieves a hypervolume close to the grid search, indicating the great coverage and quality of the Pareto frontier.

\subsubsection{Group Adaptive TTL}
\begin{figure}[t]
\centering

\begin{subfigure}[b]{0.49\columnwidth}
    \centering
    \includegraphics[width=\linewidth, trim=0 0 0 0, clip]{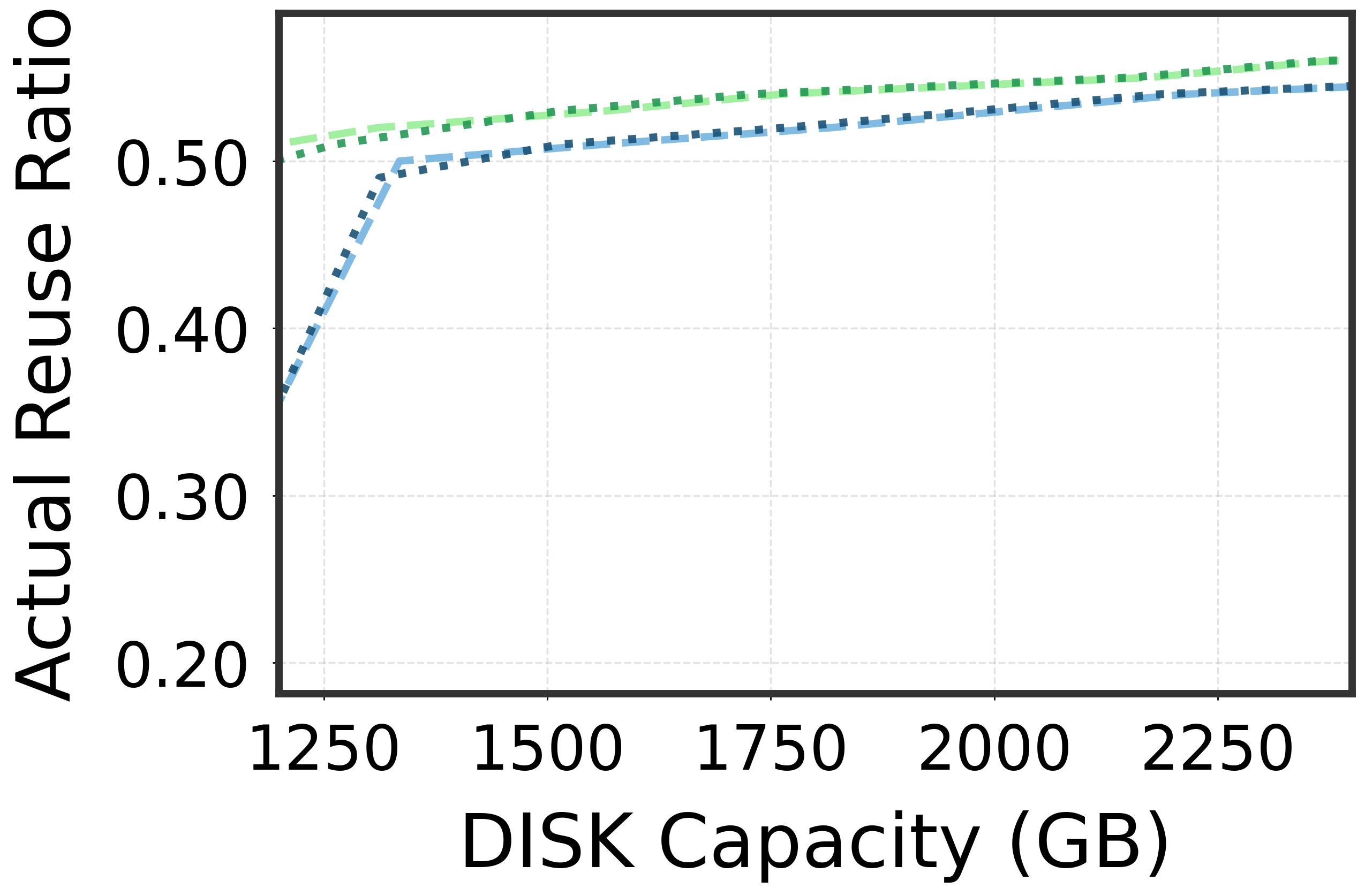}
    \caption{Trace A ins1}
    \label{fig:sub1}
\end{subfigure}
\hfill
\begin{subfigure}[b]{0.49\columnwidth}
    \centering
    \includegraphics[width=\linewidth]{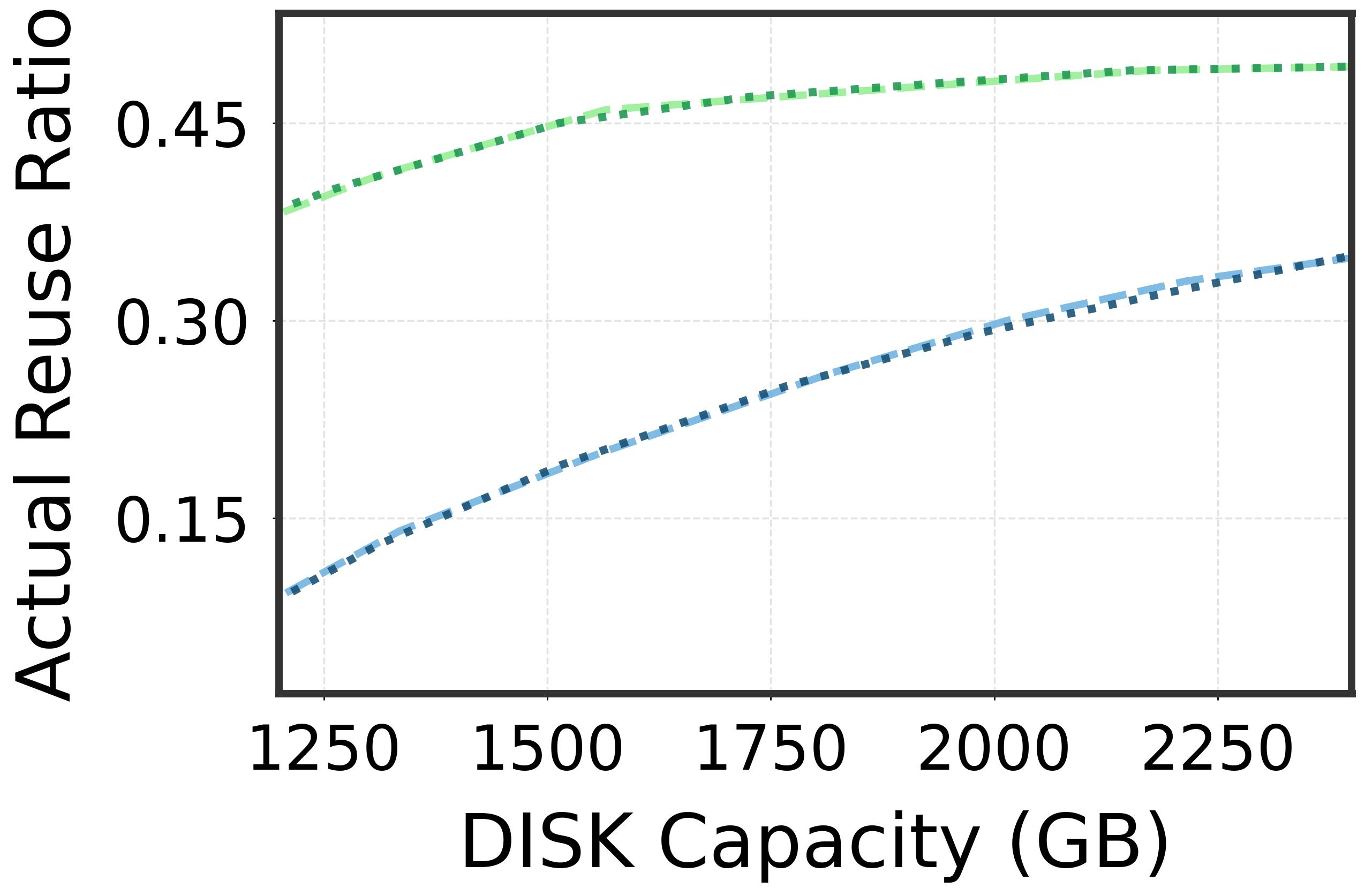}
    \caption{Trace A ins4}
    \label{fig:sub2}
\end{subfigure}

\begin{subfigure}[b]{0.49\columnwidth}
    \centering
    \includegraphics[width=\linewidth, trim=0 0 0 0, clip]{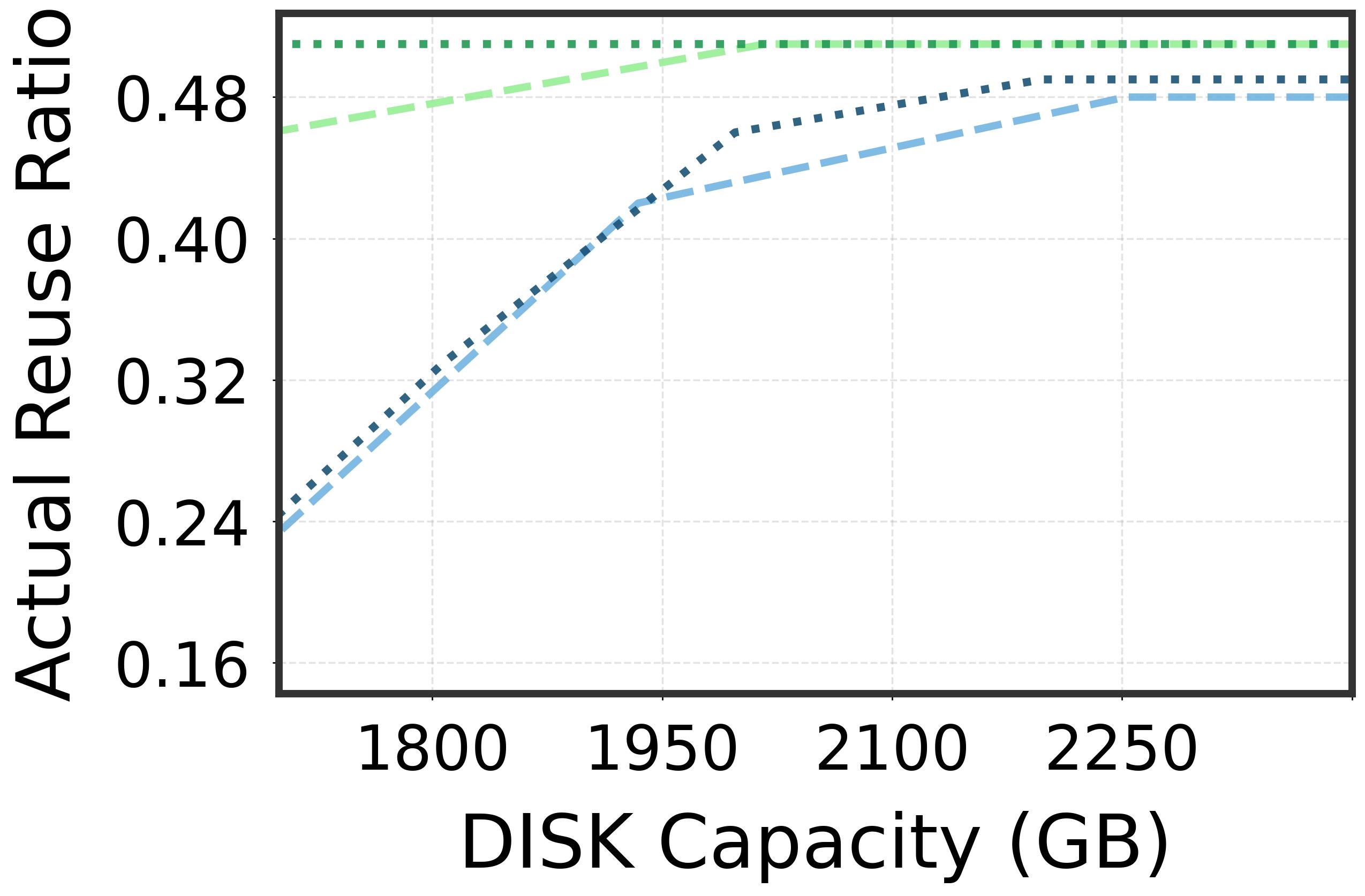}
    \caption{Trace B ins1}
    \label{fig:sub1}
\end{subfigure}
\hfill
\begin{subfigure}[b]{0.49\columnwidth}
    \centering
    \includegraphics[width=\linewidth]{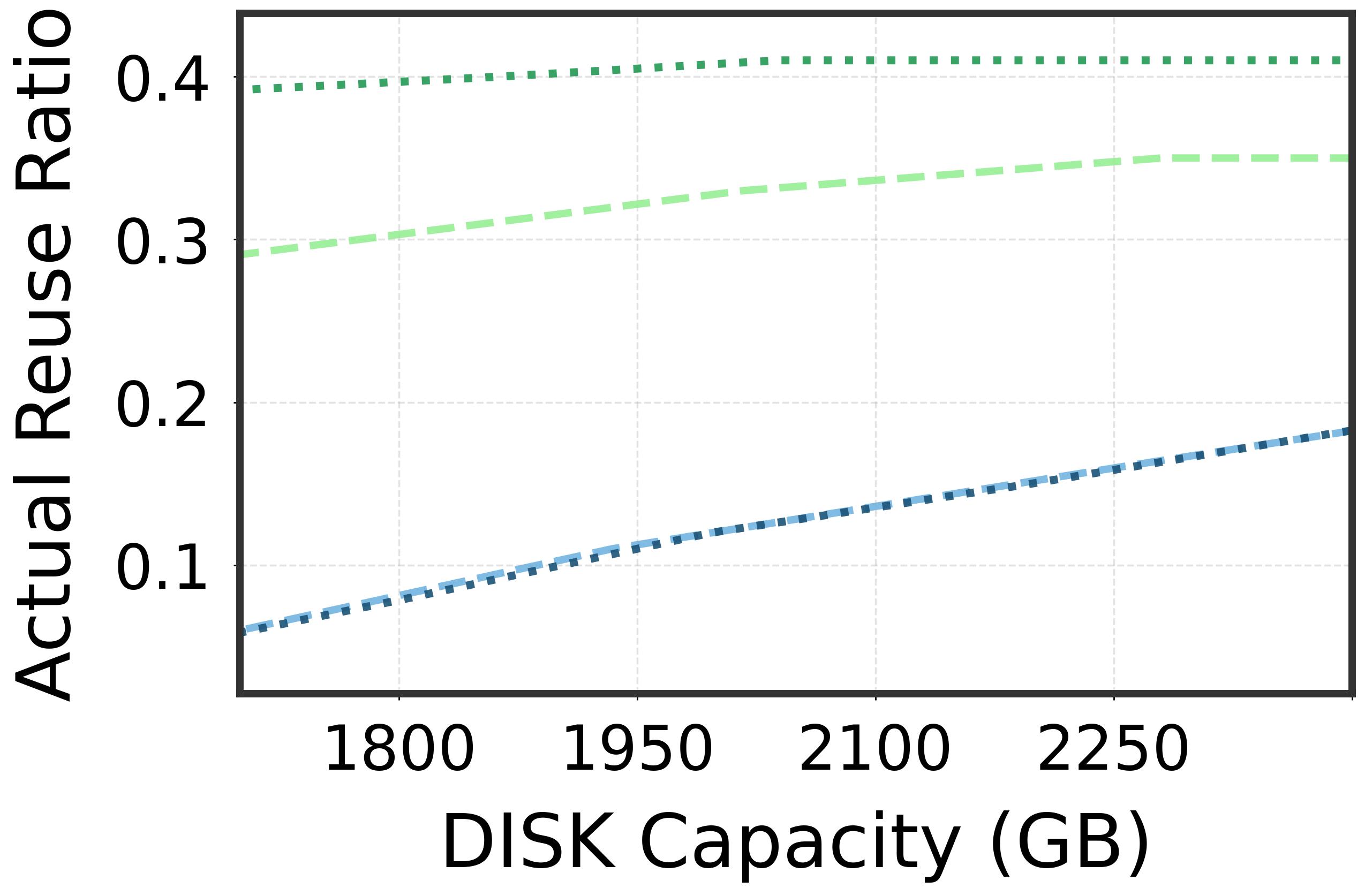}
    \caption{Trace B ins4}
    \label{fig:sub2}
\end{subfigure}

\vspace{4pt} 

\begin{subfigure}[b]{0.49\columnwidth}
    \centering
    \includegraphics[width=\linewidth]{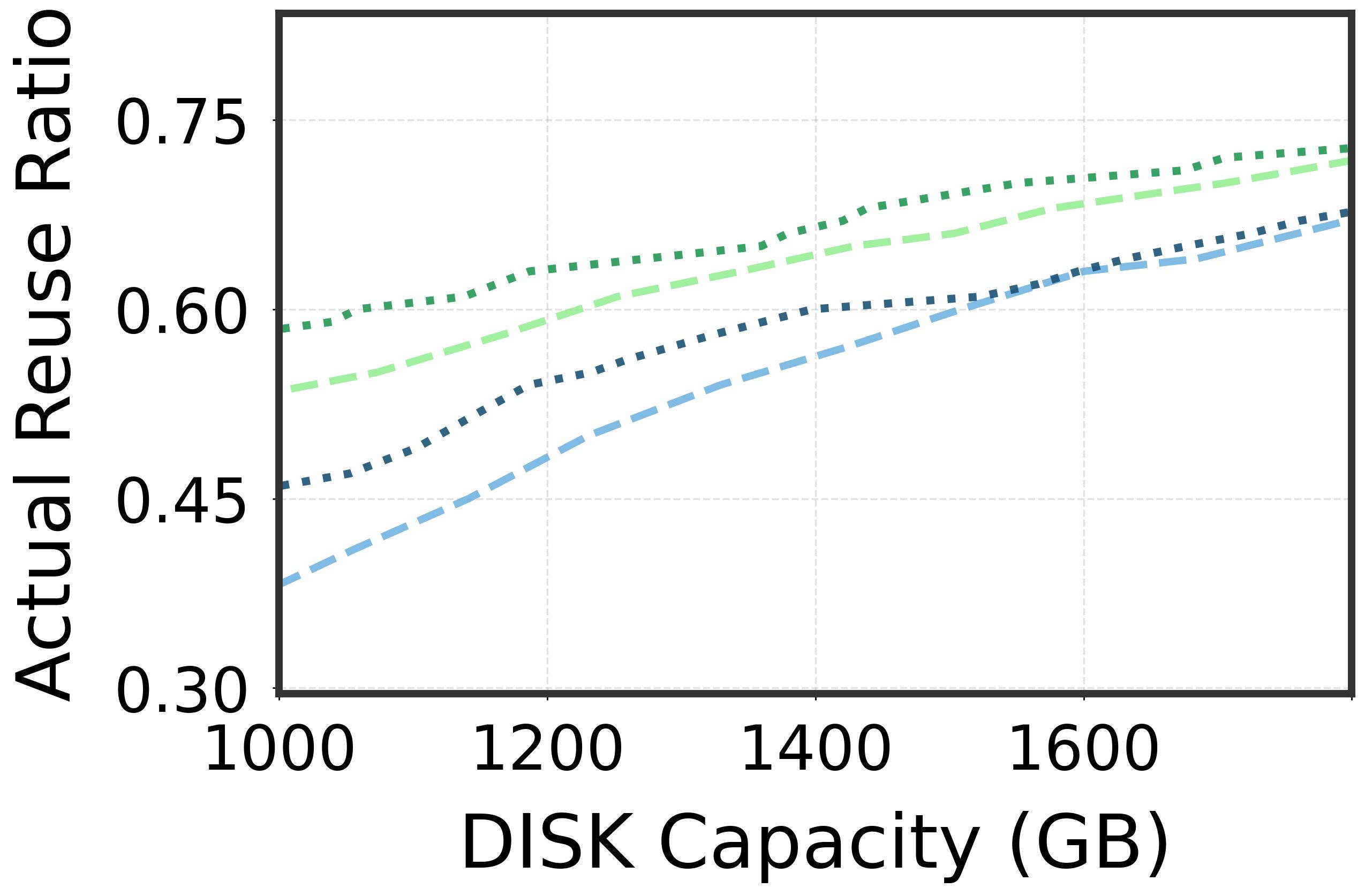}
    \caption{Trace C ins1}
    \label{fig:sub3}
\end{subfigure}
\hfill
\begin{subfigure}[b]{0.49\columnwidth}
    \centering
    \includegraphics[width=\linewidth]{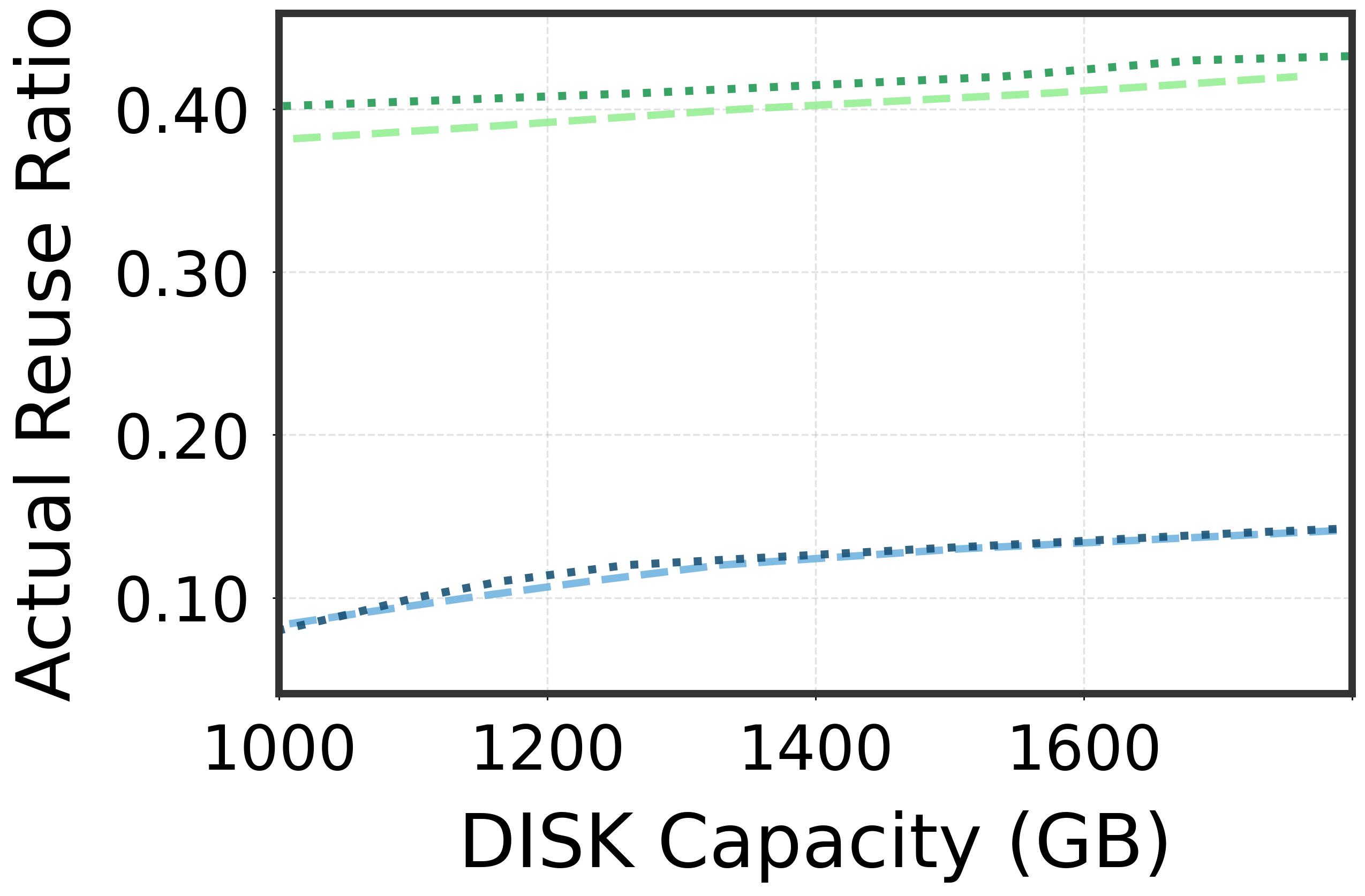}
    \caption{Trace C ins4}
    \label{fig:sub4}
\end{subfigure}

\includegraphics[width=0.7\columnwidth]{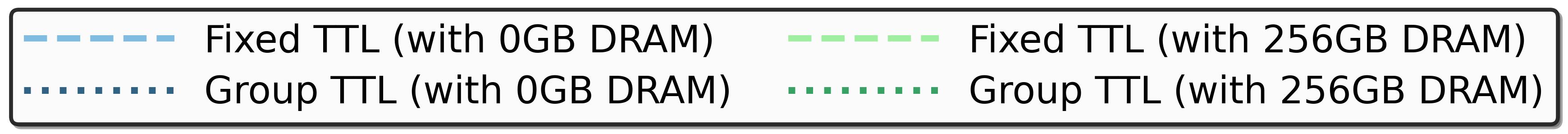}

\caption{Comparison on actual reuse ratio of group ttl and fixed ttl.}
\label{fig:new_ablation_group_reuse}
\end{figure}

\vspace{5ex}

\begin{figure*}[t]
    \centering
    
    \begin{subfigure}[b]{\textwidth}
        \centering
        \begin{subfigure}[b]{0.33\textwidth}
            \centering
            \includegraphics[width=\linewidth]{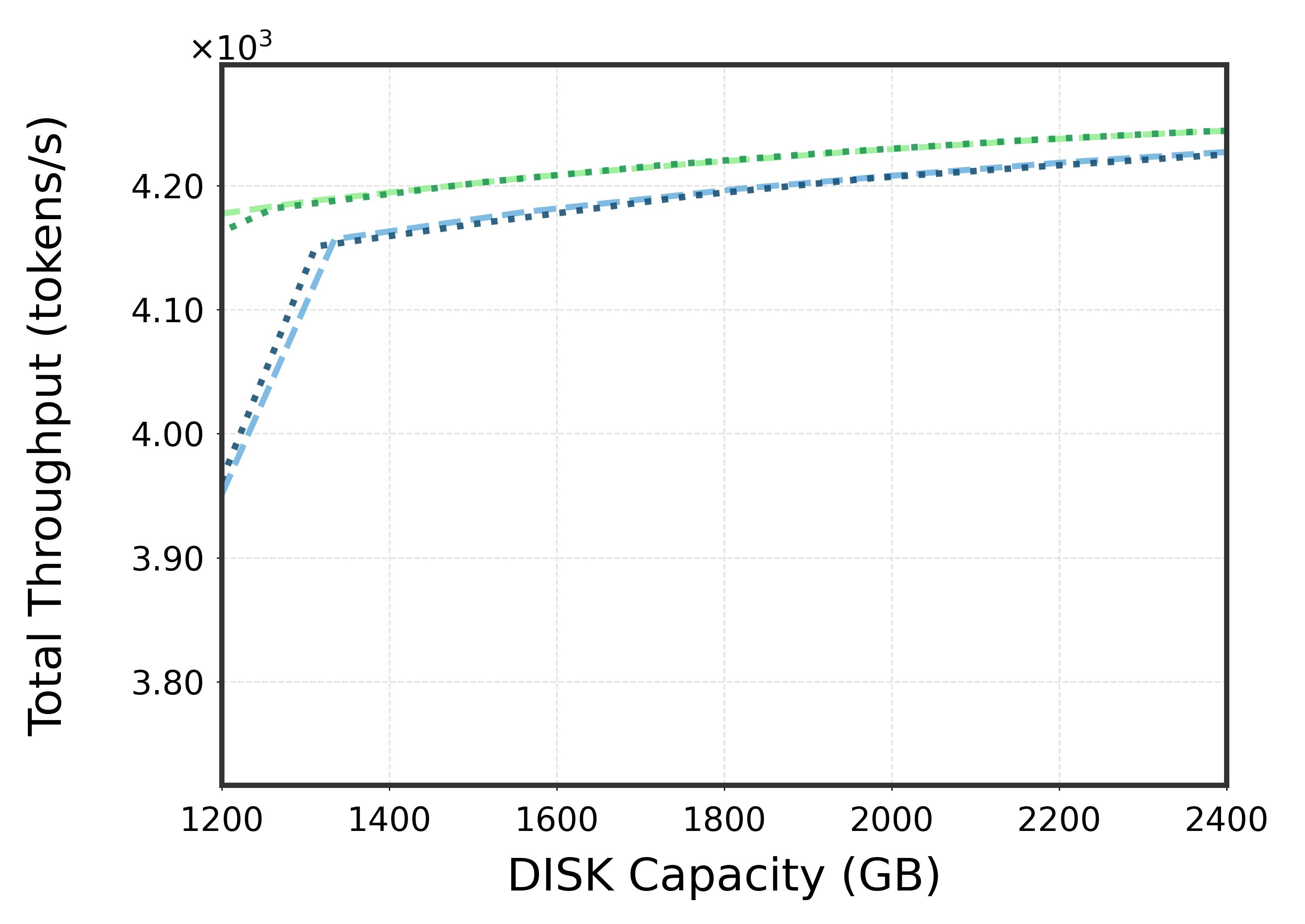}
        \end{subfigure}
        \hfill
        \begin{subfigure}[b]{0.33\textwidth}
            \centering
            \includegraphics[width=\linewidth]{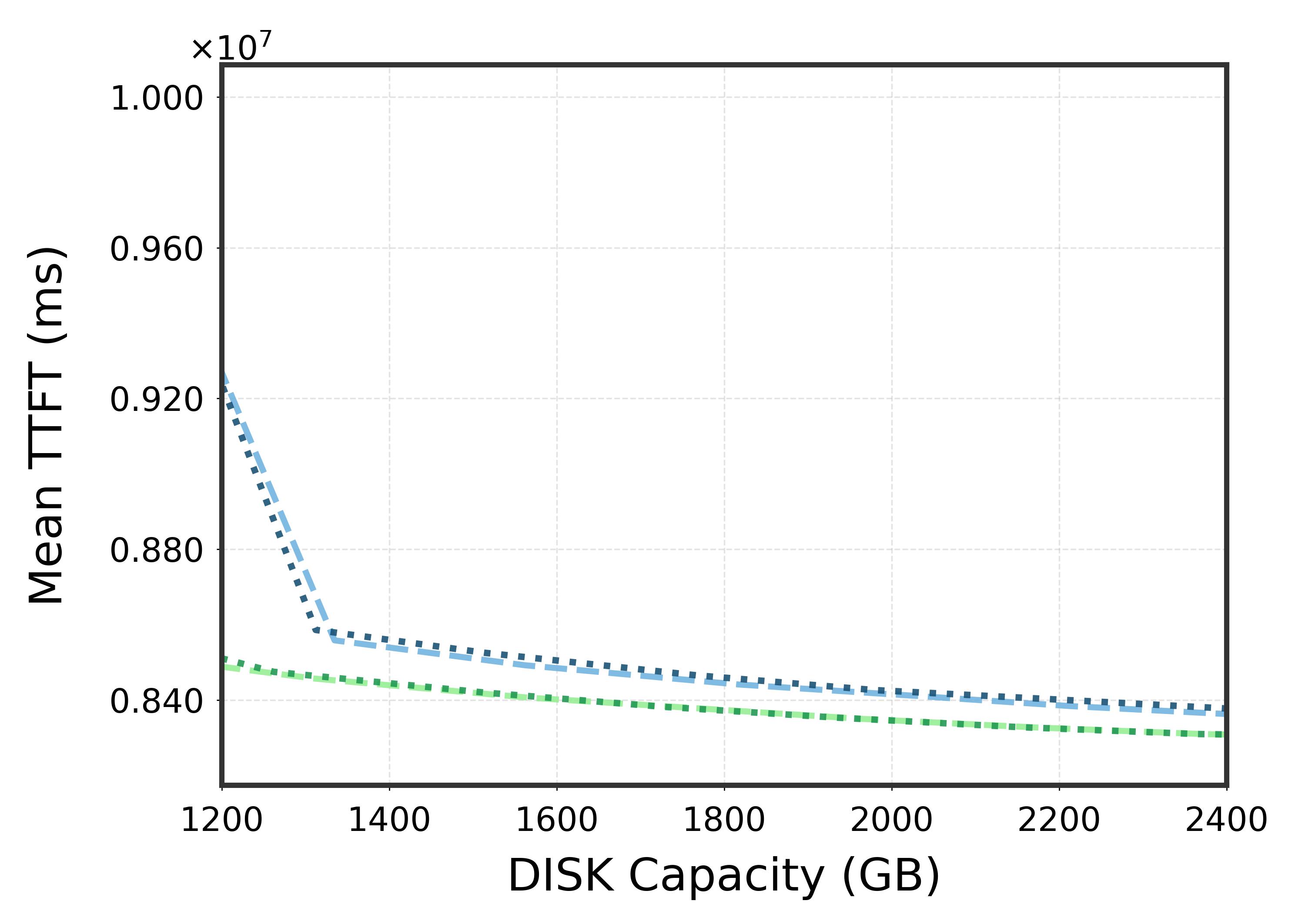}
        \end{subfigure}
        \hfill
        \begin{subfigure}[b]{0.33\textwidth}
            \centering
            \includegraphics[width=\linewidth]{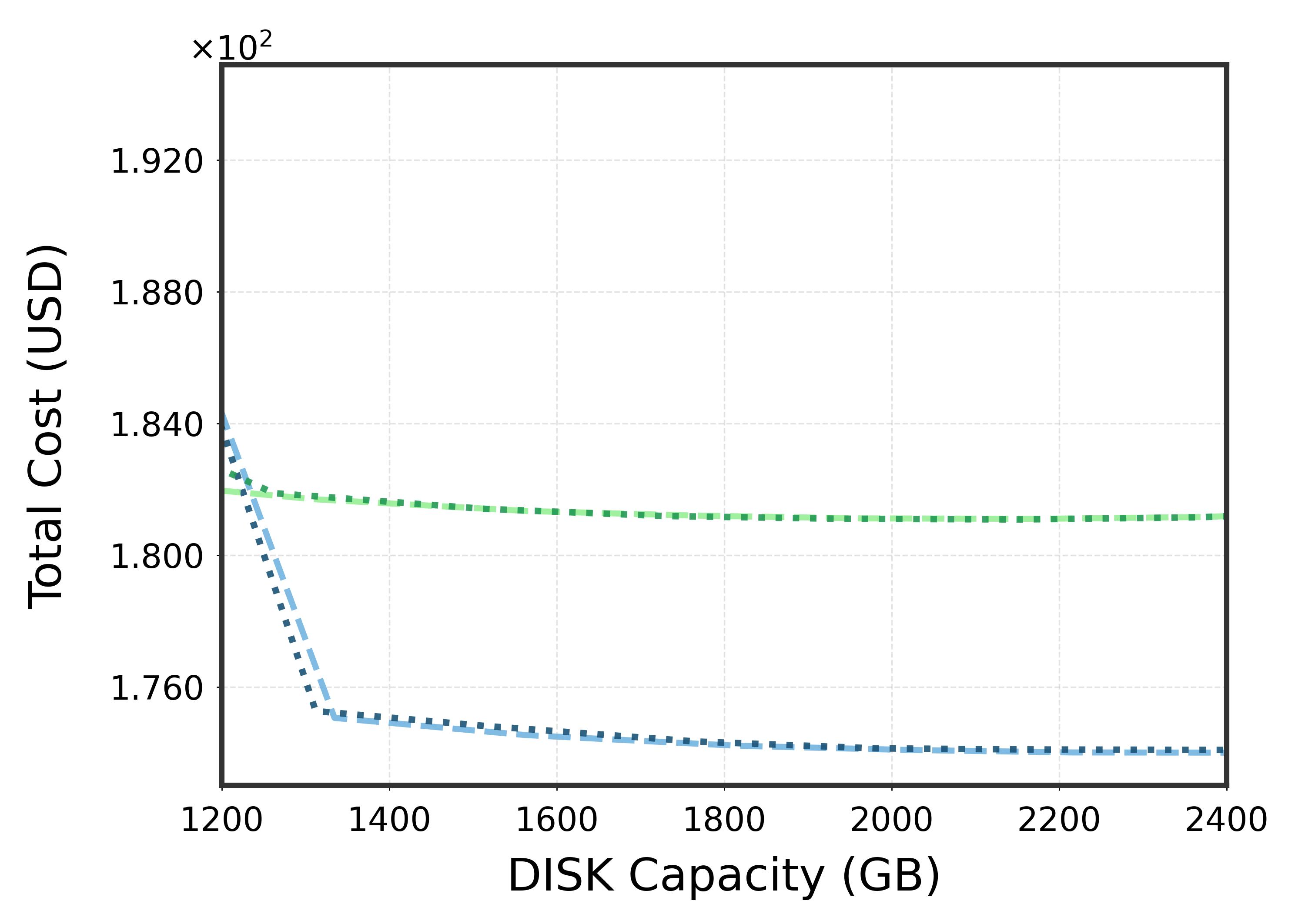}
        \end{subfigure}
        \subcaption*{\textbf{traceA ins1}}
    \end{subfigure}
    
    \vspace{3ex}
    
    \begin{subfigure}[b]{\textwidth}
        \centering
        \begin{subfigure}[b]{0.33\textwidth}
            \centering
            \includegraphics[width=\linewidth]{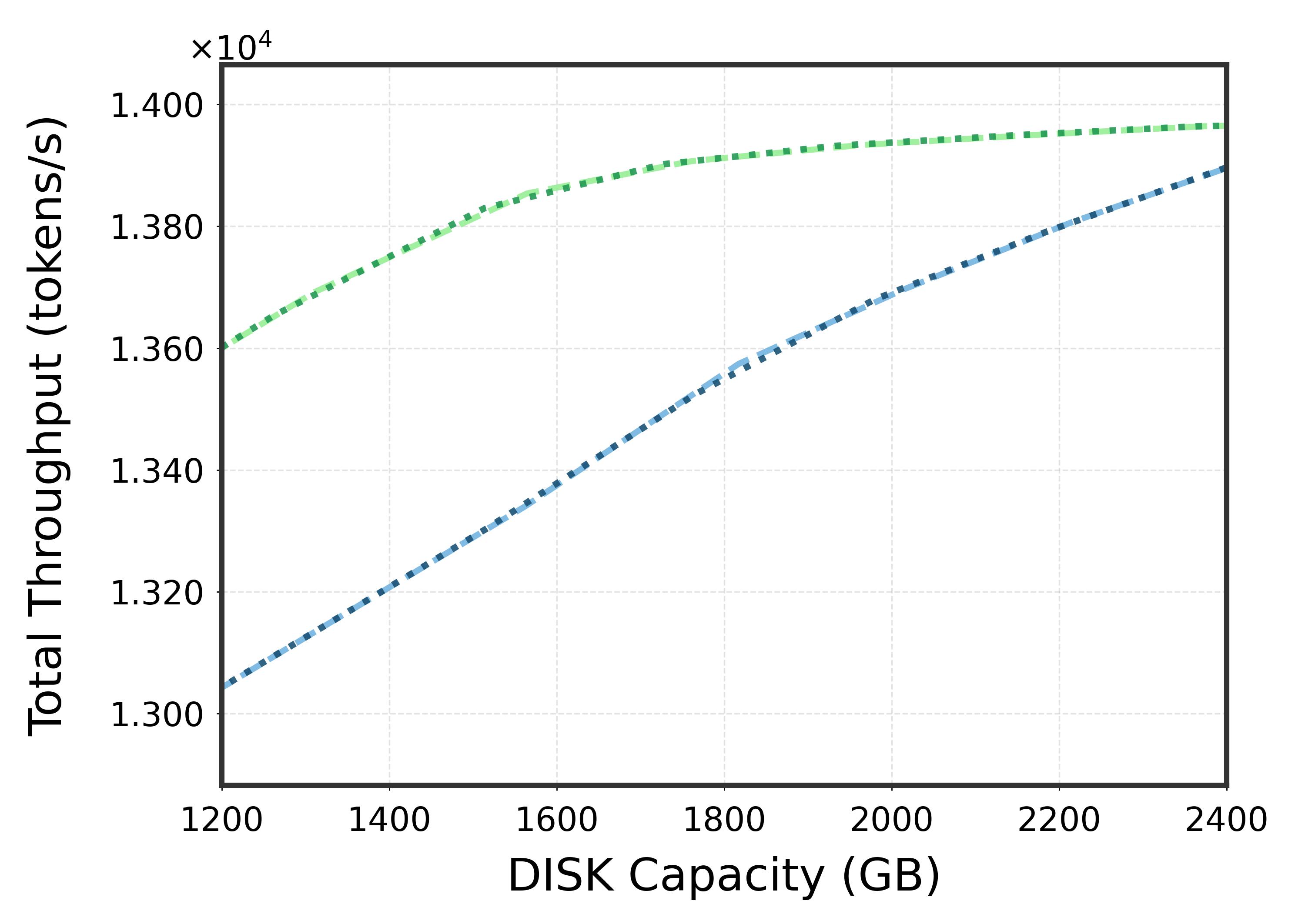}
        \end{subfigure}
        \hfill
        \begin{subfigure}[b]{0.33\textwidth}
            \centering
            \includegraphics[width=\linewidth]{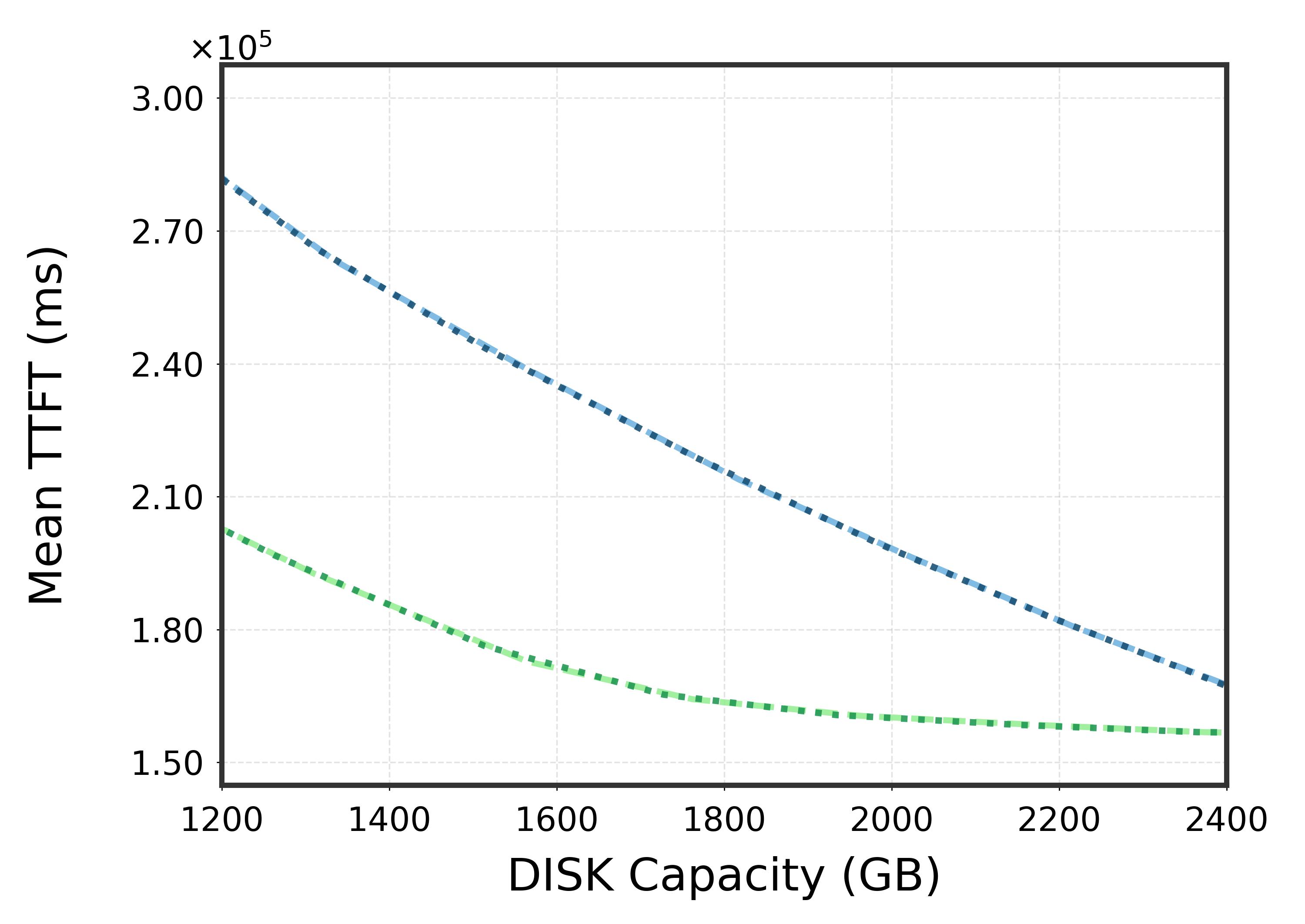}
        \end{subfigure}
        \hfill
        \begin{subfigure}[b]{0.33\textwidth}
            \centering
            \includegraphics[width=\linewidth]{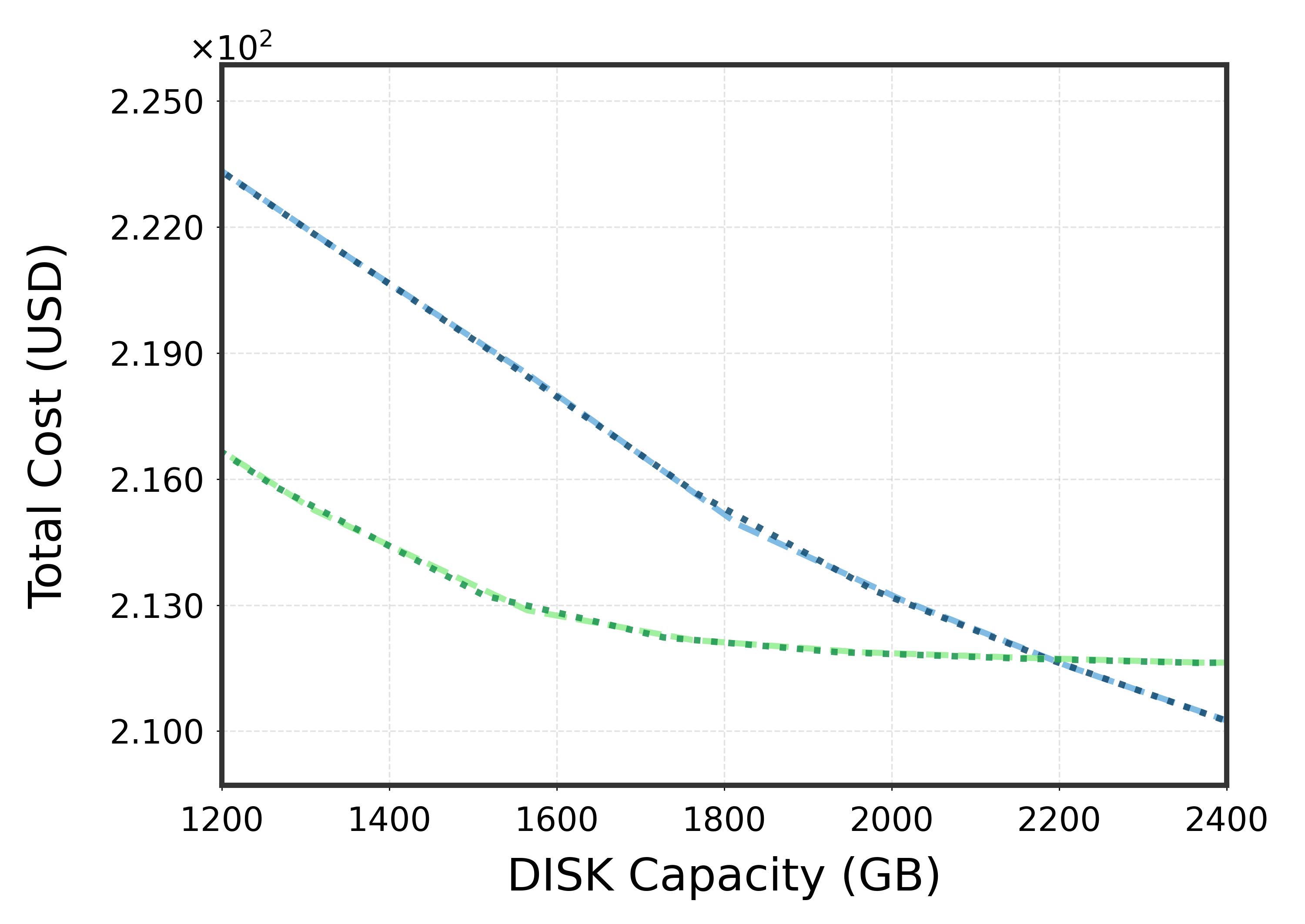}
        \end{subfigure}
        \subcaption*{\textbf{traceA ins4}}
    \end{subfigure}
    
    \vspace{3ex}
    
    \begin{subfigure}[b]{\textwidth}
        \centering
        \begin{subfigure}[b]{0.33\textwidth}
            \centering
            \includegraphics[width=\linewidth]{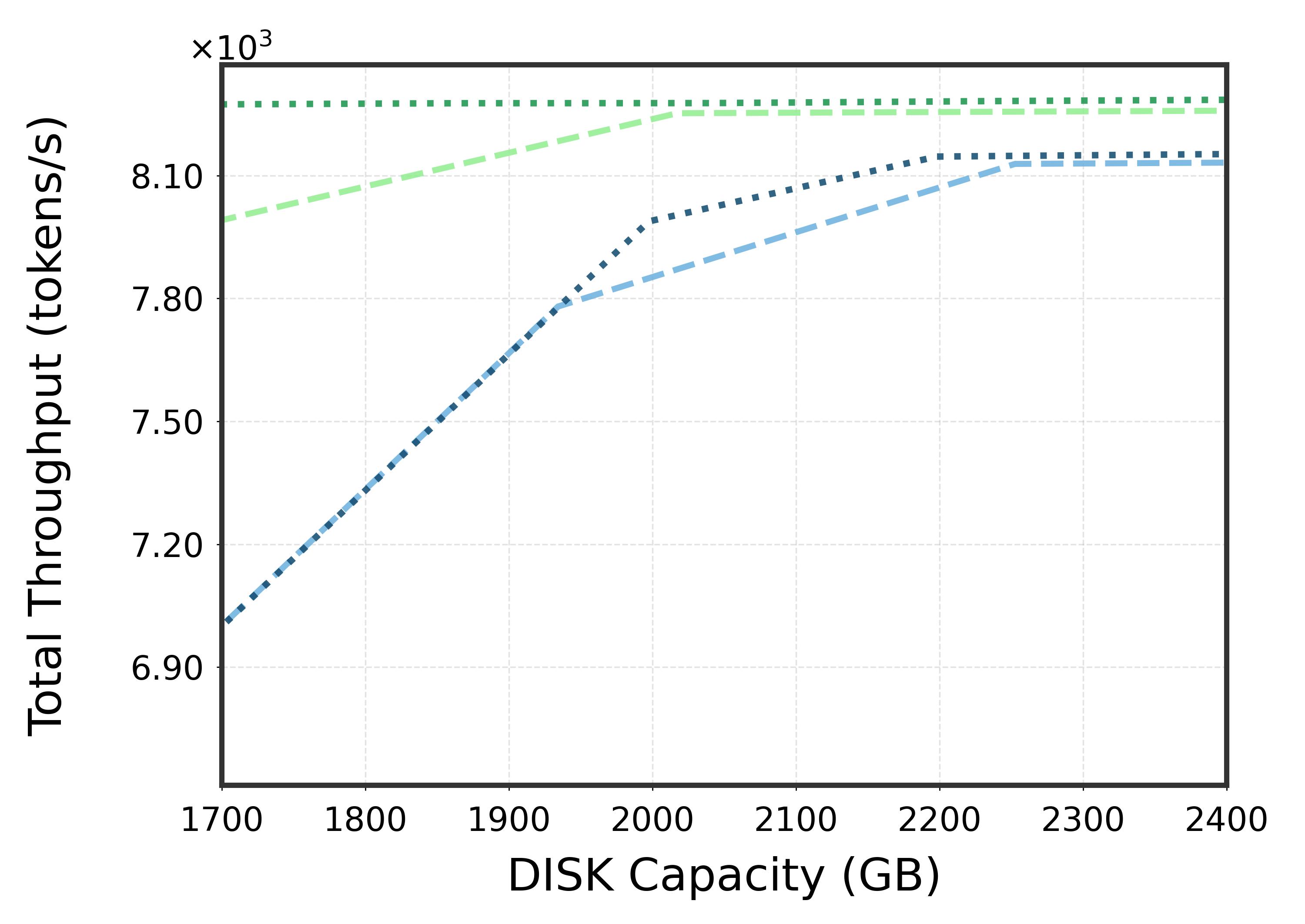}
        \end{subfigure}
        \hfill
        \begin{subfigure}[b]{0.33\textwidth}
            \centering
            \includegraphics[width=\linewidth]{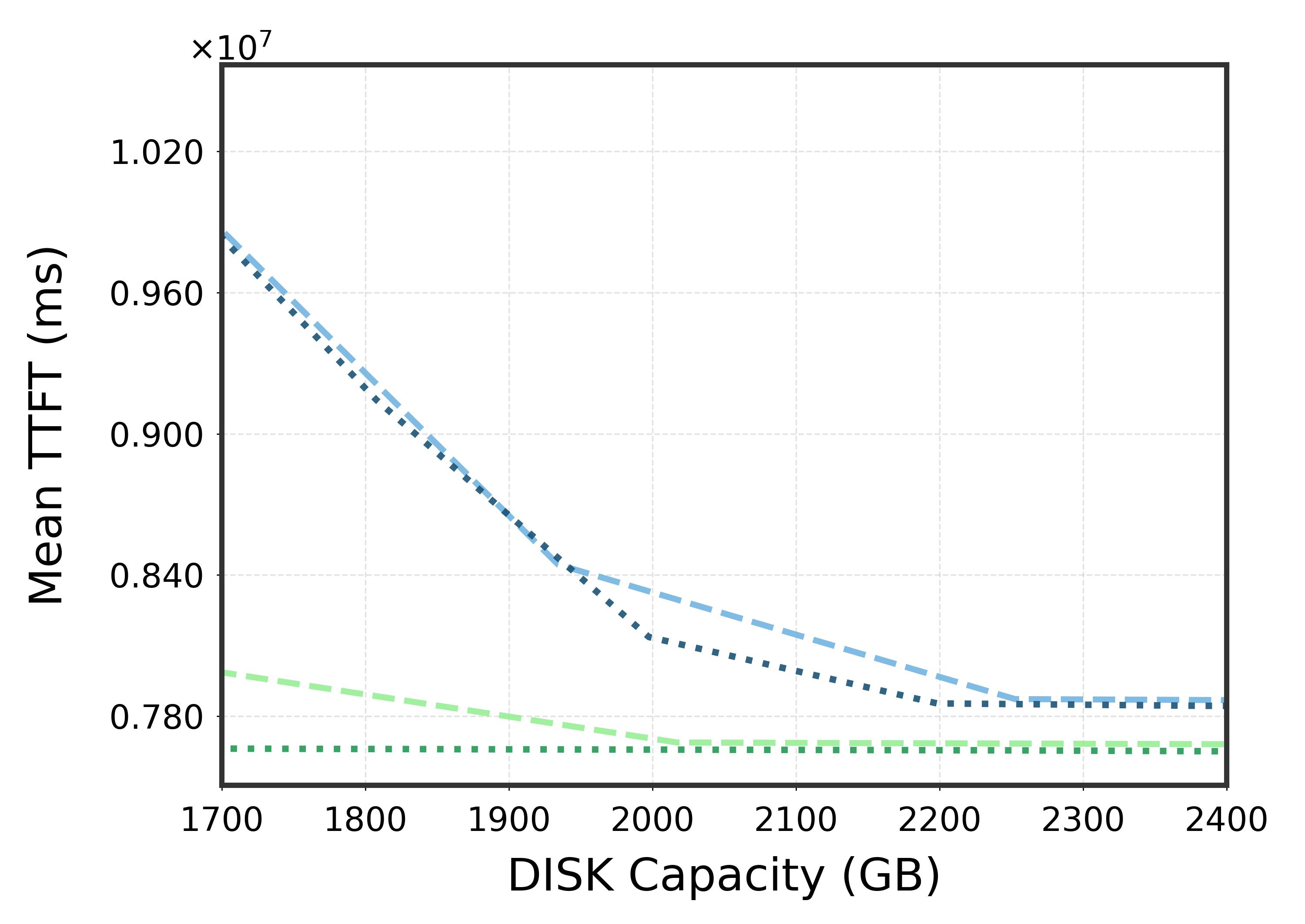}
        \end{subfigure}
        \hfill
        \begin{subfigure}[b]{0.33\textwidth}
            \centering
            \includegraphics[width=\linewidth]{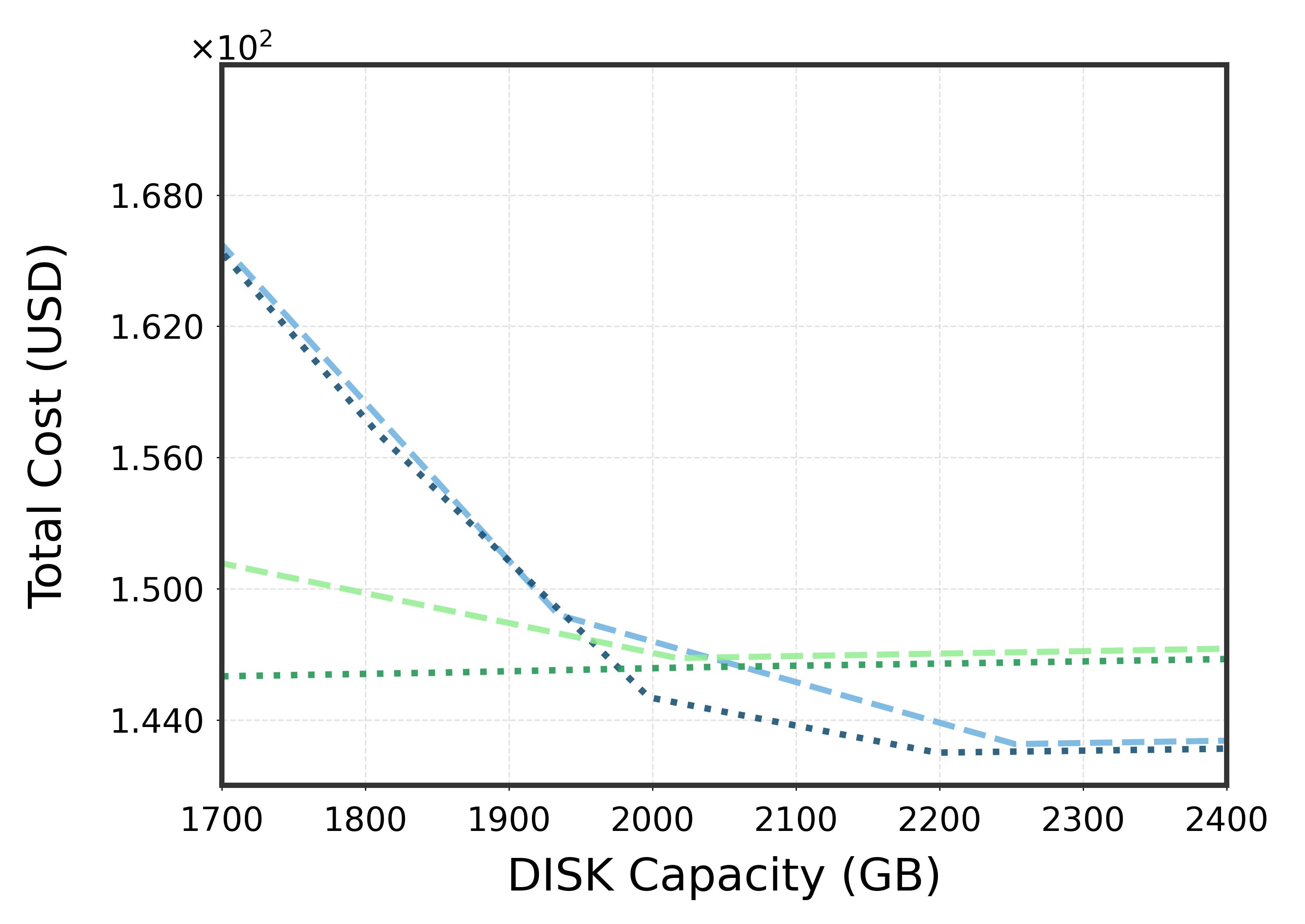}
        \end{subfigure}
        \subcaption*{\textbf{traceB ins1}}
    \end{subfigure}
    
    \vspace{3ex}
    
    \begin{subfigure}[b]{\textwidth}
        \centering
        \begin{subfigure}[b]{0.33\textwidth}
            \centering
            \includegraphics[width=\linewidth]{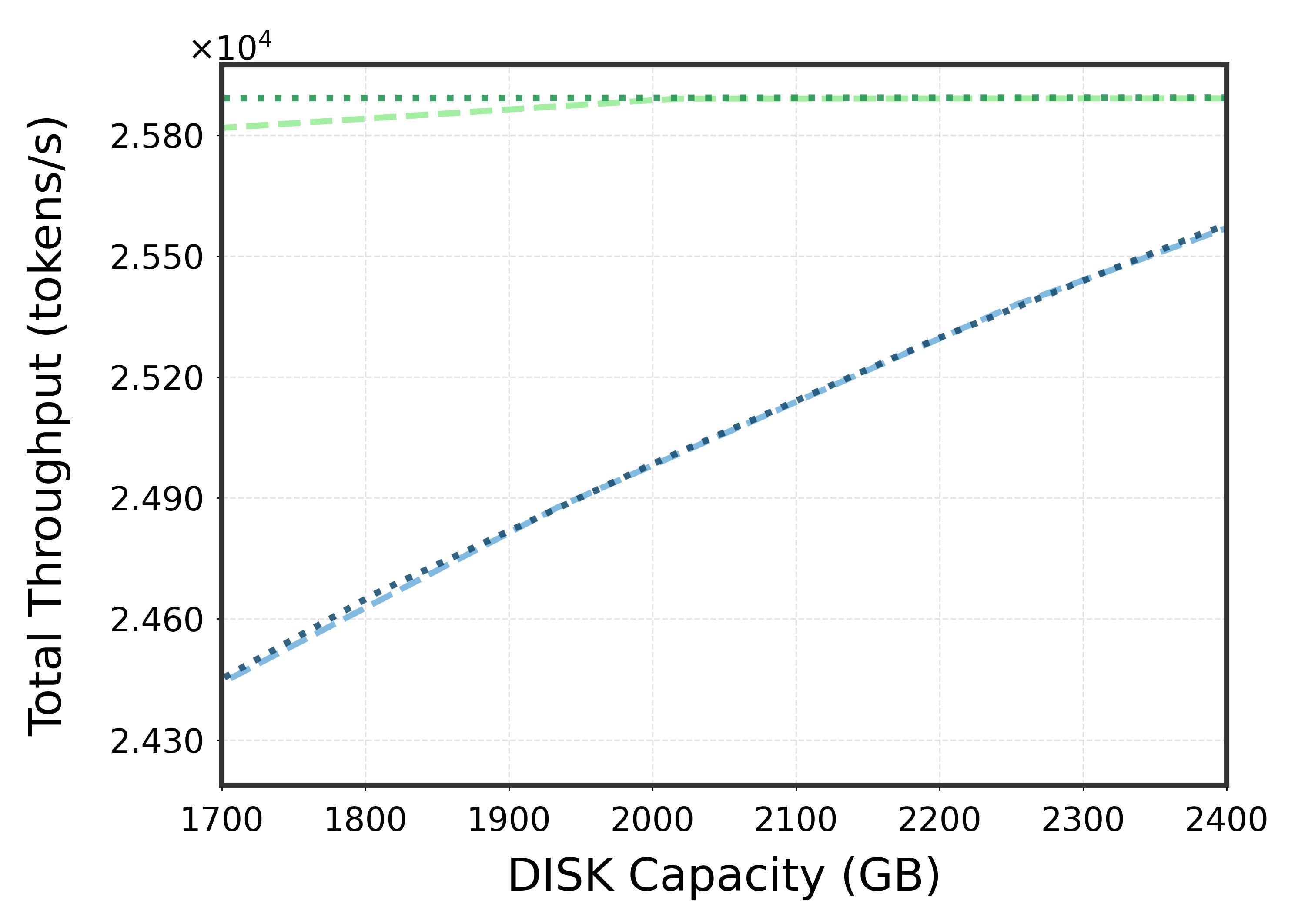}
        \end{subfigure}
        \hfill
        \begin{subfigure}[b]{0.33\textwidth}
            \centering
            \includegraphics[width=\linewidth]{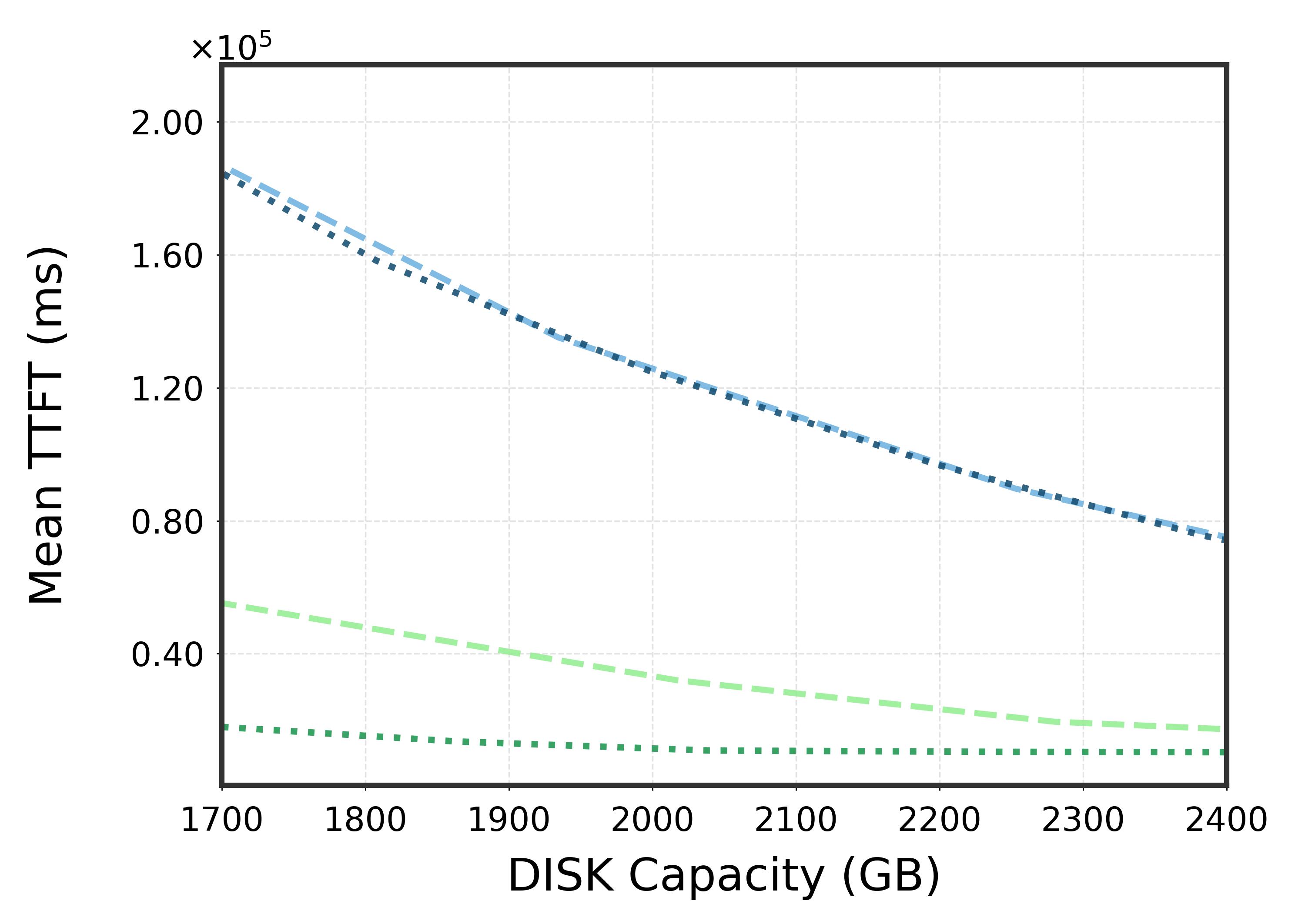}
        \end{subfigure}
        \hfill
        \begin{subfigure}[b]{0.33\textwidth}
            \centering
            \includegraphics[width=\linewidth]{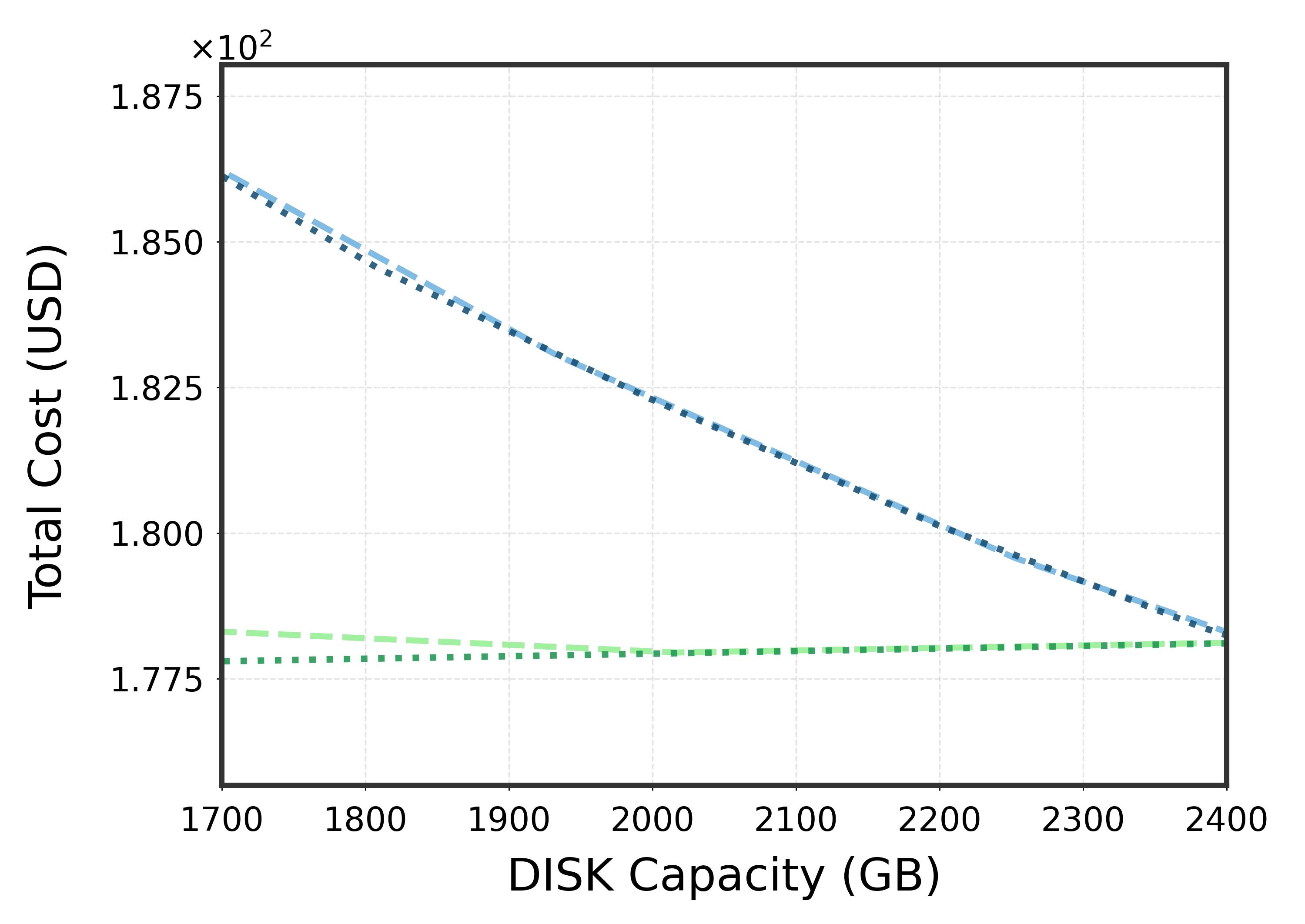}
        \end{subfigure}
        \subcaption*{\textbf{traceB ins4}}
    \end{subfigure}

    \begin{subfigure}[b]{\textwidth}
        \centering
        \includegraphics[width=0.4\linewidth]{figs/evaluation/ablation_group_ttl/legend.jpg}
    \end{subfigure}

    \caption{Ablation study: Comparison of group TTL and fixed TTL across throughput, TTFT latency, and cost under different traces and instance scales}
    \label{fig:ablation_group_ttl_A}

\end{figure*}

\begin{figure*}[t]

    \begin{subfigure}[b]{\textwidth}
        \centering
        \begin{subfigure}[b]{0.33\textwidth}
            \centering
            \includegraphics[width=\linewidth]{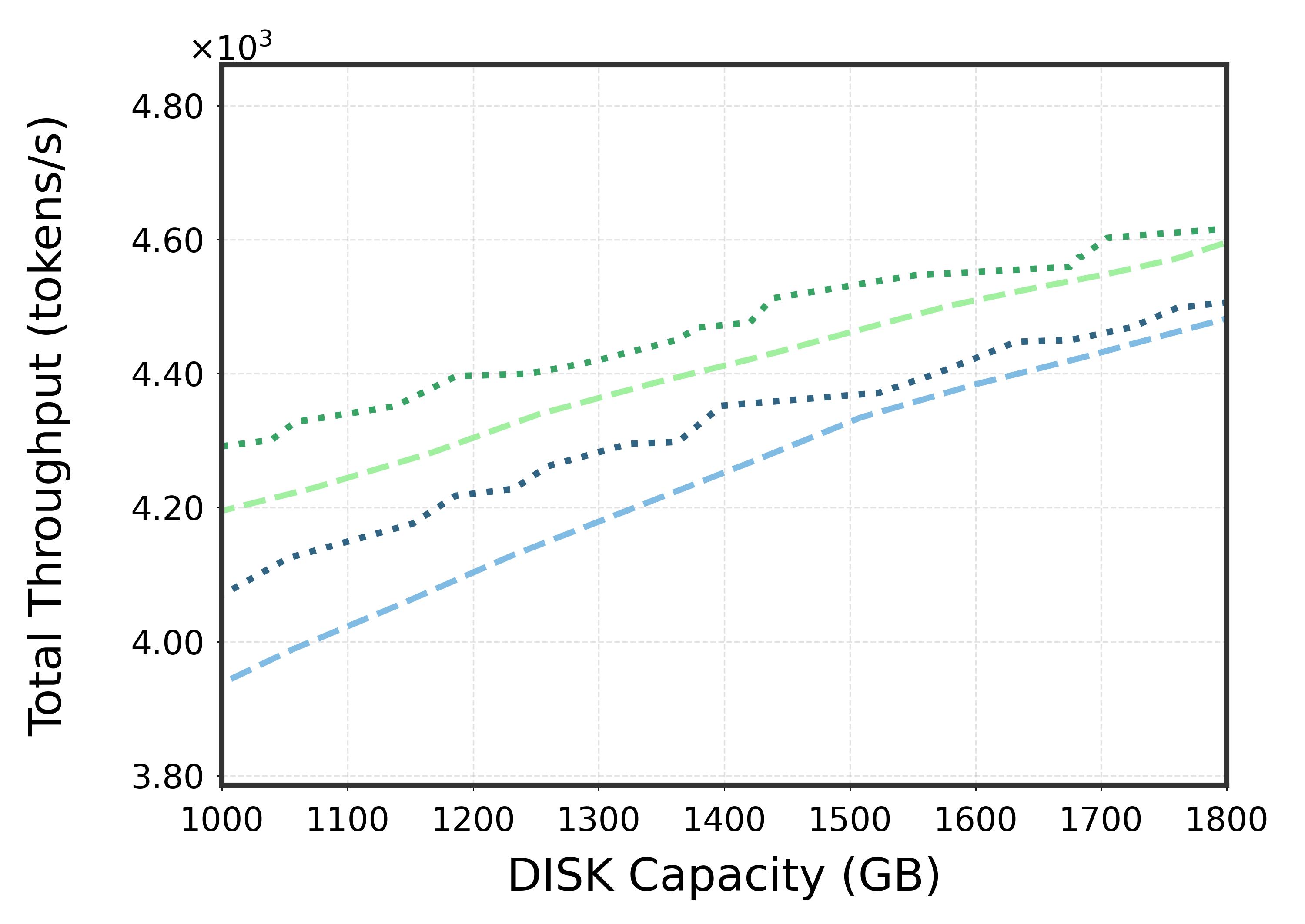}
        \end{subfigure}
        \hfill
        \begin{subfigure}[b]{0.33\textwidth}
            \centering
            \includegraphics[width=\linewidth]{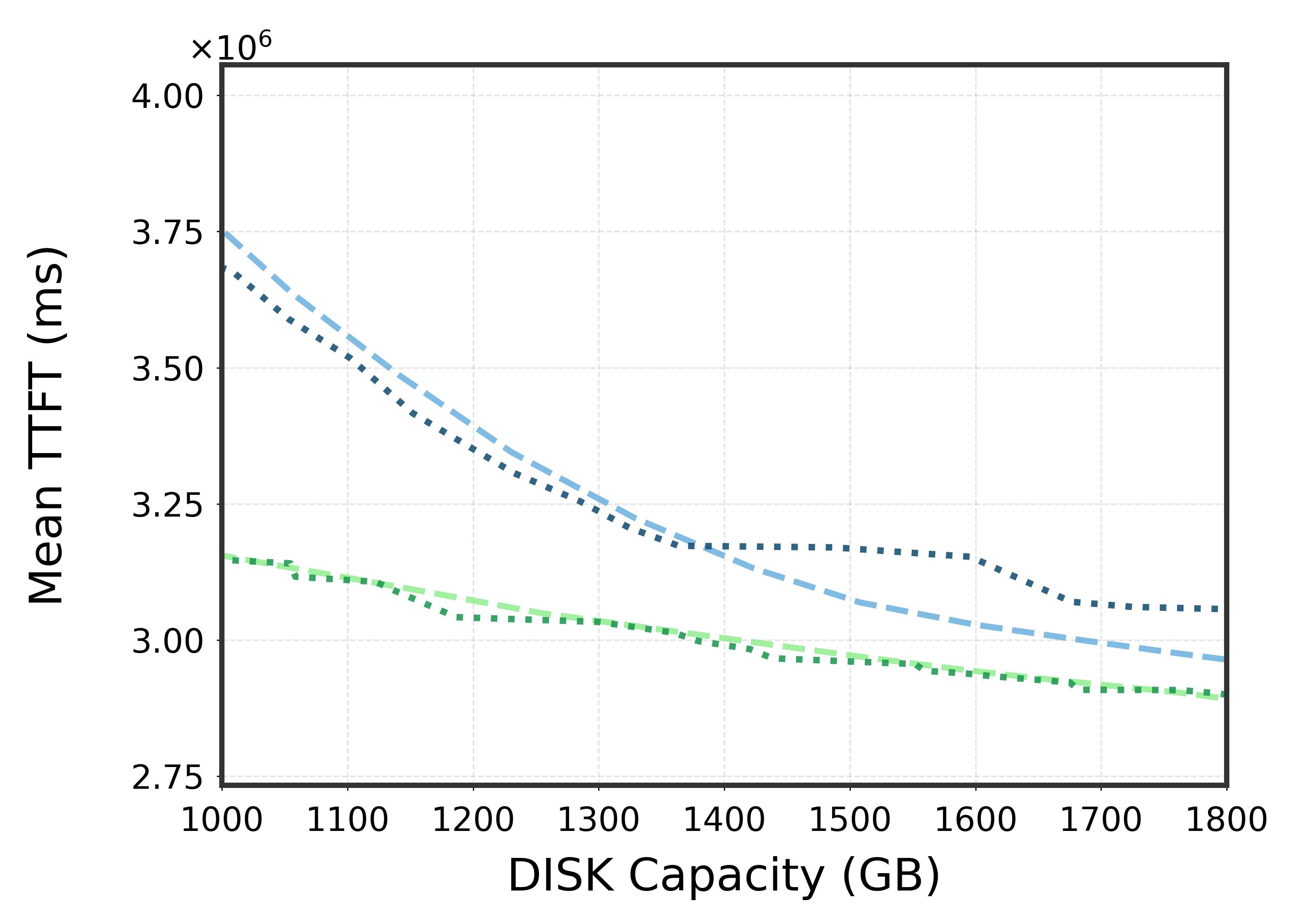}
        \end{subfigure}
        \hfill
        \begin{subfigure}[b]{0.33\textwidth}
            \centering
            \includegraphics[width=\linewidth]{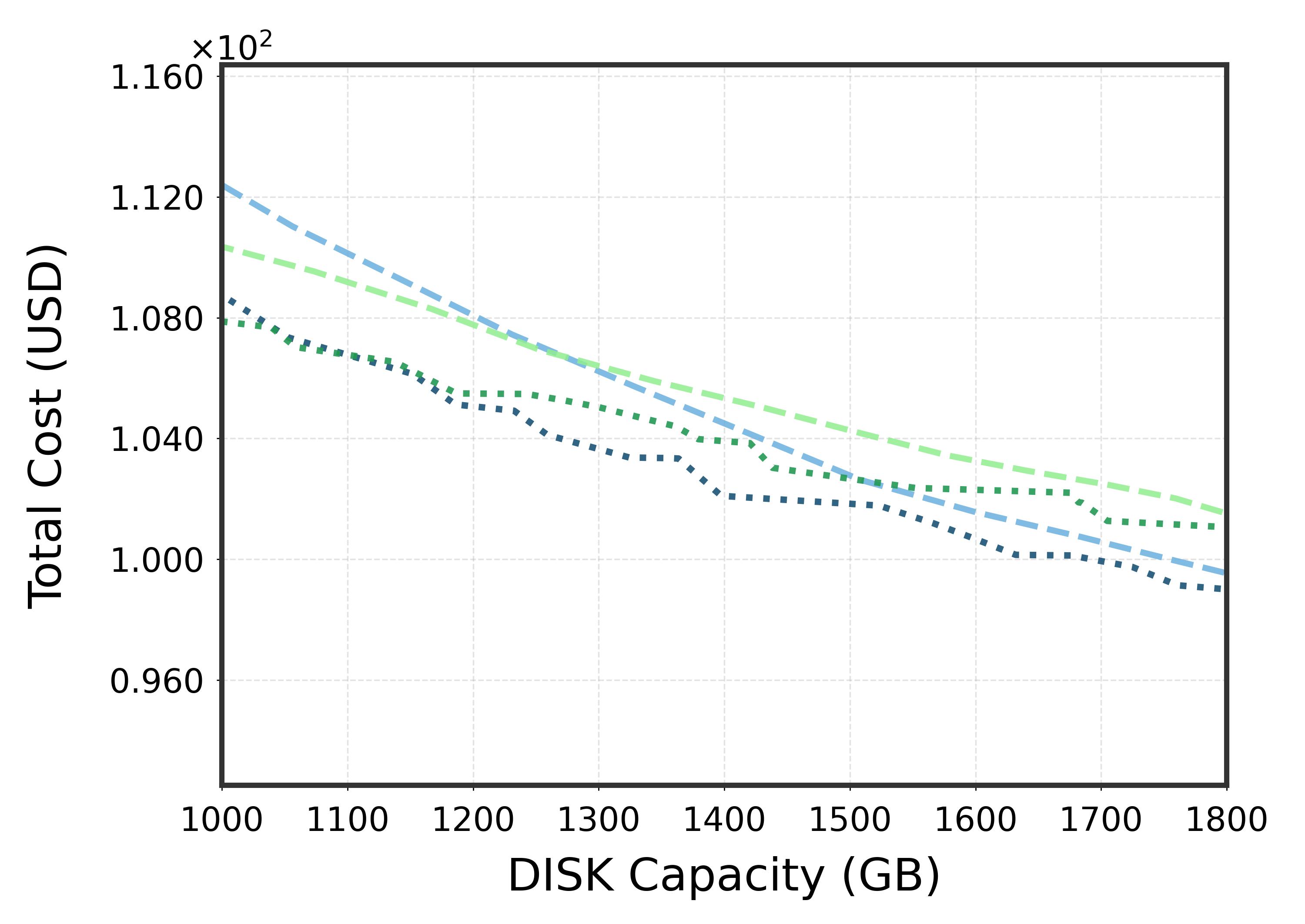}
        \end{subfigure}
        \subcaption*{\textbf{traceC ins1}}
    \end{subfigure}

    \vspace{5ex}

    \begin{subfigure}[b]{\textwidth}
        \centering
        \begin{subfigure}[b]{0.33\textwidth}
            \centering
            \includegraphics[width=\linewidth]{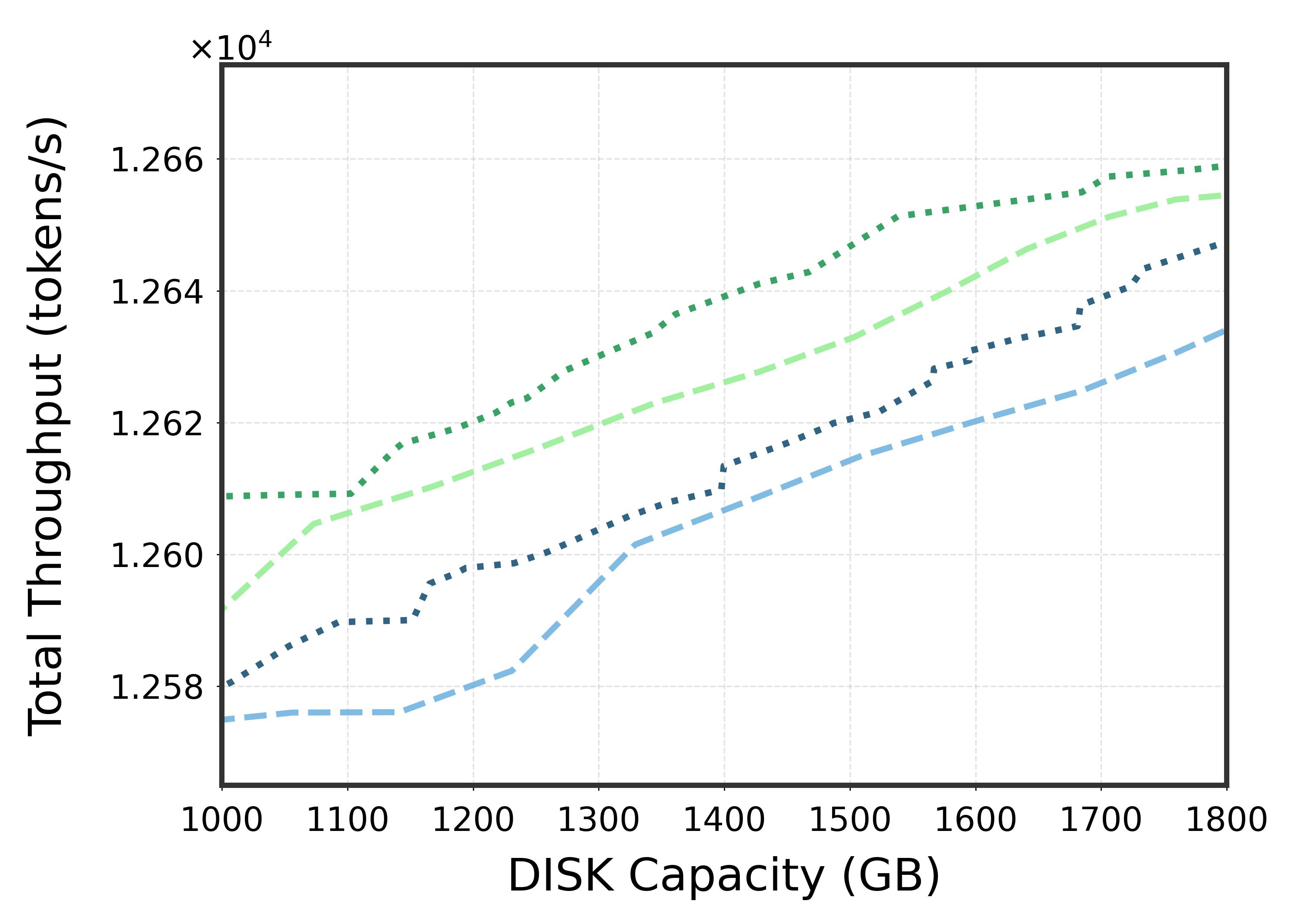}
        \end{subfigure}
        \hfill
        \begin{subfigure}[b]{0.33\textwidth}
            \centering
            \includegraphics[width=\linewidth]{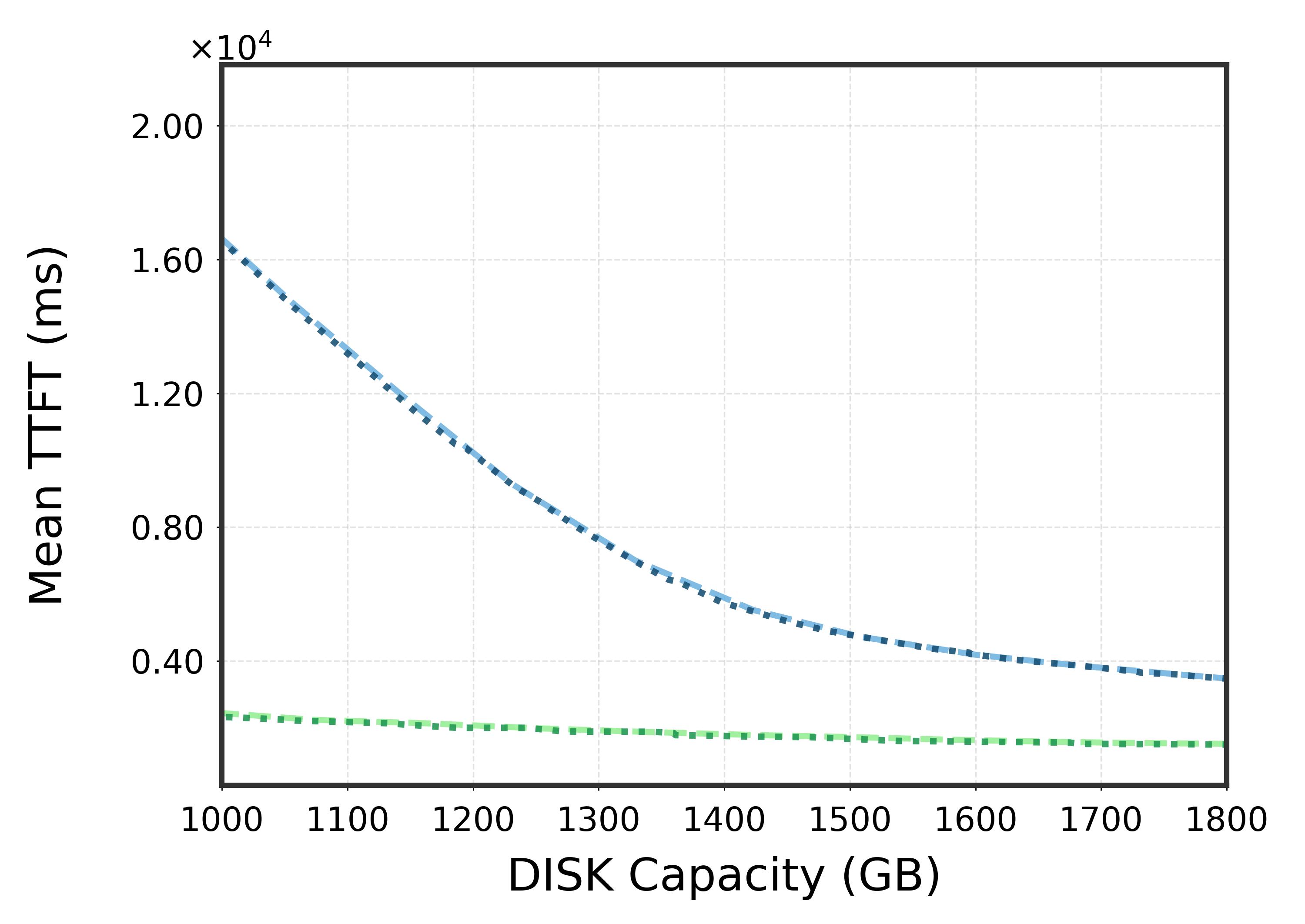}
        \end{subfigure}
        \hfill
        \begin{subfigure}[b]{0.33\textwidth}
            \centering
            \includegraphics[width=\linewidth]{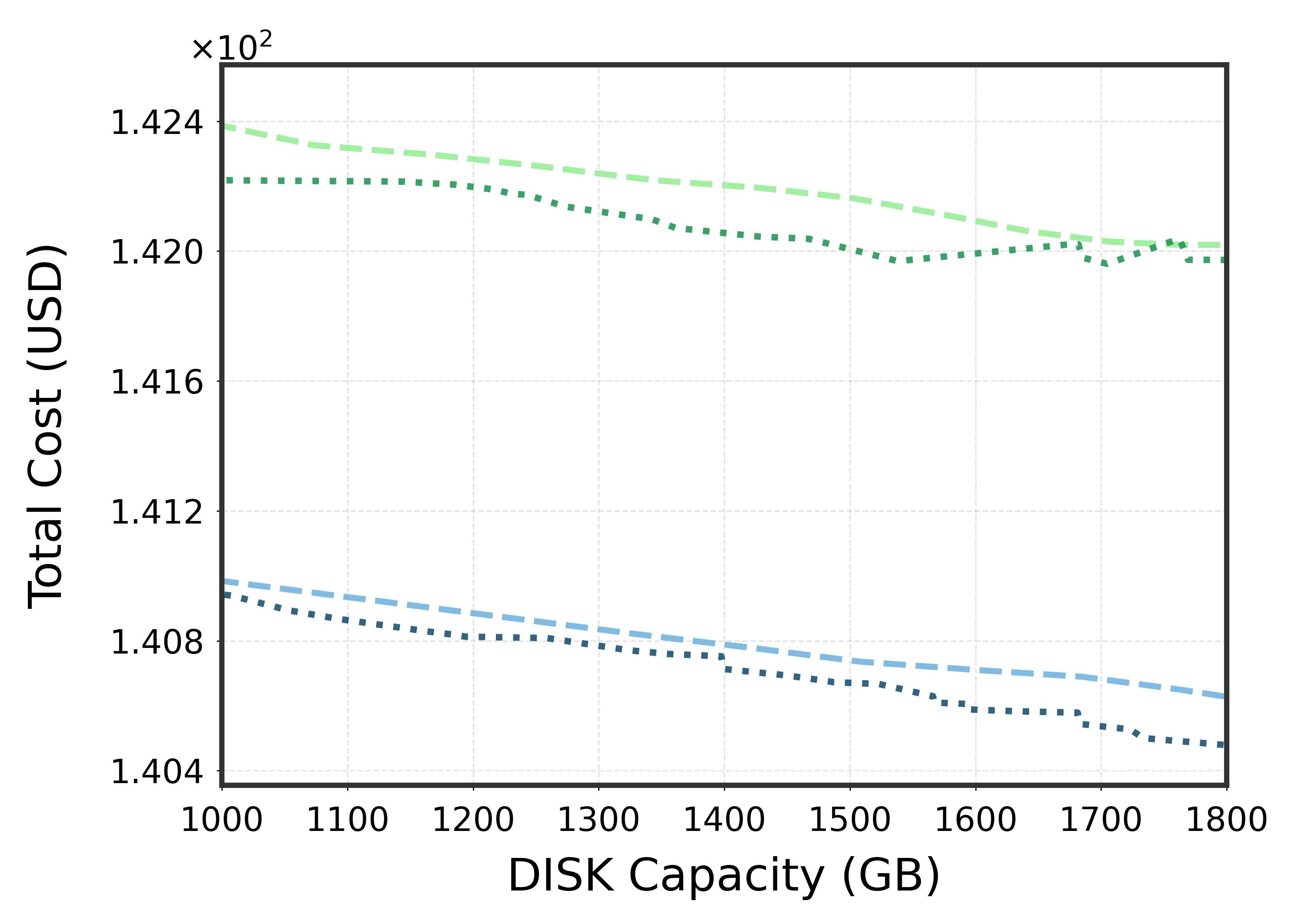}
        \end{subfigure}
        \subcaption*{\textbf{traceC ins4}}
    \end{subfigure}

    \vspace{2ex}
    
    \begin{subfigure}[b]{\textwidth}
        \centering
        \includegraphics[width=0.4\linewidth]{figs/evaluation/ablation_group_ttl/legend.jpg}
    \end{subfigure}

    \vspace{2ex}
    
    \caption{Ablation study: Comparison of group TTL and fixed TTL across throughput, TTFT latency, and cost under different traces and instance scales}
    \label{fig:ablation_group_ttl}
\end{figure*}

We compare configurations with Group TTL and with fixed TTL. Since Group TTL policy only affects retention on the disk tier, we compare how metrics vary with the average disk usage. To enable a more comprehensive evaluation, we evaluate based on two fixed DRAM capacity, 0GB and 256GB, and vary the storage budget config ($\sum_{\mathrm{block}}(Capacity_{block}\cdot TTL_{block})$) to vary the disk consumption.
The results show that, especially on trace B and C, group ttl can achieve higher reuse ratio(\autoref{fig:new_ablation_group_reuse}) and improve throughput, latency and cost(\autoref{fig:ablation_group_ttl_A} and~\autoref{fig:ablation_group_ttl}).
We also observe that group TTL yields marginal improvements under Trace A or with 4 instances. For Trace A, which is derived from an interactive chat workload, the TTL distribution is more scattered, exhibiting weak temporal

\begin{figure*}[t]
    \centering
    
    \includegraphics[width=\textwidth]{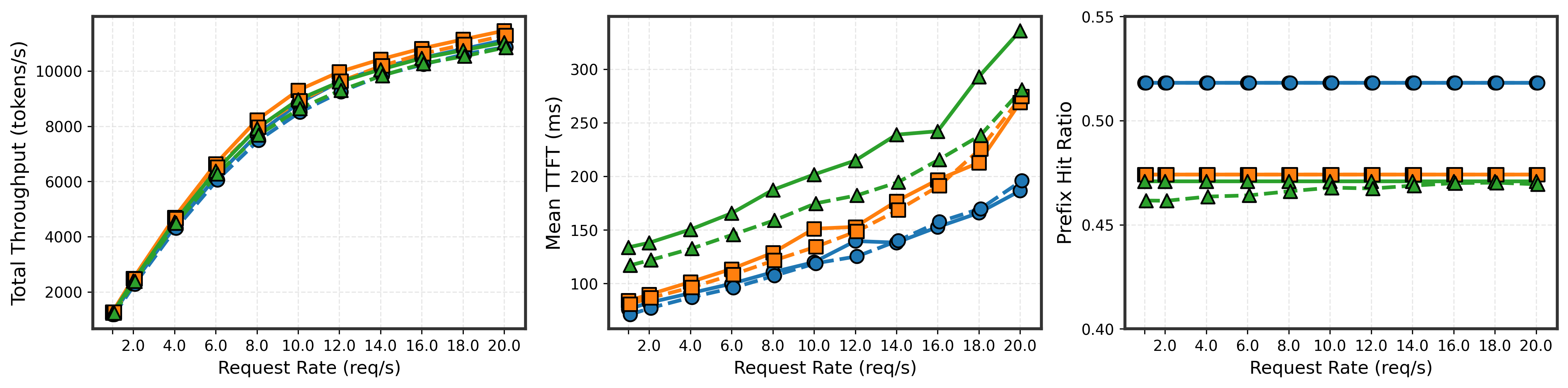}
    
    \vspace{3ex}

    \includegraphics[width=\textwidth]{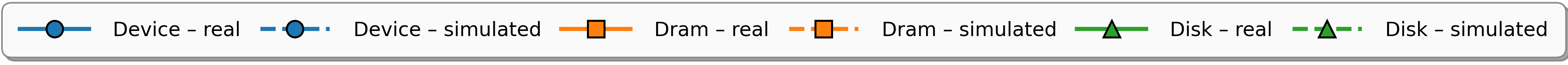}
    
    \vspace{2ex}
    
    \caption{Performance metrics under varying request rates. The legend shows real vs. simulated results for Device, DRAM, and Disk components.}
    \label{fig:performance_metrics}
\end{figure*}

\FloatBarrier
\noindent{pattern. In contrast, Trace B (API calls) and Trace C (agent-based applications) demonstrate more regular reuse interval patterns, enabling group TTL to achieve pronounced benefits. For the 4 instances set, the overall request rate is lower and queuing delays are minimal, consequently disk prefetching contributes little, diminishing the effectiveness of group TTL.}

\subsection{Simulator Accuracy}

We evaluate the accuracy of our simulator by comparing end-to-end metrics against real-system measurements, focusing on mean TTFT, total throughput, and hitrate. To isolate simulation fidelity for GPU, DRAM, and disk components, we compare three different configurations: GPU-only, GPU+host DRAM, and GPU+disk.

Our simulator achieves high fidelity across all metrics: the maximum deviations are 19.2\% for mean TTFT, 6.5\% for total throughput, and 2.0\% for hitrate as shown in~\autoref{fig:performance_metrics}. These errors are within acceptable bounds for architectural exploration and cost-performance trade-off analysis. The results confirm that our simulator provides a sufficiently accurate and efficient proxy for real-world system behavior, enabling rapid evaluation of diverse storage configurations without requiring expensive physical deployment.
\section{Conclusion}
\label{sec:Conclusion}

In this work, we present Kareto, a simulation-driven, multi-objective optimization framework for adaptive KV cache configuration in cost-effective LLM inference services. By combining a high-fidelity simulator with Pareto-based search and fine-grained, reuse-aware tuning, Kareto aims to navigate the complex trade-offs among latency, throughput, and storage cost across heterogeneous memory tiers.
Results on real-world traces suggest that Kareto can identify configurations that outperform common static baselines and do not require expert tuning. Our approach further adapts to evolving workloads through an access-pattern-aware dynamic TTL policy.
We believe that this framework offers a promising path towards autonomous and cost-performance-aware KV cache management.



\FloatBarrier

\bibliographystyle{ACM-Reference-Format}
\bibliography{citations}


\end{document}